\documentclass[aps,pra,10pt,twocolumn]{revtex4-2}%
\pdfoutput=1%
\usepackage{graphicx}%
\usepackage{amsmath, amssymb}%
\usepackage{mathphyscmd}%
\everymath{\displaystyle}%

\usepackage{hyperref}%
\hypersetup{%
	pdfinfo={%
		Author       = {David Gaspard and Arthur Goetschy},
		Title        = {Transmission eigenvalue distribution in disordered media from radiant field theory},
		CreationDate = {D:20240314182347+0100}, %
	},
	pdfborderstyle={/S/U/W 1}, %
	breaklinks=true
}

\begin{document}%
\title{Transmission eigenvalue distribution in disordered media from radiant field theory}%

\author{David Gaspard}
\email[E-mail:~]{david.gaspard@espci.psl.eu}

\author{Arthur Goetschy}
\email[E-mail:~]{arthur.goetschy@espci.psl.eu}

\affiliation{\href{https://ror.org/00kr24y60}{Institut Langevin}, \href{https://ror.org/03zx86w41}{ESPCI Paris}, \href{https://ror.org/013cjyk83}{PSL University}, \href{https://ror.org/02feahw73}{CNRS}, 75005 Paris, France}
\date{\today}

\begin{abstract}%
We develop a field-theoretic framework, called radiant field theory, to calculate the distribution of transmission eigenvalues for coherent wave propagation in disordered media.
At its core is a self-consistent transport equation for a $2\times 2$ matrix radiance, reminiscent of the radiative transfer equation but capable of capturing coherent interference effects.
This framework goes beyond the limitations of the Dorokhov-Mello-Pereyra-Kumar theory by accounting for both quasiballistic and diffusive regimes.
It also handles open geometries inaccessible to standard wave-equation solvers such as infinite slabs.
Analytical and numerical solutions are provided for these geometries, highlighting in particular the impact of the waveguide shape and the grazing modes on the transmission eigenvalue distribution in the quasiballistic regime.
By removing the macroscopic assumptions of random matrix models, this microscopic theory enables the calculation of transmission statistics in regimes previously out of reach.
It also provides a foundation for exploring more complex observables and physical effects relevant to wavefront shaping in realistic disordered systems.
\end{abstract}%
\maketitle

\section{Introduction}\label{sec:intro}

\par In 2007 it was demonstrated that it is possible to focus light behind a strongly scattering material by coherent phase manipulation of the incident wavefront \cite{Vellekoop2007}.
This coherent control technique, known today as \emph{wavefront shaping} \cite{Mosk2012, Rotter2017, Cao2022}, has led to a wide range of applications in various domains of technology including medical imaging \cite{Kubby2019}, telecommunications \cite{DiRenzo2020, Lerosey2022}, electronic microscopy \cite{YuCP2023}, and more \cite{Gigan2022a, Bliokh2023}.
The success of wavefront shaping largely relies on a powerful theoretical tool: the transmission matrix $\matr{t}$ which links the field in the input and output channels.
When the incident wavefront is matched to the conjugate of a row of this matrix, constructive interference takes place in the corresponding output channel, creating a focus point.
Alternatively, the wavefront can also be tailored to maximize the total transmitted energy \cite{Kim2012, Popoff2014}.
In this case, the achievable transmittances are given by the eigenvalues of $\herm{\matr{t}}\matr{t}$, usually referred to as transmission eigenvalues, which due to flux conservation always belong to the unit interval: $T\in[0, 1]$.
It is a famous result of quantum transport theory that, for a nonabsorbing disordered waveguide in the diffusive regime, the transmission eigenvalues are  universally distributed according to the bimodal law, $\rho(T)=\bar{T}/(2T\sqrt{1-T})$, where $\bar{T}$ is the mean transmittance \cite{Dorokhov1984, Beenakker1997}.
This distribution extends over the entire interval up to $T=1$, indicating the unintuitive existence of perfectly transmitted channels even when the disordered medium is largely opaque on average ($\bar{T}\ll 1$).

\par More generally, the eigenvectors of $\herm{\matr{t}}\matr{t}$, often called transmission eigenchannels, are transmitted with a well defined probability given by the corresponding eigenvalue.
Despite the attention that these states attracted in the literature, they currently escape theoretical description.
Indeed, since they are produced by coherent phase control, they cannot be treated by simple radiative transfer equations.
Largely for this reason, very few properties of transmission eigenchannels are known.
One emblematic feature which is not completely understood is their spatial profile in the bulk \cite{Choi2011, Davy2015b, Yilmaz2019a}.
Other unresolved issues regarding transmission eigenchannels include their spatial correlations \cite{Yilmaz2019b, Bender2020, Hong2025}, their spectral correlations \cite{Shi2015b, McIntosh2025}, or their intensity fluctuations \cite{Bender2020}.
In fact, the transmission eigenvalue distribution itself is unknown as soon as the system leaves the universality class of the bimodal law.

\par In certain circumstances, the transmission eigenvalue distribution can be described by random matrix theories.
A particularly successful theory addressing this issue is the Dorokhov-Mello-Pereyra-Kumar (DMPK) theory \cite{Dorokhov1982, Mello1988a, Beenakker1997} which is based on cutting the disordered waveguide into infinitesimal slices modeled by random scattering matrices.
The result is a Fokker-Planck-type equation, known as the DMPK equation, for the joint distribution of transmission eigenvalues.
One of its remarkable predictions is the bimodal law.
The DMPK theory has achieved great success due to the elegance of its formalism and the accuracy of its predictions, in both the diffusive and the localized regime in quasi-one dimensional systems.
A more recent example is the filtered random matrix theory \cite{Goetschy2013, Hsu2017} which captures the effect of incomplete channel control on the measured statistical properties of the scattering matrix using free probabilities \cite{Voiculescu1992, Tulino2004} and microscopic renormalization of filtering parameters \cite{Popoff2014, Hsu2015b, Hsu2017}.

\par A recurring issue with random matrix theories is the \textit{ad hoc} nature of some of their assumptions regarding the spectral or scattering properties of the system's constituents.
In the case of DMPK theory, these assumptions limit its applicability to rectilinear disordered waveguides in the multiple-scattering regime without absorption and with complete channel control.
From this point of view, the bimodal law predicted by DMPK is not as universal as often assumed.
This limitation makes it inadequate for describing realistic optical experiments where quasiballistic regime, absorption, incomplete channel control, and open geometries are common.
We will return to the assumptions behind DMPK theory in Sec.\ \ref{sec:motivations}.

\par To date, there is no microscopic theoretical framework based solely on the statistics of the disorder for the optimized states generated by wavefront shaping. %
We lay the foundations for such a framework in this article.
Instead of random matrices, our framework relies on field theory to carry out the disorder averaging, making it independent of the geometry of the disordered region, the propagation regime (quasiballistic or diffusive), and the considered observable.
Our central result is a transport equation similar to the radiative transfer equation of incoherent propagation \cite{Chandrasekhar1960} but unlike the latter able to capture the coherence of the wave due to its matrix nature.
We name our framework radiant field theory (RFT) in reference to the radiative transfer equation.
In order to validate our theory, we focus the present article on the calculation of the transmission eigenvalue distribution through a disordered waveguide and an infinite slab without absorption.
The accompanying Letter \cite{GaspardD2024-short} further extends the framework to the quasiballistic regime with absorption and incomplete channel control.
We defer the study of other properties of transmission eigenchannels, other observables, and other open geometries to future works.

\par This paper is organized as follows.
The need for a microscopic theory for transmission eigenvalues beyond DMPK theory is motivated in Sec.\ \ref{sec:motivations}.
The field theory itself is derived step by step from Secs.\ \ref{sec:model} to \ref{sec:avg-gen-fun}.
The relationship between this theory and the nonlinear sigma model is discussed in Appendix \ref{app:nonlinear-sigma}.
In Appendices \ref{app:q-smoothness} and \ref{app:convergence-and-smoothness}, it is shown that the matrix field, the main variable of the theory, slowly varies at the wavelength scale, thereby making the semiclassical approximation of Sec.\ \ref{sec:wigner} relevant.
The result of this approximation is the matrix transport equation \eqref{eq:full-eilenberger-2}, the central result of this paper.
The boundary conditions of this equation are obtained in Sec.\ \ref{sec:boundaries}.
An analytical solution in the quasiballistic regime is derived in Appendix \ref{app:quasiballistic-solution}.
Numerical results are presented in Sec.\ \ref{sec:results}.
The matrix transport equation is solved numerically in two geometries, the waveguide in Sec.\ \ref{sec:numerical-waveguide} and the infinite slab in Sec.\ \ref{sec:numerical-slab}, using the integration method of Appendices \ref{app:modal-eilenberger} and \ref{app:integration}.
Finally, conclusions are drawn in Sec.\ \ref{sec:conclusions}.

\section{Field theory for the transmission statistics}\label{sec:field-theory}

\subsection{Motivations}\label{sec:motivations}

\par Like most random matrix theories, DMPK theory relies on assumptions that govern the statistical ensembles of random matrices used to describe wave-matter interactions.
The most restrictive and least well-controlled of these is the so-called isotropy hypothesis \cite{Mello1988a, Mello1991a, Mello1992, Beenakker1997}.
According to this hypothesis, each infinitesimal slice of the disordered medium along the waveguide scatters the wave uniformly in every direction.
This assumption implies that the statistical distribution of the slice's transfer matrix is independent of its singular value decomposition (SVD) and depends only on the transmission eigenvalues.
In addition, this statistical isotropy is preserved under the stacking of slices \cite{Mello1988a}.
As a result, the random transfer matrices belong to a well-defined rotation-invariant ensemble reminiscent of the Gaussian ensembles.
The evolution of this ensemble with increasing system length is governed by a Fokker-Planck-type equation [Eq.\ (5) of Ref.\ \cite{Mello1991a}], which can be closed on the transmission eigenvalues only. This leads to the DMPK equation [Eq.\ (2) of Ref.\ \cite{Mello1991a}].

\par The issue with the isotropy hypothesis is that it does not hold in the thin-slab limit.
A straightforward perturbative calculation based on the Born approximation shows that the scattering probabilities of an infinitesimal disordered slice inevitably depend on the input and output directions, even in the case of delta-correlated disorder \cite{Mello1991a, Mello1992}.
Consequently, the actual ensemble of transfer matrices is not microscopically independent of its SVD, i.e., it is not isotropic.
This anisotropy induces statistical correlations between the transmission eigenvalues and eigenvectors which prevent the closure of the Fokker-Planck-type equation on the eigenvalues alone.
Neglecting these correlations is only justified in the diffusive regime, where the flux is approximately uniformly distributed across channels. However, in the quasiballistic regime, this approximation fails, rendering the predictions of the DMPK equation inaccurate.
A similar issue arises when absorption or amplification is introduced \cite{Misirpashaev1997, Brouwer1998a}, or when the system features obstacles, non-ideal waveguiding, or open geometries.
While several generalizations of the DMPK equation have been proposed to account for some of these effects \cite{Muttalib1999, Douglas2014, Suslov2018}, a fully satisfactory and comprehensive solution remains elusive.

\par Another important limitation of DMPK theory stems from its mathematical formulation based on stacking infinitesimal disordered slices.
As a consequence of this construction, the theory does not naturally give access to observables inside the disordered medium, such as the spatial profile of transmission eigenchannels, without substantial modifications of the formalism.

\par To address the limitations of random matrix theories, one can turn to a microscopic description that links the scattering matrix directly to the statistical properties of the disorder.
A promising framework for this is field theory.
The first field-theoretic approaches to wave propagation in random media emerged in the late 1970s with the works of Wegner and Schäfer \cite{Wegner1979, Schafer1980}, inspired by concurrent developments in coherent electron transport \cite{Edwards1975a, Thouless1975, Nitzan1977, Aharony1977}.
This framework was extended to include Anderson localization by Efetov and collaborators through the introduction of supersymmetry \cite{Efetov1982, Efetov1983a, Efetov1997}.
An interesting connection was soon identified between the field theory of disordered conductors and the quasiclassical theory of nonequilibrium superconductivity \cite{Efetov1980}, originally developed at the inception of Bardeen-Cooper-Schrieffer (BCS) theory.
Key contributions in this domain include the kinetic equations of Gorkov \cite{Gorkov1959a}, Eilenberger \cite{Eilenberger1968}, Usadel \cite{Usadel1970}, and Larkin and Ovchinnikov \cite{Larkin1968}, who analyzed how disorder, introduced by impurities, affects the superconducting coherence length in type-II superconductors.
This analogy proved fruitful: in the 1990s, Nazarov exploited it to derive the transmission eigenvalue distribution through a disordered conductor in the diffusive regime \cite{Nazarov1994a}.
His derivation of the bimodal law via field theory is an important result in mesoscopic physics and forms the foundation of his circuit theory \cite{Nazarov2009}.
In modern terms, this approach belongs to the class of nonlinear sigma models \cite{Lerner2003, Kamenev2023}, and, as such, remains confined to the diffusive regime.

\par The field-theoretical framework developed in this work is inspired by Nazarov's technique but goes beyond the diffusion approximation, making it applicable to the quasiballistic regime.
At its core is a matrix transport equation structurally similar to the Eilenberger equation of nonequilibrium superconductivity \cite{Eilenberger1968, Usadel1970, Kopnin2001}.
We derive this equation not only for rectilinear waveguide geometries but also for open systems such as an infinite disordered slab---a configuration inaccessible to direct numerical simulation of the wave equation.
Our approach thus provides, to the best of our knowledge, the only available theoretical framework capable of describing such systems.
This framework is further extended in the accompanying Letter \cite{GaspardD2024-short} to incorporate additional physical effects, including absorption and incomplete channel control.

\par Since the equations of this paper are valid in a space of arbitrary dimension, the volume and surface area of the unit ball in the space $\mathbb{R}^d$ frequently appear.
They are respectively given by
\begin{equation}\label{eq:ball-surf-vol}
V_d = \frac{\pi^{\frac{d}{2}}}{\Gamma(\frac{d}{2}+1)}  \:,\quad\text{and}\quad
S_d = dV_d = \frac{2\pi^{\frac{d}{2}}}{\Gamma(\frac{d}{2})}  \:,
\end{equation}
where $\Gamma(z)$ stands for the gamma function \cite{Olver2010}.
Furthermore, regarding the notations, we will use the hat ($\op{A}$) for operators acting in the position space and Dirac's bra-ket notations for the corresponding vectors.
In addition, $\Tr(\cdot)$ will represent the trace over a continuous basis, and $\tr(\cdot)$ the trace over a discrete basis.

\subsection{Presentation of the model}\label{sec:model}
In this section, we present the physical model studied in this paper, along with the key quantities of interest, including the transmission matrix.
The system involves a generic scalar field $\psi(\vect{r})$ obeying the stationary wave equation
\begin{equation}\label{eq:wave-equation}
\left[ \lapl_{\vect{r}} + k^2 - U(\vect{r}) \right] \psi(\vect{r}) = 0  \:,
\end{equation}
where $k$ is the wavenumber and $U(\vect{r})$ is a random potential related to the refractive index $n(\vect{r})$ by $U(\vect{r})=k^2[1-n(\vect{r})^2]$.
We assume that the random realizations of this potential are distributed according to the normal distribution:
\begin{equation}\label{eq:gaussian-disorder}
\mathcal{P}[U(\vect{r})] = \nfac \exp\left( -\int_{\mathbb{R}^d} \D\vect{r}\: \frac{U(\vect{r})^2}{2\alpha(\vect{r})} \right)  \:.
\end{equation}
The arbitrary factor $\nfac$ in Eq.\ \eqref{eq:gaussian-disorder} is defined such that the functional integral of $\mathcal{P}[U(\vect{r})]$ over $U(\vect{r})$ equals $1$.
The local variance $\alpha(\vect{r})$ in Eq.\ \eqref{eq:gaussian-disorder} is referred to as the disorder strength.
The potential $U(\vect{r})$ obeying the distribution \eqref{eq:gaussian-disorder} is characterized by a Dirac-delta autocorrelation function in space \cite{Akkermans2007},
\begin{equation}\label{eq:disorder-correlation}
\avg{U(\vect{r}) U(\vect{r}')} = \alpha(\vect{r}) \delta(\vect{r}-\vect{r}')  \:,
\end{equation}
where $\tavg{\cdots}$ stands for the average over the realizations of $U(\vect{r})$ weighted by the distribution \eqref{eq:gaussian-disorder}.
The characteristic distance between two successive scatterings is measured by the mean free path $\ell$ which is related to the disorder strength by
\begin{equation}\label{eq:def-lscat}
\frac{1}{\ell(\vect{r})} = \frac{\pi\nu}{k}\alpha(\vect{r})  \:.
\end{equation}
Since the autocorrelation length prescribed by Eq.\ \eqref{eq:disorder-correlation} is infinitely smaller than the wavelength, each scattering event caused by $U(\vect{r})$ fully randomizes the wave's direction of propagation.
This scattering isotropy is different from the DMPK isotropy hypothesis mentioned earlier because the latter corresponds to the scattering by an infinitesimal slice of disorder and not by the fluctuations of $U(\vect{r})$. %
This scattering isotropy makes both the scattering and transport mean free paths equal to $\ell$.
We will assume that the disordered region has a finite spatial extent of length $L$:
\begin{equation}\label{eq:def-disordered-region}
\alpha(\vect{r}) = \begin{cases}
\alpha  & \text{if}~x\in[0, L]  \:,\\
0       & \text{otherwise} \:.
\end{cases}
\end{equation}
According to Eq.\ \eqref{eq:def-disordered-region}, the potential $U(\vect{r})$ vanishes identically outside the disordered region.
From now on, we will omit the spatial dependency of $\ell(\vect{r})$ and $\alpha(\vect{r})$ but all following equations remain valid for any disorder profile.
In Eq.\ \eqref{eq:def-lscat}, $\nu$ is the density of states defined by
\begin{equation}\label{eq:def-dos-general}
\nu(k) \defeq \frac{1}{V} \Tr[\delta(k^2 - \op{H})]  \:,
\end{equation}
where $V$ is the quantization volume of the system, $\op{H}$ the Hamiltonian corresponding to Eq.\ \eqref{eq:wave-equation},
\begin{equation}\label{eq:def-hamiltonian}
\op{H} = \op{\vect{p}}^2 + U(\op{\vect{r}})  \:,
\end{equation}
and $\op{\vect{p}}=-\I\grad_{\vect{r}}$ the momentum operator.
We assume that the disorder strength is not too large in order to prevent the potential from significantly affecting the density of states.
This assumption is typically valid in the weak scattering regime: $k\ell\gg 1$. 
In this regime, the density of states \eqref{eq:def-dos-general} is given in fairly good precision by
\begin{equation}\label{eq:cavity-dos}
\nu(k) = \frac{1}{V} \sum_{\vect{p}} \delta(k^2 - \vect{p}^2)  \:.
\end{equation}
However, the value of $\nu(k)$ in Eq.\ \eqref{eq:cavity-dos} still relies on the geometry of the system, especially the boundary conditions.
If the wavelength is much smaller than the system size in all directions, then the density of states \eqref{eq:cavity-dos} reduces to the free-space density of states
\begin{equation}\label{eq:free-dos}
\nu_0(k) = \int_{\mathbb{R}^d} \frac{\D\vect{p}}{(2\pi)^d} \:\delta(k^2 - \vect{p}^2) = \frac{S_dk^{d-2}}{2(2\pi)^d}  \:.
\end{equation}
\par
\begin{figure}[ht]%
\includegraphics{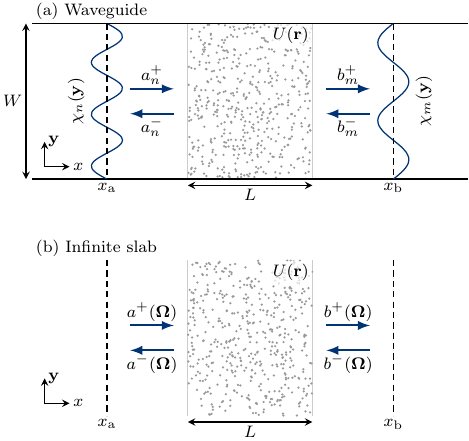}%
\caption{Schematic representation of the two geometries considered in this paper, namely the waveguide (a) and the infinite slab (b).
Both geometries contain a disordered medium of thickness $L$ characterized by a random potential $U(\vect{r})$.
The dashed lines at $x_\porta$ and $x_\portb$ represent the emission and reception surfaces in the Fisher and Lee formula \eqref{eq:fisher-lee-1}.}%
\label{fig:waveguide-geometry}%
\end{figure}

 In this paper, we assume that the wave described by Eq.\ \eqref{eq:wave-equation} propagates in a waveguide containing a disordered medium and connected to the outside by two leads, as shown in Fig.\ \ref{fig:waveguide-geometry}(a).
We also consider the limit of an infinitely wide waveguide, that we refer to as the infinite slab, depicted in Fig.\ \ref{fig:waveguide-geometry}(b).
In each lead, the field $\psi(\vect{r})$ can be decomposed on the basis of transverse eigenmodes by %
\begin{equation}\label{eq:modal-decomposition}
\psi(\vect{r}) = \sum_{n=1}^{N_{\rm p}} \left( a^+_{n} \E^{\I k_{\para,n}x} + a^-_{n} \E^{-\I k_{\para,n}x} \right) \frac{\chi_{n}(\vect{y})}{\sqrt{k_{\para,n}}}  \:.
\end{equation}
The transverse eigenmodes $\chi_{n}(\vect{y})$ in Eq.\ \eqref{eq:modal-decomposition} are defined by
\begin{equation}\label{eq:def-transverse-modes}
\left( \lapl_{\vect{y}} + \vect{p}_{\perp,n}^2 \right) \chi_{n}(\vect{y}) = 0 , \quad
\int \D\vect{y}\: \cc{\chi}_{m}(\vect{y}) \chi_{n}(\vect{y}) = \delta_{mn} .
\end{equation}
If we gather all the coefficients in the vectors $\vect{a}^\pm=(a^\pm_{1},a^\pm_{2},\ldots,a^\pm_{N_{\rm p}})$ and $\vect{b}^\pm=(b^\pm_{1},b^\pm_{2},\ldots,b^\pm_{N_{\rm p}})$, then the incident wave amplitudes are related to the outgoing wave amplitudes by the scattering matrix, or $\matr{S}$ matrix, according to
\begin{equation}\label{eq:def-s-matrix}
\begin{pmatrix}\vect{a}^-\\ \vect{b}^+\end{pmatrix} = \underbrace{\begin{pmatrix}\matr{r} & \matr{t}'\\ \matr{t} & \matr{r}'\end{pmatrix}}_{\matr{S}} \begin{pmatrix}\vect{a}^+\\ \vect{b}^-\end{pmatrix}  \:,
\end{equation}
where $\matr{t},\matr{t}'$ are transmission matrices, and $\matr{r},\matr{r}'$ reflection matrices \cite{Beenakker1997}.
\par The ability to transmit more or less signal from lead ``$\porta$'' to ``$\portb$'' is characterized by the singular values of the transmission matrix.
Indeed, if we take the singular value decomposition
\begin{equation}\label{eq:transmission-svd}
\matr{t} = \matr{V}'\,\sqrt{\matr{T}}\,\herm{\matr{V}}  \:,
\end{equation}
where $\matr{V},\matr{V}'$ are unitary matrices and $\matr{T}$ is a positive-definite diagonal matrix, then each amplitude in lead ``$\portb$'' is related to an amplitude in lead ``$\porta$'' by
\begin{equation}\label{eq:transmission-b-from-a}
\tilde{\vect{b}}^+ = \sqrt{\matr{T}}\, \tilde{\vect{a}}^+  \:,
\end{equation}
using $\tilde{\vect{b}}^+ = \herm{\matr{V}'}\vect{b}^+$ and $\tilde{\vect{a}}^+ = \herm{\matr{V}}\vect{a}^+$.
Equation \eqref{eq:transmission-b-from-a} shows that the relevant mode-per-mode transmission coefficients are the elements of $\sqrt{\matr{T}}$.
The corresponding transmittances are thus simply the elements of
\begin{equation}\label{eq:transmission-diag}
\matr{T} = \diag(T_1, T_2, \ldots, T_{N_{\rm p}})  \:,
\end{equation}
which also turn out to be the eigenvalues of the product $\herm{\matr{t}}\matr{t}$ according to Eq.\ \eqref{eq:transmission-svd}: $\herm{\matr{t}}\matr{t} = \matr{V}\,\matr{T}\,\herm{\matr{V}}$.
The values $T_1,T_2,\ldots,T_{N_{\rm p}}$ are generally known as the transmission eigenvalues \cite{Beenakker1997}. %
Since the $\matr{S}$ matrix is unitary, the transmission eigenvalues are smaller than $1$ and are thus contained in the unit interval: $T_{n}\in[0, 1]~\forall n$.
\par When considering transmission through a random medium, the $T_{n}$'s fluctuate so that it is useful to look for their average distribution defined by
\begin{equation}\label{eq:def-transmission-distrib}
\rho(T) \defeq \frac{1}{N_{\rm p}} \sum_{n=1}^{N_{\rm p}} \avg{\delta(T - T_n)} 
 = \frac{1}{N_{\rm p}} \avg{\tr\delta(T - \herm{\matr{t}}\matr{t})}  \:,
\end{equation}
and normalized according to $\textstyle\int_0^1 \D T\:\rho(T)=1$.
Finding the distribution $\rho(T)$ is the main purpose of this paper.
\par The question now is to relate the transmission matrix $\matr{t}$ to the properties of the disordered potential $U(\vect{r})$.
To this end, we resort to the Fisher and Lee formula \cite{Fisher1981}
\begin{equation}\label{eq:fisher-lee-1}
t_{mn} = 2\I\sqrt{k_{\para,m} k_{\para,n}} \, \bra{\chi_{m},x_\portb} \op{G}^+ \ket{\chi_{n},x_\porta}  \:,
\end{equation}
which provides the matrix element corresponding to the transmission from the mode $\chi_{n}$ in lead ``$\porta$'' to the mode $\chi_{m}$ in lead ``$\portb$''.
In Eq.\ \eqref{eq:fisher-lee-1}, $x_\porta$ and $x_\portb$ are the coordinates of the emission and reception surfaces, respectively, and $\op{G}^+$ is the Green's operator defined by
\begin{equation}\label{eq:def-full-green-op}
\op{G}^\pm \defeq \frac{1}{k^2 \pm\I\varepsilon - \op{H}}  \:,
\end{equation}
in terms of the Hamiltonian \eqref{eq:def-hamiltonian} of the full problem.
The symbol $\varepsilon$ in Eq.\ \eqref{eq:def-full-green-op} represents a small but nonzero positive quantity. %
The Fisher and Lee formula \eqref{eq:fisher-lee-1} can be used to express any function of the transmission matrix, $f(\herm{\matr{t}} \matr{t})$, in terms of the Green's operator \eqref{eq:def-full-green-op}:
\begin{equation}\label{eq:fisher-lee-2}
\tr f(\herm{\matr{t}} \matr{t}) = \Tr f(\op{G}^+ \op{K}_\porta \op{G}^- \op{K}_\portb)  \:.
\end{equation}
The operators $\op{K}_\porta$ and $\op{K}_\portb$ in Eq.~\eqref{eq:fisher-lee-2} are essentially projectors on the surfaces at $x_\porta$ and $x_\portb$, respectively.
According to Eq.\ \eqref{eq:fisher-lee-1}, they are given by
\begin{equation}\label{eq:contact-op-ambiguous}
\op{K}(x) = \sum_{n=1}^{N_{\rm p}} 2k_{\para,n} \ket{\chi_{n},x} \bra{\chi_{n},x}  \:,
\end{equation}
with $\op{K}_\porta=\op{K}(x_\porta)$ and $\op{K}_\portb=\op{K}(x_\portb)$.
Several important remarks should be made regarding Eqs.\ \eqref{eq:fisher-lee-2}--\eqref{eq:contact-op-ambiguous}.
\par First, the fact that Eq.\ \eqref{eq:fisher-lee-2} holds for an arbitrary function $f(x)$ implies that $\herm{\matr{t}}\matr{t}$ and $\op{G}^+\op{K}_\porta\op{G}^-\op{K}_\portb$ are related by a similarity transformation, and thus share the same eigenvalues.
This similarity is not trivial because they do not act on the same space.
While $\herm{\matr{t}}\matr{t}$ acts on the modal space defined on the surfaces at $x_\porta$ and $x_\portb$, the operator $\op{G}^+\op{K}_\porta\op{G}^-\op{K}_\portb$ acts on the whole volume of the waveguide.
The surface to volume extension provided by Eq.\ \eqref{eq:fisher-lee-2} is a key ingredient of our field-theoretic approach since it provides a way to relate the sought transmission statistics to the characteristics of the random potential $U(\vect{r})$.
\par Second, we notice that the operator $\op{K}(x)$ in Eq.\ \eqref{eq:contact-op-ambiguous} is very close to a current operator, but not quite because it is positive definite ($k_{\para,n}>0$) in contradiction with the possible existence of negative currents. %
In order to restore the current interpretation of $\op{K}(x)$, it is possible to redefine it as
\begin{equation}\label{eq:contact-op-current}
\op{K}(x) = \op{p}_{\para} \delta(\op{x}-x) + \delta(\op{x}-x) \op{p}_{\para}  \:.
\end{equation}
This redefinition coincides with Eq.\ \eqref{eq:contact-op-ambiguous} when inserted into Eq.\ \eqref{eq:fisher-lee-2} if $x_\porta\rightarrow -\infty$ and $x_\portb\rightarrow +\infty$.
However, the advantage of Eq.\ \eqref{eq:contact-op-current} over \eqref{eq:contact-op-ambiguous} is that the current is conserved in the absence of absorption ($\partial_{x}\op{K}(x)=0$).
Therefore, the positions $x_\porta$ and $x_\portb$ can be moved to any place, including in the disordered region, without altering the spectrum of $\herm{\matr{t}}\matr{t}$.
The latter will always correspond to the transmission from $x\rightarrow -\infty$ to $x\rightarrow +\infty$.
On the other hand, the pseudo-current \eqref{eq:contact-op-ambiguous} makes the spectrum of $\herm{\matr{t}}\matr{t}$ dependent on $x_\porta$ and $x_\portb$ when they are located in the disordered region.
This leads to the concept of deposition matrix \cite{Bender2022a, McIntosh2024, McIntosh2025} which is beyond the scope of the present paper and will be addressed in a future work.

\subsection{Generating function and duplicated space}\label{sec:duplication}
In this section, we construct a statistical field theory for the transmission eigenvalue distribution \eqref{eq:def-transmission-distrib} inspired by the work of Nazarov \cite{Nazarov1994a, Nazarov2009}.

\par Since the transmission matrix $\matr{t}$ is inversely proportional to the Gaussian random potential $U(\op{\vect{r}})$, it is more convenient to consider the eigenvalues of $(\herm{\matr{t}}\matr{t})^{-1}$ rather than those of $\herm{\matr{t}}\matr{t}$, although they are trivially related.
We thus define the generating function 
\begin{equation}\label{eq:def-gen-fun}
F(\gamma) \defeq \frac{1}{N_{\rm p}} \avg{ \tr\left( \frac{1}{(\herm{\matr{t}}\matr{t})^{-1} - \gamma} \right) } \:,
\end{equation}
which depends on the variable $\gamma\in\mathbb{C}$.
According to the property $\textstyle\Im(\frac{1}{x-\I 0^+})=\pi\delta(x)$, the sought transmission eigenvalue density \eqref{eq:def-transmission-distrib} can be extracted from the imaginary part of the generating function \eqref{eq:def-gen-fun} up to the change of variable $\gamma=T^{-1}$:
\begin{equation}\label{eq:rho-from-gen-fun}
\rho(T) = \frac{1}{\pi T^2} \Im F(\tfrac{1}{T} + \I 0^+)  \:.
\end{equation}
\par The generating function \eqref{eq:def-gen-fun} can also be derived from the function
\begin{equation}\label{eq:partition-from-transmission}
Z(\gamma) \defeq \nfac \det(1 - \gamma \herm{\matr{t}}\matr{t})^{-1}  \:,
\end{equation}
using
\begin{equation}\label{eq:gen-fun-from-partition}
F(\gamma) = \frac{1}{N_{\rm p}} \der{}{\gamma} \avg{\ln Z(\gamma)}  \:,
\end{equation}
and the determinantal relation $\ln\det(\matr{A})=\tr\ln(\matr{A})$.
The symbol $\nfac$ in Eq.\ \eqref{eq:partition-from-transmission} is a $\gamma$-independent immaterial prefactor which is eliminated anyway by the logarithmic derivative.
The relation \eqref{eq:fisher-lee-2} implies that $\herm{\matr{t}}\matr{t}$ can be replaced by $\op{G}^+ \op{K}_\porta \op{G}^- \op{K}_\portb$ under a trace or a determinant. Equation \eqref{eq:partition-from-transmission} thus becomes
\begin{equation}\label{eq:partition-from-det-1}
Z(\gamma) = \nfac \det(1 - \gamma \op{G}^+ \op{K}_\porta \op{G}^- \op{K}_\portb)^{-1}  \:.
\end{equation}
In order to split the product of operators in the determinant \eqref{eq:partition-from-det-1}, we resort to the duplication technique proposed by Nazarov \cite{Nazarov1994a} and other authors before \cite{Efetov1982, Efetov1983a, Efetov1997, Lerner2003}.
We notice that the determinant \eqref{eq:partition-from-det-1} closely resembles the determinant of a two-by-two block matrix:
\begin{equation}\label{eq:det-two-by-two}
\det\begin{pmatrix}\op{A} & \op{B}\\ \op{C} & \op{D}\end{pmatrix} = \det(\op{A}\op{D}) \det(1 - \op{A}^{-1}\op{B}\op{D}^{-1}\op{C})  \:.
\end{equation}
The correspondence between Eqs.\ \eqref{eq:partition-from-det-1} and \eqref{eq:det-two-by-two} suggests to define the Hamiltonian-type matrix
\begin{equation}\label{eq:def-matrix-w}
\op{\matr{W}} \defeq \begin{pmatrix}k^2 + \I\varepsilon - \op{H} & \gamma_\porta\op{K}_\porta\\ \gamma_\portb\op{K}_\portb & k^2 - \I\varepsilon - \op{H}\end{pmatrix}  \:,
\end{equation}
such that the function $Z(\gamma)$ is given by
\begin{equation}\label{eq:partition-from-det-2}
Z(\gamma) = \nfac \det(\op{\matr{W}})^{-1}  \:.
\end{equation}
The remaining factor $\det(\op{G}^+\op{G}^-)$ in the right-hand side of Eq.\ \eqref{eq:det-two-by-two} does not depend on $\gamma$ and can thus be absorbed in $\nfac$.
The parameters $\gamma_\porta$ and $\gamma_\portb$ associated with the contact interactions in Eq.\ \eqref{eq:def-matrix-w} must obey the relation
\begin{equation}\label{eq:gamma-product}
\gamma = \gamma_\porta\gamma_\portb  \:.
\end{equation}
They can be equal ($\gamma_\porta=\gamma_\portb$) but do not have to. %
We will see the consequences of this degree of freedom in Sec.\ \ref{sec:avg-gen-fun}.
Last but not least, the determinant in Eq.\ \eqref{eq:partition-from-det-2} can be expressed as a Gaussian functional integral
\begin{equation}\label{eq:partition-integral}
Z(\gamma) = \int \fD{\vect{\phi}(\vect{r})} \E^{ \I \int \D\vect{r}\: \herm{\vect{\phi}}(\vect{r}) \op{\matr{W}} \vect{\phi}(\vect{r}) }  \:,
\end{equation}
where $\vect{\phi}(\vect{r})=\tran{[\phi_1(\vect{r}), \phi_2(\vect{r})]}$ is a two-component complex field.
By analogy with statistical field theory, $Z(\gamma)$ will be referred to as the partition function.

\subsection{Disorder averaging and the replica method}\label{sec:disorder-averaging}
In this section, we average the transmission eigenvalue distribution over the random potential $U(\vect{r})$.
According to Eqs.\ \eqref{eq:rho-from-gen-fun} and \eqref{eq:gen-fun-from-partition}, the calculation of $\rho(T)$ boils down to the calculation of $\avg{\ln Z}$.
However, due to the presence of the logarithm, this calculation is not straightforward.
To overcome this issue, we exploit the replica method which is based on the formula:
\begin{equation}\label{eq:replica-trick}
\avg{\ln Z} = \lim_{R\rightarrow 0} \pder{\avg{Z^R}}{R}  \:.
\end{equation}
The point is that $\tavg{Z^R}$ can be evaluated by replicating the fields $R$ times, hence the name of this method.
Using Eq.\ \eqref{eq:partition-integral} and averaging over the potential distribution \eqref{eq:gaussian-disorder}, we get
\begin{equation}\label{eq:zr-1}
\avg{Z^R} = \int \fD{U} \E^{-\int \D\vect{r} \frac{U(\vect{r})^2}{2\alpha(\vect{r})}} \int \fD{\vect{\Phi}} \E^{\I \int \D\vect{r}\: \herm{\vect{\Phi}}(\vect{r})\op{\matr{W}}\vect{\Phi}(\vect{r})}  \:,
\end{equation}
where $\vect{\Phi}(\vect{r})=\tran{[\vect{\phi}_1(\vect{r}), \vect{\phi}_2(\vect{r}), \ldots, \vect{\phi}_R(\vect{r})]}$ is the replicated field, and $\op{\matr{W}}$ acts identically on all replicas.
The normalization factor $\nfac$ introduced in Eq.\ \eqref{eq:gaussian-disorder} has been absorbed into the differential element $\fD{U}$ in \eqref{eq:zr-1}.
Henceforth, all $\gamma$-independent prefactors will be systematically omitted from the expression of $\avg{Z^R}$, because they will be eliminated by the logarithmic derivative \eqref{eq:gen-fun-from-partition}, anyway.
In order to achieve the integral over the disorder $U(\vect{r})$ in Eq.\ \eqref{eq:zr-1}, we have to separate the potential term from the Hamiltonian $\op{\matr{W}}$.
According to Eq.\ \eqref{eq:def-matrix-w}, the latter can be expressed as
\begin{equation}\label{eq:matrix-w-expanded}
\op{\matr{W}} = \op{\matr{W}}_0 - U(\op{\vect{r}}) \matr{1}_{2R}  \:,
\end{equation}
where $\op{\matr{W}}_0$ contains all other terms independent of the disorder:
\begin{equation}\label{eq:def-matrix-w-zero}
\op{\matr{W}}_0 \defeq \begin{pmatrix}k^2 - \op{\vect{p}}^2 + \I\varepsilon & \gamma_\porta\op{K}_\porta\\ \gamma_\portb\op{K}_\portb & k^2 - \op{\vect{p}}^2 - \I\varepsilon \end{pmatrix}  \:.
\end{equation}
Just like $\op{\matr{W}}$, the Hamiltonian $\op{\matr{W}}_0$ is proportional to the identity in the replica space.
In the following equations, we omit identity matrices for brevity.
Substituting Eq.\ \eqref{eq:matrix-w-expanded} into Eq.\ \eqref{eq:zr-1} leads to
\begin{equation}\label{eq:zr-2}
\avg{Z^R} = \int \fD{U} \fD{\vect{\Phi}} \E^{\int \D\vect{r} \left( -\frac{U^2}{2\alpha} + \I \herm{\vect{\Phi}} (\op{\matr{W}}_0 - U) \vect{\Phi} \right)}  \:,
\end{equation}
which is a Gaussian functional integral over the potential $U(\vect{r})$.
It is thus possible to evaluate it by completing the square as follows:
\begin{equation}\label{eq:square-completion}
-\frac{U^2}{2\alpha} - \I U\herm{\vect{\Phi}}\vect{\Phi} = -\frac{(U + \I\alpha\herm{\vect{\Phi}}\vect{\Phi})^2}{2\alpha} - \frac{\alpha}{2} (\herm{\vect{\Phi}}\vect{\Phi})^2  \:.
\end{equation}
Substituting Eq.\ \eqref{eq:square-completion} into Eq.\ \eqref{eq:zr-2} and integrating out $U(\vect{r})$ yields
\begin{equation}\label{eq:zr-3}
\avg{Z^R} = \int \fD{\vect{\Phi}} \E^{\int \D\vect{r} \left( \I \herm{\vect{\Phi}} \op{\matr{W}}_0 \vect{\Phi} - \frac{\alpha}{2} (\herm{\vect{\Phi}} \vect{\Phi})^2 \right)}  \:.
\end{equation}
The occurrence of a $\vect{\Phi}^4$ term in Eq.\ \eqref{eq:zr-3} suggests a first analogy with the superconductivity literature, where such terms model electron-electron interactions.
\par In order to get rid of the $\vect{\Phi}^4$ term in the Lagrangian of Eq.\ \eqref{eq:zr-3}, we perform a Hubbard-Stratonovich transformation \cite{Efetov1997, Kamenev2023}.
This transformation is based on the Gaussian functional integral over a $2R\times 2R$ complex matrix field $\matr{Q}(\vect{r})$:
\begin{equation}\label{eq:hubbard-strato-1}
\int \fD{\matr{Q}} \E^{\int \D\vect{r} \frac{\alpha}{2} \tr\left( \left( \matr{Q}(\vect{r}) - \vect{\Phi}(\vect{r})\herm{\vect{\Phi}}(\vect{r}) \right)^2 \right)} = 1  \:.
\end{equation}
The matrix $\matr{Q}(\vect{r})$ is not necessarily Hermitian, but the contour integral over the matrix elements of $\matr{Q}$ is suppose to follow the steepest descent so that there is no convergence issue.
Expanding the square in Eq.\ \eqref{eq:hubbard-strato-1} and isolating the $\vect{\Phi}^4$ term provides
\begin{equation}\label{eq:hubbard-strato-2}
\E^{-\int \D\vect{r} \frac{\alpha}{2} (\herm{\vect{\Phi}} \vect{\Phi})^2}
 = \int \fD{\matr{Q}} \E^{ \int \D\vect{r} \left( -\alpha\herm{\vect{\Phi}}\matr{Q}\vect{\Phi} + \frac{\alpha}{2} \tr[\matr{Q}(\vect{r})^2] \right) }  \:.
\end{equation}
Then, Eq.\ \eqref{eq:hubbard-strato-2} can be used to replace the $\vect{\Phi}^4$ term in Eq.\ \eqref{eq:zr-3} by two non-quartic terms:
\begin{equation}\label{eq:zr-4}
\avg{Z^R} = \int \fD{\matr{Q}} \fD{\vect{\Phi}} \E^{\int \D\vect{r} \left( \I \herm{\vect{\Phi}} (\op{\matr{W}}_0 + \I\alpha\matr{Q}) \vect{\Phi} + \frac{\alpha}{2} \tr[\matr{Q}(\vect{r})^2] \right)}  \:.
\end{equation}
Physically, the Hubbard-Stratonovich transformation resembles a mean-field approximation, with the matrix $\matr{Q}(\vect{r})\simeq\tavg{\vect{\Phi}(\vect{r})\herm{\vect{\Phi}}(\vect{r})}$ playing the role of the mean field.
However, unlike a mean-field approximation, the Hubbard-Stratonovich transformation is exact.
Given the integral over $\vect{\Phi}$ in Eq.\ \eqref{eq:zr-4} is now Gaussian, it can be evaluated exactly in terms of a determinant:
\begin{equation}\label{eq:zr-5}
\avg{Z^R} = \int \fD{\matr{Q}} \det(\op{\matr{W}}_0 + \I\alpha\op{\matr{Q}})^{-1} \E^{\int \D\vect{r} \frac{\alpha}{2} \tr[\matr{Q}(\vect{r})^2]}  \:.
\end{equation}
This determinant can also be expressed by a logarithm using the general property
\begin{equation}\label{eq:det-operator}
\det(\op{\matr{A}}) = \E^{\Tr\ln(\op{\matr{A}})} = \E^{\int \D\vect{r} \bra{\vect{r}} \tr\ln(\op{\matr{A}}) \ket{\vect{r}}}  \:.
\end{equation}
The result is
\begin{equation}\label{eq:zr-6}
\avg{Z^R} = \int \fD{\matr{Q}} \E^{\mathcal{L}^{(R)}[\matr{Q}]}  \:,
\end{equation}
with the Lagrangian
\begin{equation}\label{eq:full-lagrangian}\begin{split}
& \mathcal{L}^{(R)}[\matr{Q}] \defeq \\ 
& \int \D\vect{r} \left( \frac{\alpha}{2} \tr[\matr{Q}(\vect{r})^2] - \bra{\vect{r}} \tr\ln(\op{\matr{W}}_0 + \I\alpha\op{\matr{Q}}) \ket{\vect{r}} \right)  \:.
\end{split}\end{equation}

\subsection{Single-replica approximation in the nonlocalized regime}\label{sec:single-replica}
In this section, we approximate the exact Lagrangian \eqref{eq:full-lagrangian} in the nonlocalized regime.
To this end, we assume that the field $\matr{Q}(\vect{r})$ can be diagonalized in the replica space by a position-independent similarity transformation $\matr{Q}\rightarrow\matr{U}\matr{Q}\matr{U}^{-1}$.
This assumption only holds if the replica fields $\matr{Q}_{rr'}(\vect{r})$ for $r,r'\in\{1,\ldots,R\}$ are linear combinations of at most $R$ fields.
In this way, the partition function \eqref{eq:zr-6} can be written as
\begin{equation}\label{eq:zr-7}\begin{split}
& \avg{Z^R} =  \\
& \int \fD{\matr{Q}_1}\cdots\fD{\matr{Q}_R} \abs{\Delta(\matr{Q}_1, \ldots, \matr{Q}_R)} \E^{ \sum_{r=1}^{R} \mathcal{L}^{(1)}[\matr{Q}_{r}] }  \:.
\end{split}\end{equation}
In Eq.\ \eqref{eq:zr-7}, $\matr{Q}_1(\vect{r}), \ldots, \matr{Q}_R(\vect{r})$ are the $2\times 2$ matrix fields from individual replicas, $\Delta(\matr{Q}_1, \ldots, \matr{Q}_R)$ is a Jacobian involving interactions between replicas, and $\mathcal{L}^{(1)}$ is the Lagrangian \eqref{eq:full-lagrangian} corresponding to a single replica ($R=1$).
We want to show that these single-replica Lagrangians $\mathcal{L}^{(1)}$ dominate over the contribution of the Jacobian in the nonlocalized regime. To do this,
we momentarily focus on the multiple-scattering regime ($L\gg\ell$), where $\mathcal{L}^{(1)}$ can be reduced to the nonlinear sigma model Lagrangian (see Appendix \ref{app:nonlinear-sigma} for the derivation),
\begin{equation}\label{eq:nlsigma-approx-1}
\mathcal{L}^{(1)}[\matr{Q}] \simeq \int \D\vect{r}\, \pi\nu \tr\left( \frac{D}{2} (\grad_{\vect{r}}\tilde{\matr{Q}})^2 + \cdots \right)  \:,
\end{equation}
where the trailing dots represent smaller terms.
Since our goal is only to reveal the order of magnitude of $\mathcal{L}^{(1)}$, the present reasoning remains general.
The matrix field $\tilde{\matr{Q}}(\vect{r})$ in Eq.\ \eqref{eq:nlsigma-approx-1} obeys the normalization constraint $\tilde{\matr{Q}}(\vect{r})^2=\matr{1}_{2}$, where $\matr{1}_{2}$ is the $2\times 2$ identity matrix, and $D$ is the (dimensionless) diffusivity
\begin{equation}\label{eq:def-diffusivity}
D \defeq \frac{k\ell}{d}  \:.
\end{equation}
If we restrict ourselves to a waveguide geometry free of transverse localization ($W\ll N_{\rm p}\ell$) and use the normalized coordinate $\tilde{x}=x/L$, then Eq.\ \eqref{eq:nlsigma-approx-1} becomes
\begin{equation}\label{eq:nlsigma-approx-2}
\mathcal{L}^{(1)}[\matr{Q}] \simeq A \int_{0}^{1} \D\tilde{x}\, \tr\left( \frac{1}{4} (\partial_{\tilde{x}}\tilde{\matr{Q}})^2 + \cdots \right)  \:,
\end{equation}
where the factor $A$, proportional to the dimensionless conductance, reads
\begin{equation}\label{eq:nlsigma-large-factor}
A = \frac{2\pi\nu D S_\perp}{L} = \frac{V_d}{2V_{d-1}} \frac{N_{\rm p}\ell}{L}  \:,
\end{equation}
$S_\perp$ being the surface area of the waveguide cross section.
The second expression in Eq.\ \eqref{eq:nlsigma-large-factor} is based on Eqs.\ \eqref{eq:free-dos}, \eqref{eq:def-diffusivity}, and the number of propagating modes
\begin{equation}\label{eq:number-of-modes}
N_{\rm p} = \frac{V_{d-1}k^{d-1}S_\perp}{(2\pi)^{d-1}}  \:.
\end{equation}
In the absence of strong localization in the longitudinal direction, that is when $L\ll N_{\rm p}\ell$, the factor $A$ in Eq.\ \eqref{eq:nlsigma-approx-2} is much larger than $1$.
Therefore, in this regime, we expect that the single-replica Lagrangians in Eq.\ \eqref{eq:zr-7} dominate the contribution of the Jacobian, so that the interaction between replicas can be neglected:
\begin{equation}\label{eq:zr-8}
\avg{Z^R} \simeq \int \fD{\matr{Q}_1}\cdots\fD{\matr{Q}_R} \E^{ \sum_{r=1}^{R} \mathcal{L}^{(1)}[\matr{Q}_{r}] }  \:.
\end{equation}
We can thus write
\begin{equation}\label{eq:zr-8-bis}
\avg{Z^R} \simeq \left( \int \fD{\matr{Q}} \E^{ \mathcal{L}^{(1)}[\matr{Q}] } \right)^R  \:.
\end{equation}
Then, in the limit $A\gg 1$, the integral \eqref{eq:zr-8-bis} can be approximated by the contribution at the saddle point
\begin{equation}\label{eq:zr-9}
\avg{Z^R} \simeq \E^{ R\mathcal{L}^{(1)}[\matr{Q}_\saddle] }  \:,
\end{equation}
where $\matr{Q}_\saddle(\vect{r})$ is the single-replica $2\times 2$ matrix field at the saddle point defined by the equation
\begin{equation}\label{eq:def-saddle-point}
\fder{\mathcal{L}^{(1)}}{\matr{Q}(\vect{r})}[\matr{Q}_\saddle] = \matr{0}  \:.
\end{equation}
Substituting Eq.\ \eqref{eq:zr-9} into the replica limit \eqref{eq:replica-trick} yields
\begin{equation}\label{eq:logz-approx}
\avg{\ln Z} \simeq \mathcal{L}^{(1)}[\matr{Q}_\saddle]  \:.
\end{equation}
Therefore, we have just shown that in the nonlocalized regime the full problem reduces to a single replica. %
It is clear from the above calculation that, in the strongly localized regime (i.e. for $L\gg N_{\rm p}\ell$), the approximation \eqref{eq:zr-8} does not hold anymore since $A$ is then small, and the interaction between replicas cannot be neglected.
The emergence of strong Anderson localization in the replica formalism thus requires special attention (see Ref.\ \cite{Altland2005}), but this issue is not addressed in the present paper.
In the following sections, we will drop the subscript ``$\saddle$'' referring to the saddle-point solution.

\subsection{Saddle-point equation}\label{sec:saddle-point}
According to Eqs.\ \eqref{eq:full-lagrangian} and \eqref{eq:def-saddle-point}, the saddle point equation reads
\begin{equation}\label{eq:saddle-point-1}
\matr{Q}(\vect{r}) = \bra{\vect{r}} \frac{\I}{\op{\matr{W}}_0 + \I\alpha\matr{Q}(\op{\vect{r}})} \ket{\vect{r}}  \:.
\end{equation}
Equation \eqref{eq:saddle-point-1} means that the field $\matr{Q}(\vect{r})$ can be interpreted as the return probability at $\vect{r}=\vect{r}'$ of a matrix Green's function $\matr{\Gamma}(\vect{r}, \vect{r}')=\bra{\vect{r}} \op{\matr{\Gamma}} \ket{\vect{r}'}$ formally defined by
\begin{equation}\label{eq:def-gamma}
\op{\matr{\Gamma}} \defeq \frac{\I}{\op{\matr{W}}_0 + \I\alpha\matr{Q}(\op{\vect{r}})}  \:.
\end{equation}
Using this Green's function and Eq.\ \eqref{eq:def-matrix-w-zero}, the saddle-point equation \eqref{eq:saddle-point-1} can be represented in position space by
\begin{equation}\label{eq:saddle-point-2}\begin{split}
\Big[ & \lapl_{\vect{r}} + k^2 + \I\varepsilon\lmat_3 + \I\alpha\matr{Q}(\vect{r})  \\
 & + \gamma_\porta\op{K}_\porta\lmat_+ + \gamma_\portb\op{K}_\portb\lmat_- \Big] \matr{\Gamma}(\vect{r}, \vect{r}') 
 = \I\matr{1}_2\delta(\vect{r}-\vect{r}')  \:, 
\end{split}\end{equation}
with the self-consistent condition
\begin{equation}\label{eq:self-consistency-1}
\matr{Q}(\vect{r}) = \matr{\Gamma}(\vect{r}, \vect{r})  \:.
\end{equation}
In Eq.\ \eqref{eq:saddle-point-2}, the symbols $\lmat_{3,\pm}$ stand for the Pauli matrices acting in the duplicated space,
\begin{equation}\label{eq:def-lambda}
\lmat_+ \defeq \begin{pmatrix}0 & 1\\ 0 & 0\end{pmatrix} , \quad
\lmat_- \defeq \begin{pmatrix}0 & 0\\ 1 & 0\end{pmatrix} , \quad
\lmat_3 \defeq \begin{pmatrix}1 & 0\\ 0 & -1\end{pmatrix} \:.
\end{equation}
\par Notably, Eqs.\ \eqref{eq:saddle-point-2}--\eqref{eq:self-consistency-1} are analogous to the Gorkov equation for type-II superconductors with impurities, where the duplicated space is interpreted as the Nambu (electron-hole) space \cite{Gorkov1959a, Eilenberger1968, Kopnin2001}.
In particular, we identify a similar self-consistent term $\I\alpha\matr{Q}(\vect{r})$ that accounts for impurity scattering.
This explains the strong connection with superconductivity theory pointed out by Nazarov \cite{Nazarov1994a, Nazarov2009}.
The main difference is the absence in Eq.\ \eqref{eq:saddle-point-2} of the superconducting order parameter $\matr{\Delta}(\vect{r}) = \Delta(\vect{r})\lmat_+ + \cc{\Delta}(\vect{r})\lmat_-$ where $\Delta(\vect{r})$ is a function of the matrix element $Q_{12}(\vect{r})$.
This term $\matr{\Delta}(\vect{r})$, derived from the mean-field approximation of the BCS Hamiltonian, is central to the self-consistency of the Gorkov equation.
Unlike the field $\matr{Q}(\vect{r})$ in Eqs.\ \eqref{eq:saddle-point-2}--\eqref{eq:self-consistency-1}, the superconducting term $\matr{\Delta}(\vect{r})$ lacks diagonal  elements.
Another difference is the presence of contact interactions $\gamma_\porta\op{K}_\porta\lmat_+$ and $\gamma_\portb\op{K}_\portb\lmat_-$, which have no direct equivalent in superconductivity.

\subsection{Disorder-averaged generating function}\label{sec:avg-gen-fun}
Once the field $\matr{Q}(\vect{r})$ satisfying the saddle-point equations \eqref{eq:saddle-point-2}--\eqref{eq:self-consistency-1} is found, the disorder-averaged generating function can be obtained from Eqs.\ \eqref{eq:gen-fun-from-partition} and \eqref{eq:logz-approx}:
\begin{equation}\label{eq:avg-gen-fun-1}
F(\gamma) = \frac{1}{N_{\rm p}} \der{}{\gamma}\mathcal{L}^{(1)}[\matr{Q}(\vect{r})]  \:.
\end{equation}
The total derivative in Eq.\ \eqref{eq:avg-gen-fun-1} reads
\begin{equation}\label{eq:action-total-der}
\der{\mathcal{L}^{(1)}}{\gamma} = \pder{\mathcal{L}^{(1)}}{\gamma} + \int \D{\vect{r}} \fder{\mathcal{L}^{(1)}}{\matr{Q}(\vect{r})} \pder{\matr{Q}(\vect{r})}{\gamma}  \:.
\end{equation}
The last term in the right-hand side of Eq.\ \eqref{eq:action-total-der} is identically zero by definition of the saddle point so that only the explicit dependency of $\mathcal{L}^{(1)}$ in $\gamma$ matters.
Since this dependency comes solely from the second term in the Lagrangian \eqref{eq:full-lagrangian}, Eq.\ \eqref{eq:avg-gen-fun-1} becomes
\begin{equation}\label{eq:avg-gen-fun-2}
F(\gamma) = -\frac{1}{N_{\rm p}} \pder{}{\gamma} \Tr\ln(\op{\matr{W}}_0 + \I\alpha\matr{Q}(\op{\vect{r}}))  \:.
\end{equation}
Then, using the Green's operator \eqref{eq:def-gamma} and the fact that the only dependency of $\op{\matr{W}}_0$ on the parameter $\gamma$ comes from the interaction terms $\gamma_\porta\op{K}_\porta$ and $\gamma_\portb\op{K}_\portb$, we get
\begin{equation}\label{eq:avg-gen-fun-3}
F(\gamma) = \frac{\I}{N_{\rm p}} \Tr\left[ \op{\matr{\Gamma}} \big( \gamma'_\porta\op{K}_\porta\lmat_+ + \gamma'_\portb\op{K}_\portb\lmat_- \big) \right]  \:,
\end{equation}
where the primes refer to derivatives with respect to $\gamma$.
It should be noted that, due to the constraint \eqref{eq:gamma-product}, we must have
\begin{equation}\label{eq:gamma-product-der}
\gamma_\porta \gamma'_\portb + \gamma'_\porta \gamma_\portb = 1  \:.
\end{equation}
In the numerical simulations of Sec.\ \ref{sec:results}, we will set $\gamma_\porta=\gamma_\portb=\sqrt{\gamma}$, so that we have in this case
\begin{equation}\label{eq:gamma-primes-simu}
\gamma'_\porta = \gamma'_\portb = \frac{1}{2\sqrt{\gamma}}  \:.
\end{equation}
Another valid choice, made in the Letter \cite{GaspardD2024-short}, could be $\gamma_\porta=\gamma$, $\gamma_\portb=1$, and thus $\gamma'_\porta=1$, $\gamma'_\portb=0$.
More explicitly, the generating function \eqref{eq:avg-gen-fun-3} is given by
\begin{equation}\label{eq:avg-gen-fun-4}
F(\gamma) = \frac{\I}{N_{\rm p}} \tr \int \D\vect{r}\: \bra{\vect{r}} \op{\matr{\Gamma}} \big( \gamma'_\porta\op{K}_\porta\lmat_+ + \gamma'_\portb\op{K}_\portb\lmat_- \big) \ket{\vect{r}}  \:.
\end{equation}
Using the expression \eqref{eq:contact-op-current} of the current operator $\op{K}(x)$, it can be shown that
\begin{equation}\label{eq:gamma-proj-current-1}
\int \D\vect{r}\, \bra{\vect{r}} \op{\matr{\Gamma}} \op{K}(x_0) \ket{\vect{r}} = 2\int \D\vect{y}\, \matr{J}_{\para}(x_0,\vect{y})  \:,
\end{equation}
where $\matr{J}_{\para}$ is the longitudinal component of the matrix current defined by
\begin{equation}\label{eq:def-matrix-current}
\vect{\matr{J}}(\vect{r}) \defeq \lim_{\vect{r}'\rightarrow\vect{r}} \frac{\grad_{\vect{r}}\matr{\Gamma}(\vect{r}, \vect{r}') - \grad_{\vect{r}'}\matr{\Gamma}(\vect{r}, \vect{r}')}{2\I}  \:.
\end{equation}
Each of the components of the length-$d$ vector $\vect{\matr{J}}(\vect{r})$ is a $2\times 2$ complex matrix.
Since the current in a rectilinear waveguide is invariant with respect to the transverse coordinate $\vect{y}$, Eq.\ \eqref{eq:gamma-proj-current-1} even reduces to
\begin{equation}\label{eq:gamma-proj-current-2}
\int \D\vect{r}\, \bra{\vect{r}} \op{\matr{\Gamma}} \op{K}(x_0) \ket{\vect{r}} = 2S_\perp \matr{J}_{\para}(x_0)  \:.
\end{equation}
Using the property \eqref{eq:gamma-proj-current-2}, Eq.\ \eqref{eq:avg-gen-fun-4} becomes
\begin{equation}\label{eq:avg-gen-fun-5}
F(\gamma) = \frac{2\I S_\perp}{N_{\rm p}} \tr\big( \gamma'_\porta\matr{J}_{\para}(x_\porta) \lmat_+ + \gamma'_\portb\matr{J}_{\para}(x_\portb)\lmat_- \big)  \:.
\end{equation}
The remaining trace over the duplicated space in Eq.\ \eqref{eq:avg-gen-fun-5} selects the off-diagonal components of the matrix current.
Therefore, we obtain
\begin{equation}\label{eq:avg-gen-fun-6}
F(\gamma) = \frac{2\I S_\perp}{N_{\rm p}} \left( \gamma'_\porta J^{21}_{\para}(x_\porta) + \gamma'_\portb J^{12}_{\para}(x_\portb) \right)  \:.
\end{equation}
The result \eqref{eq:avg-gen-fun-6} suggests to define for later use a normalized matrix current absorbing the prefactors:
\begin{equation}\label{eq:def-normalized-current}
\tilde{\vect{\matr{J}}}(\vect{r}) \defeq \frac{2S_\perp}{N_{\rm p}} \vect{\matr{J}}(\vect{r})  \:.
\end{equation}
Note that, according to Eq.\ \eqref{eq:number-of-modes}, the ratio $N_{\rm p}/S_\perp$ only depends on the wavelength and thus remains finite in the limit of an infinitely wide waveguide ($N_{\rm p}\rightarrow\infty$).
In the notation \eqref{eq:def-normalized-current}, Eq.\ \eqref{eq:avg-gen-fun-6} becomes
\begin{equation}\label{eq:avg-gen-fun-7}
F(\gamma) = \I \left( \gamma'_\porta \tilde{J}^{21}_{\para}(x_\porta) + \gamma'_\portb \tilde{J}^{12}_{\para}(x_\portb) \right)  \:.
\end{equation}
Here, we recall that $\tilde{\matr{J}}_{\para}(x)$ in Eq.\ \eqref{eq:avg-gen-fun-7} is the matrix current at the saddle point.
Of course, the values of $\tilde{\matr{J}}_{\para}(x)$ at $x_\porta$ and $x_\portb$ still depend on the choice of $\gamma_\porta$ and $\gamma_\portb$.

\section{Semiclassical approximation}\label{sec:eilenberger-theory}
In order to determine the generating function from Eq.\ \eqref{eq:avg-gen-fun-7}, we have to solve the saddle-point equations \eqref{eq:saddle-point-2}--\eqref{eq:self-consistency-1}.
This is, however, a nontrivial task because in addition to the matrix nature of this system of equations, the self-consistency condition \eqref{eq:self-consistency-1} makes the problem nonlinear.
Nevertheless, it is possible to effectively approach the solution of this problem in the semiclassical approximation, that is when the wavelength is vanishingly small compared to the other characteristic lengths of the model. %
As we will see in this section, this approximation gives rise to a matrix transport equation similar to the Eilenberger equation initially developed for superconductivity \cite{Eilenberger1968, Usadel1970, Kopnin2001}.

\subsection{Matrix transport equation}\label{sec:wigner}
\par There are several ways to proceed to the semiclassical approximation.
Perhaps the most rigorous way is to exploit the phase-envelope decomposition \eqref{eq:zaitsev-decomposition} of Appendix \ref{app:q-smoothness}, as proposed by Refs.\ \cite{Zaitsev1984, Nazarov1999b}.
This way is however rather long and technical due to the four carrier waves ($\E^{\I k_{\para,n}(\sigma x+\sigma'x')}$ for all $\sigma=\pm$ and $\sigma'=\pm$), and leads to the same results as the Wigner transform approach presented in this section.
The Wigner transform of a operator $\op{A}$ acting on the position space is defined by \cite{Case2008, Cohen2020-vol3}
\begin{equation}\label{eq:def-wigner-transform}
\wigner{\op{A}}_{\vect{p},\vect{r}} \defeq \int_{\mathbb{R}^d} \D\vect{s} \bra{\vect{r}+\tfrac{\vect{s}}{2}} \op{A} \ket{\vect{r}-\tfrac{\vect{s}}{2}} \E^{-\I\vect{p}\cdot\vect{s}}  \:.
\end{equation}
In particular, the Wigner transform of the Green's operator $\op{\matr{\Gamma}}$ reads
\begin{equation}\label{eq:def-gamma-wigner}
\matr{\Gamma}(\vect{p},\vect{r}) \defeq \wigner{\op{\matr{\Gamma}}}_{\vect{p},\vect{r}}  \:.
\end{equation}
In order to obtain a transport equation for $\matr{\Gamma}(\vect{p},\vect{r})$, we consider the commutation relation $[\op{\matr{\Gamma}}^{-1},\op{\matr{\Gamma}}]=\matr{0}$.
According to the definition \eqref{eq:def-gamma}, this relation is equivalent to
\begin{equation}\label{eq:w-gamma-commut-1}
[\op{\matr{W}}_0 + \I\alpha\matr{Q}(\op{\vect{r}}), \op{\matr{\Gamma}}] = \matr{0}  \:.
\end{equation}
The definition \eqref{eq:def-matrix-w-zero} allows us to rewrite Eq.\ \eqref{eq:w-gamma-commut-1} as
\begin{equation}\label{eq:w-gamma-commut-2}
[-\op{\vect{p}}^2 + \I\varepsilon\lmat_3 + \I\alpha\matr{Q}(\op{\vect{r}}) + \gamma_\porta\op{K}_\porta\lmat_+ + \gamma_\portb\op{K}_\portb\lmat_-, \op{\matr{\Gamma}}] = \matr{0}  \:.
\end{equation}
Then, we take the Wigner transform of Eq.\ \eqref{eq:w-gamma-commut-2} using Eq.\ \eqref{eq:def-wigner-transform}.
The first transform is given exactly by
\begin{equation}\label{eq:wigner-com-p2}
\wigner{[\op{\vect{p}}^2, \op{\matr{\Gamma}}]} = -2\I\vect{p}\cdot\grad_{\vect{r}}\matr{\Gamma}(\vect{p},\vect{r})  \:.
\end{equation}
The transform of $[\varepsilon\lmat_3,\op{\matr{\Gamma}}]$ is trivial because $\varepsilon\lmat_3$ does not depend on the position.
The third transform in Eq.\ \eqref{eq:w-gamma-commut-2} reads
\begin{equation}\label{eq:wigner-com-q}
\wigner{[\matr{Q}(\op{\vect{r}}), \op{\matr{\Gamma}}]} = [\matr{Q}(\vect{r}), \matr{\Gamma}(\vect{p},\vect{r})] + \tfrac{\I}{2} \{\grad_{\vect{r}}\matr{Q}, \grad_{\vect{p}}\matr{\Gamma}\} + \cdots  \:,
\end{equation}
where $\{\matr{A}, \matr{B}\}=\matr{A}\matr{B}+\matr{B}\matr{A}$ denotes the anticommutator.
Note the implicit dot product of the two gradients in Eq.\ \eqref{eq:wigner-com-q}.
The trailing dots in Eq.\ \eqref{eq:wigner-com-q} represent terms with higher-order derivatives which can be neglected as long as $\matr{Q}(\vect{r})$ smoothly varies in space.
As discussed in Appendix \ref{app:q-smoothness} (see especially Fig.\ \ref{fig:qfield1d-dirac}), this condition is satisfied in the absence of strong variations of the potential on the wavelength scale.
It is interesting to note that, if $\matr{Q}(\vect{r})$ and $\matr{\Gamma}(\vect{p},\vect{r})$ were scalar quantities, the first term in the right-hand side of Eq.\ \eqref{eq:wigner-com-q} would be zero, and the expansion would be dominated by the drift term $\I\grad_{\vect{r}}\matr{Q}\cdot\grad_{\vect{p}}\matr{\Gamma}$.
The preponderance of the commutator $[\matr{Q}(\vect{r}), \matr{\Gamma}(\vect{p},\vect{r})]$ over the other terms in Eq.\ \eqref{eq:wigner-com-q} stems from the matrix nature of these quantities.
More generally, we understand that this matrix nature makes the Wigner transform transparent in the sense that the formal operators $\op{\vect{r}}$ and $\op{\vect{p}}$ are simply replaced by their corresponding eigenvalues $\vect{r}$ and $\vect{p}$.
The last transforms with the current operators in Eq.\ \eqref{eq:w-gamma-commut-2} are given by
\begin{equation}\label{eq:wigner-com-current}
\wigner{[\op{K}(x_0)\lmat_\pm, \op{\matr{\Gamma}}]} = 2p_{\para}\delta(x-x_0)[\lmat_\pm, \matr{\Gamma}(\vect{p},\vect{r})]  \:,
\end{equation}
according to Eq.\ \eqref{eq:contact-op-current}.
The result \eqref{eq:wigner-com-current} can be obtained by smoothing the Dirac delta on a much greater length scale than the wavelength, as required by the semiclassical approximation discussed in Appendix \ref{app:q-smoothness}, and then by letting the length scale tend to zero after the Wigner transform.
Although this approach seems approximate, the result \eqref{eq:wigner-com-current} turns out to be exact because it can be proved rigorously in the representation \eqref{eq:zaitsev-decomposition}.
Gathering Eqs.\ \eqref{eq:wigner-com-p2}--\eqref{eq:wigner-com-current} into Eq.\ \eqref{eq:w-gamma-commut-2} leads to a matrix transport equation for $\matr{\Gamma}(\vect{p},\vect{r})$,
\begin{equation}\label{eq:full-eilenberger-1}\begin{split}
& 2\I\vect{p}\cdot\grad_{\vect{r}}\matr{\Gamma} + \I\varepsilon[\lmat_3, \matr{\Gamma}] + \I\alpha[\matr{Q}(\vect{r}), \matr{\Gamma}] \\
& + 2\gamma_\porta p_{\para}\delta(x-x_\porta)[\lmat_+, \matr{\Gamma}] + 2\gamma_\portb p_{\para}\delta(x-x_\portb)[\lmat_-, \matr{\Gamma}] = \matr{0}  \:,
\end{split}\end{equation}
Equation \eqref{eq:full-eilenberger-1} is supplemented by the self-consistent condition \eqref{eq:self-consistency-1}, which, in the Wigner representation \eqref{eq:def-wigner-transform}, becomes
\begin{equation}\label{eq:self-consistency-2}
\matr{Q}(\vect{r}) = \int \frac{\D\vect{p}}{(2\pi)^d} \, \matr{\Gamma}(\vect{p},\vect{r})  \:.
\end{equation}
\par The momentum integral in Eq.\ \eqref{eq:self-consistency-2} suggests the definition of a directional version of $\matr{\Gamma}(\vect{p},\vect{r})$, that we refer to as the matrix radiance by analogy with radiative transfer,
\begin{equation}\label{eq:def-g-radiance}
\matr{g}(\vect{\Omega},\vect{r}) \defeq \frac{S_d}{\pi\nu} \int \frac{\D{\vect{p}}}{(2\pi)^d} \matr{\Gamma}(\vect{p}, \vect{r}) \delta(\tfrac{\vect{p}}{\norm{\vect{p}}} - \vect{\Omega})  \:,
\end{equation}
where $\vect{\Omega}$ is a unit directional vector ($\norm{\vect{\Omega}}=1$).
The integral in Eq.\ \eqref{eq:def-g-radiance} suffers from ultraviolet divergence ($\vect{p}\rightarrow\infty$) in dimensions $d\geq 2$, and thus needs regularization.
To this end, we restrict the integration domain to the momentum shell $\norm{\vect{p}}\in[k-\delta{k}, k+\delta{k}]$.
The momentum cutoff $\delta{k}$ should be much smaller than $k$, but still significantly larger than both $\tfrac{\varepsilon}{k}$ and $\tfrac{1}{\ell}$.
This ensures that we capture the full contributions from absorption, represented by a finite $\varepsilon$, and scattering, which correspond to the terms $\I\varepsilon\lmat_3$ and $\I\alpha\matr{Q}$ in Eq.\ \eqref{eq:def-gamma}:
\begin{equation}\label{eq:momentum-cutoff}
\max\left( \frac{\varepsilon}{k}, \frac{1}{\ell} \right) \ll \delta{k} \ll k  \:.
\end{equation}
Exploiting the smallness of the cutoff $\delta{k}$ with respect to $k$ to let appear the free-space density of states \eqref{eq:free-dos}, Eq.\ \eqref{eq:def-g-radiance} becomes
\begin{equation}\label{eq:def-g-radiance-alt}
\matr{g}(\vect{\Omega},\vect{r}) = \frac{2k\nu_0}{\pi\nu} \dashint_{k} \D{p} \,\matr{\Gamma}(p\vect{\Omega},\vect{r})  \:,
\end{equation}
where
\begin{equation}\label{eq:def-cauchy-pv}
\dashint_{k} f(p)\D{p} \defeq \lim_{\delta{k}\rightarrow +\infty} \int_{k-\delta{k}}^{k+\delta{k}} f(p)\D{p}  
\end{equation}
denotes the Cauchy principal value of the integral on the momentum shell at $k$.
We will see in Sec.\ \ref{sec:boundaries} that the normalization factor in the definition \eqref{eq:def-g-radiance} ensures the property $\matr{g}(\vect{\Omega},\vect{r})^2=\matr{1}_2$.
According to Eq.\ \eqref{eq:def-g-radiance}, the expression \eqref{eq:self-consistency-2} becomes $\textstyle\matr{Q}(\vect{r}) = \pi\nu \oint \frac{\D\vect{\Omega}}{S_d} \matr{g}(\vect{\Omega},\vect{r})$.
It is then convenient to absorb the prefactor $\pi\nu$ in a normalized version of the $\matr{Q}$ field defined by
\begin{equation}\label{eq:def-normalized-q}
\tilde{\matr{Q}}(\vect{r}) \defeq \frac{1}{\pi\nu} \matr{Q}(\vect{r})  \:.
\end{equation}
Therefore, the self-consistent condition \eqref{eq:self-consistency-2} now reads
\begin{equation}\label{eq:self-consistency-3}
\tilde{\matr{Q}}(\vect{r}) = \oint \frac{\D\vect{\Omega}}{S_d} \matr{g}(\vect{\Omega},\vect{r})  \:.
\end{equation}
The $\tilde{\matr{Q}}$ field can thus be interpreted as the directional average of the matrix radiance.
Furthermore, the matrix current $\vect{\matr{J}}$ can also be expressed from $\matr{g}(\vect{\Omega},\vect{r})$.
By applying the same approach used for $\matr{Q}$ to the definition \eqref{eq:def-matrix-current}, we find
\begin{equation}\label{eq:j-matrix-from-g}\begin{split}
\vect{\matr{J}}(\vect{r}) & = \int \frac{\D\vect{p}}{(2\pi)^d} \vect{p}\,\matr{\Gamma}(\vect{p},\vect{r})  \\
 & = \pi\nu k \oint \frac{\D\vect{\Omega}}{S_d} \vect{\Omega} \:\matr{g}(\vect{\Omega},\vect{r})  \:.
\end{split}\end{equation}
Moreover, according to Eqs.\ \eqref{eq:number-of-modes} and \eqref{eq:def-normalized-current}, the normalized matrix current can be written as
\begin{equation}\label{eq:jn-from-g}
\tilde{\vect{\matr{J}}}(\vect{r}) = \oint \frac{\D\vect{\Omega}}{2V_{d-1}} \vect{\Omega} \:\matr{g}(\vect{\Omega},\vect{r})  \:.
\end{equation}
\par It should be noted that the relations \eqref{eq:self-consistency-3}--\eqref{eq:jn-from-g} implicitly assume that the number of transverse modes tends to infinity ($N_{\rm p}\rightarrow\infty$). %
Some of these relations will have to be adapted when the number of modes is finite, as explained in Appendix\ \ref{app:modal-eilenberger}.
In anticipation of these future changes, it is convenient to define the directional mean of any quantity $A(\vect{\Omega})$ by
\begin{equation}\label{eq:directional-mean-continuous}
\avg{A}^\pm_\kappa = \frac{\int_{\pm\Omega_{\para}>0} \D{\vect{\Omega}} \abs{\Omega_{\para}}^\kappa A(\vect{\Omega})}{\int_{\pm\Omega_{\para}>0} \D{\vect{\Omega}} \abs{\Omega_{\para}}^\kappa}  \:,
\end{equation}
where the sign refers to the forward and backward hemispheres, and $\kappa$ is the order of the moment.
The normalization factor in Eq.\ \eqref{eq:directional-mean-continuous} is given by
\begin{equation}\label{eq:directional-moments}
\int_{\pm\Omega_{\para}>0} \D{\vect{\Omega}} \abs{\Omega_{\para}}^\kappa = \frac{V_{d+\kappa-2}}{V_{\kappa-1}} 
 = \begin{cases}
\tfrac{1}{2}S_{d} & (\kappa=0)  \:,\\
V_{d-1} & (\kappa=1)  \:.
\end{cases}
\end{equation}
Using the notation \eqref{eq:directional-mean-continuous}, the matrix field and currents become
\begin{equation}\label{eq:qn-jn-from-g-unified}
\tilde{\matr{Q}}(x) = \frac{\avg{\matr{g}}^+_0 + \avg{\matr{g}}^-_0}{2}  \:,\quad
\tilde{\matr{J}}_{\para}(x) = \frac{\avg{\matr{g}}^+_1 - \avg{\matr{g}}^-_1}{2}  \:.
\end{equation}
\par By integrating it over the momentum shell, the matrix transport equation \eqref{eq:full-eilenberger-1} now reads
\begin{equation}\label{eq:full-eilenberger-2}\begin{split}
& \vect{\Omega}\cdot\grad_{\vect{r}}\matr{g} + \tfrac{\varepsilon}{2k}[\lmat_3, \matr{g}] + \tfrac{1}{2\ell}[\tilde{\matr{Q}}(\vect{r}), \matr{g}] \\
& - \I\gamma_\porta\Omega_{\para}\delta(x-x_\porta)[\lmat_+, \matr{g}] - \I\gamma_\portb\Omega_{\para}\delta(x-x_\portb)[\lmat_-, \matr{g}] = \matr{0}  \:,
\end{split}\end{equation}
which is reminiscent of the Eilenberger equation of superconductivity \cite{Eilenberger1968, Usadel1970, Kopnin2001}.
The Eilenberger equation was first derived in Ref.\ \cite{Eilenberger1968} using a diagrammatic approach and is the semiclassical approximation of the Gorkov equation of type-II superconductors with impurities \cite{Gorkov1959a}, mentioned previously below Eq.\ \eqref{eq:saddle-point-2}.
The differences between Eq.\ \eqref{eq:full-eilenberger-2} and the Eilenberger equation is the absence of the superconducting term $\matr{\Delta}(\vect{r})$ and the presence of the contact interactions at $x_\porta$ and $x_\portb$.
\par The first three terms in Eq.\ \eqref{eq:full-eilenberger-2} govern propagation in the disordered waveguide, while the last two account for contact interactions at $x_\porta$ and $x_\portb$.
Let us integrate the equation over infinitesimal intervals surrounding $x_\porta$ and $x_\portb$ to determine the effect of these two terms.
The integration over $x$ can be carried out rigorously by replacing the Dirac deltas by window functions.
The results are the following discontinuity conditions
\begin{equation}\label{eq:contact-cond-g}\begin{aligned}
\matr{g}(\vect{\Omega},x_\porta^+) & = \E^{\I\gamma_\porta\lmat_+} \matr{g}(\vect{\Omega},x_\porta^-) \E^{-\I\gamma_\porta\lmat_+}  \:,\\
\matr{g}(\vect{\Omega},x_\portb^+) & = \E^{\I\gamma_\portb\lmat_-} \matr{g}(\vect{\Omega},x_\portb^-) \E^{-\I\gamma_\portb\lmat_-}  \:,
\end{aligned}\end{equation}
where $x^\pm=x\pm\epsilon$ for $\epsilon\xrightarrow{>}0$.
It should be noted that the discontinuity conditions \eqref{eq:contact-cond-g} do not depend on the direction, so that they can be integrated over $\vect{\Omega}$ exactly.
From that, we deduce the corresponding discontinuity conditions for the $\tilde{\matr{Q}}$ field,
\begin{equation}\label{eq:contact-cond-qn}\begin{aligned}
\tilde{\matr{Q}}(x_\porta^+) & = \E^{\I\gamma_\porta\lmat_+} \tilde{\matr{Q}}(x_\porta^-) \E^{-\I\gamma_\porta\lmat_+}  \:,\\
\tilde{\matr{Q}}(x_\portb^+) & = \E^{\I\gamma_\portb\lmat_-} \tilde{\matr{Q}}(x_\portb^-) \E^{-\I\gamma_\portb\lmat_-}  \:,
\end{aligned}\end{equation}
and for the matrix current
\begin{equation}\label{eq:contact-cond-jn}\begin{aligned}
\tilde{\vect{\matr{J}}}(x_\porta^+) & = \E^{\I\gamma_\porta\lmat_+} \tilde{\vect{\matr{J}}}(x_\porta^-) \E^{-\I\gamma_\porta\lmat_+}  \:,\\
\tilde{\vect{\matr{J}}}(x_\portb^+) & = \E^{\I\gamma_\portb\lmat_-} \tilde{\vect{\matr{J}}}(x_\portb^-) \E^{-\I\gamma_\portb\lmat_-}  \:.
\end{aligned}\end{equation}
The effect of the discontinuity conditions \eqref{eq:contact-cond-qn} on the $\tilde{\matr{Q}}$ field is visible in Fig.\ \ref{fig:qfield1d-dirac}.
Furthermore, an important consequence of Eq.\ \eqref{eq:contact-cond-jn} is that the matrix current is not conserved at the contact points $x_\porta$ and $x_\portb$.
It is nevertheless conserved everywhere else in the absence of absorption ($\varepsilon=0$) since the integral over all the directions of the matrix transport equation \eqref{eq:full-eilenberger-2} without the contact terms immediately provides
\begin{equation}\label{eq:matrix-current-conservation}
\grad_{\vect{r}}\cdot\tilde{\vect{\matr{J}}}(\vect{r}) = \matr{0}  \:.
\end{equation}

\subsection{Boundary conditions at infinity}\label{sec:boundaries}
In this section, the boundary conditions at infinity ($\vect{r}\rightarrow\infty$) are derived for the matrix radiance $\matr{g}(\vect{\Omega},\vect{r})$.
We will see that these boundary conditions are somewhat unusual for transport equations due to the matrix nature of the variables.
These boundary conditions can be found in the expansion of the Green's operator \eqref{eq:def-gamma} which is
\begin{equation}\label{eq:gamma-t-expansion-1}
\op{\matr{\Gamma}} = \op{\matr{G}}_0 + \op{\matr{G}}_1  \:,\qquad
\op{\matr{G}}_1 \defeq \op{\matr{G}}_0\op{\matr{T}}\op{\matr{G}}_0  \:,
\end{equation}
where $\op{\matr{G}}_0$ is the free-space Green's operator defined by
\begin{equation}\label{eq:def-gamma-zero}
\op{\matr{G}}_0 \defeq \frac{\I}{k^2 - \op{\vect{p}}^2 + \I\varepsilon\lmat_3}  \:,
\end{equation}
and $\op{\matr{T}}$ is the transition operator collecting all scattering diagrams involving $\I\alpha\matr{Q}(\op{\vect{r}})$, $\gamma_\porta\op{K}_\porta\lmat_+$, and $\gamma_\portb\op{K}_\portb\lmat_-$ in Eq.\ \eqref{eq:saddle-point-2}.
In order to determine the boundary conditions, we seek to know the far-field asymptotic behavior of $\matr{\Gamma}(\vect{r}, \vect{r}')$, or more precisely of
\begin{equation}\label{eq:radiance-from-g0-g1}
\matr{g}(\vect{\Omega},\vect{r}) = \matr{g}_0(\vect{\Omega},\vect{r}) + \matr{g}_1(\vect{\Omega},\vect{r})  \:,
\end{equation}
where $\matr{g}_0$ and $\matr{g}_1$ are related to the Wigner transforms of $\op{\matr{G}}_0$ and $\op{\matr{G}}_1$ in Eq.\ \eqref{eq:gamma-t-expansion-1}.
\par Let us first consider the $\op{\matr{G}}_0$ term in Eq.\ \eqref{eq:gamma-t-expansion-1}.
According to Eqs.\ \eqref{eq:def-g-radiance-alt}, we get
\begin{equation}\label{eq:asym-g-zero-1}
\matr{g}_0(\vect{\Omega},\vect{r}) = \frac{2k}{\pi} \dashint_{k} \D{p} \,\matr{G}_0(p\vect{\Omega},\vect{r}) 
 = \frac{1}{\I\pi} \dashint_0 \frac{\D{u}}{u - \I\varepsilon\lmat_3}  \:,
\end{equation}
since the Wigner transform of $\op{\matr{G}}_0$ is just $\matr{G}_0(\vect{p},\vect{r})=\I/(k^2 - \vect{p}^2 + \I\varepsilon\lmat_3)$. This yields
\begin{equation}\label{eq:asym-g-zero-2}
\matr{g}_0(\vect{\Omega},\vect{r}) = \lmat_3  \:.
\end{equation}
The result \eqref{eq:asym-g-zero-2} is the contribution of the free Green's operator $\op{\matr{G}}_0$ to the boundary conditions at infinity.
\par We still have to take into account the multiple-scattering term $\op{\matr{G}}_1$ in Eq.\ \eqref{eq:gamma-t-expansion-1}.
In order to compact the notations, we split the $\matr{T}$ matrix into its four components in the duplicated space:
\begin{equation}\label{eq:t-matrix-decomposition}\begin{aligned}
& \op{\matr{T}} = \op{\matr{T}}_{11} + \op{\matr{T}}_{12} + \op{\matr{T}}_{21} + \op{\matr{T}}_{22}  \:,\\
& \op{\matr{T}}_{11} = \begin{pmatrix}\op{T}_{11} & 0\\ 0 & 0\end{pmatrix}  \:,\quad
  \op{\matr{T}}_{12} = \begin{pmatrix}0 & \op{T}_{12}\\ 0 & 0\end{pmatrix}  \:,\\
& \op{\matr{T}}_{21} = \begin{pmatrix}0 & 0\\ \op{T}_{21} & 0\end{pmatrix}  \:,\quad
  \op{\matr{T}}_{22} = \begin{pmatrix}0 & 0\\ 0 & \op{T}_{22}\end{pmatrix}  \:.
\end{aligned}\end{equation}
Using the fact that $\op{\matr{G}}_0$ is diagonal in the duplicated space,
\begin{equation}\label{eq:gamma-zero-explicit}
\matr{G}_0(\vect{r}, \vect{r}') = \begin{pmatrix}G_0^+(\vect{r}, \vect{r}') & 0\\ 0 & G_0^-(\vect{r}, \vect{r}')\end{pmatrix} \:, 
\end{equation}
the Green's operator $\op{\matr{G}}_1$, expressed in the position basis, then reads 
\begin{equation}\label{eq:gamma-one-1}\begin{split}
\matr{G}_1(\vect{r}, \vect{r}') & = \iint\D\vect{r}_1\D\vect{r}_2 \\
& \times \Big( G_0^+(\vect{r}, \vect{r}_1) \matr{T}_{11}(\vect{r}_1, \vect{r}_2) G_0^+(\vect{r}_2, \vect{r}')  \\
& + G_0^+(\vect{r}, \vect{r}_1) \matr{T}_{12}(\vect{r}_1, \vect{r}_2) G_0^-(\vect{r}_2, \vect{r}') \\
& + G_0^-(\vect{r}, \vect{r}_1) \matr{T}_{21}(\vect{r}_1, \vect{r}_2) G_0^+(\vect{r}_2, \vect{r}') \\
& + G_0^-(\vect{r}, \vect{r}_1) \matr{T}_{22}(\vect{r}_1, \vect{r}_2) G_0^-(\vect{r}_2, \vect{r}') \Big) \:.
\end{split}\end{equation}
We notice that the $G_0^+G_0^+$ and $G_0^-G_0^-$ terms in Eq.\ \eqref{eq:gamma-one-1} oscillate at the wavelength scale, even at arbitrarily large distances from the system ($\vect{r},\vect{r}'\rightarrow\infty$).
Since Appendix \ref{app:q-smoothness} established that the $\matr{Q}$ field varies slowly at this scale, the contribution of these terms must be negligible. This requirement enforces that the diagonal elements of the $\matr{T}$ matrix are zero:
$\matr{T}_{11}=\matr{T}_{22}=\matr{0}$.
To evaluate the contribution of the other elements in Eq.\ \eqref{eq:gamma-one-1}, we consider the following change of variable motivated by the assumed relative proximity between $\vect{r}$ and $\vect{r}'$ on the one hand and between $\vect{r}_1$ and $\vect{r}_2$ on the other hand:
\begin{equation}\label{eq:asym-var-change}
\begin{cases}
\vect{r}  = \bar{\vect{r}} + \tfrac{\vect{s}}{2}  \:,\\
\vect{r}' = \bar{\vect{r}} - \tfrac{\vect{s}}{2}  \:,
\end{cases} \quad
\begin{cases}
\vect{r}_1 = \vect{R} + \tfrac{\vect{\Delta}}{2}  \:,\\
\vect{r}_2 = \vect{R} - \tfrac{\vect{\Delta}}{2}  \:.
\end{cases}
\end{equation}
Assuming that $\norm{\bar{\vect{r}}-\vect{R}}\gg\norm{\vect{s}-\vect{\Delta}}$, which is true for $\bar{\vect{r}}$ outside the scattering region, we have the approximation
\begin{equation}\label{eq:asym-g0p-g0m}
G_0^\pm(\vect{r}, \vect{r}_1)G_0^\mp(\vect{r}_2, \vect{r}') \simeq \abs{G_0^+(\bar{\vect{r}}, \vect{R})}^2 \E^{\pm\I k(\vect{s} - \vect{\Delta})\cdot\vect{e}_{\bar{\vect{r}},\vect{R}}}  \:,
\end{equation}
where $\vect{e}_{\bar{\vect{r}},\vect{R}}$ is the unit vector defined by
\begin{equation}\label{eq:asym-unit-vector}
\vect{e}_{\bar{\vect{r}},\vect{R}} \defeq \frac{\bar{\vect{r}} - \vect{R}}{\norm{\bar{\vect{r}} - \vect{R}}}  \:.
\end{equation}
Note that the approximation \eqref{eq:asym-g0p-g0m} does not require the far-field assumption, $\vect{r},\vect{r}'\rightarrow\infty$, and is thus valid at finite distance from the system.
Using Eqs.\ \eqref{eq:asym-var-change} and \eqref{eq:asym-g0p-g0m}, Eq.\ \eqref{eq:gamma-one-1} becomes
\begin{equation}\label{eq:gamma-one-2}\begin{split}
\matr{G}_1(\vect{r}, \vect{r}') & = \iint\D\vect{R}\D\vect{\Delta} \abs{G_0^+(\bar{\vect{r}}, \vect{R})}^2 \\
& \times \Big( \matr{T}_{12}(\vect{R}+\tfrac{\vect{\Delta}}{2},  \vect{R}-\tfrac{\vect{\Delta}}{2}) \E^{\I k(\vect{s} - \vect{\Delta})\cdot\vect{e}_{\bar{\vect{r}},\vect{R}}}  \\
& + \matr{T}_{21}(\vect{R}+\tfrac{\vect{\Delta}}{2},  \vect{R}-\tfrac{\vect{\Delta}}{2}) \E^{-\I k(\vect{s} - \vect{\Delta})\cdot\vect{e}_{\bar{\vect{r}},\vect{R}}} \Big)  \:.
\end{split}\end{equation}
In Eq.\ \eqref{eq:gamma-one-2}, we identify the integral over $\vect{\Delta}$ as the Wigner transform $\matr{T}(\vect{p},\vect{r})=\mathcal{W}(\op{\matr{T}})_{\vect{p},\vect{r}}$.
In addition, we calculate the Wigner transform of $\op{\matr{G}}_1$, which consists of a Fourier transform of Eq.\ \eqref{eq:gamma-one-2} in the variable $\vect{s}$. We get
\begin{equation}\label{eq:gamma-one-3}\begin{split}
\matr{G}_1(\vect{p},\bar{\vect{r}}) & = (2\pi)^d \int\D\vect{R} \abs{G_0^+(\bar{\vect{r}}, \vect{R})}^2 \\
& \times \Big ( \matr{T}_{12}(k\vect{e}_{\bar{\vect{r}},\vect{R}},\vect{R}) \delta(\vect{p} - k\vect{e}_{\bar{\vect{r}},\vect{R}})  \\
& + \matr{T}_{21}(-k\vect{e}_{\bar{\vect{r}},\vect{R}},\vect{R}) \delta(\vect{p} + k\vect{e}_{\bar{\vect{r}},\vect{R}}) \Big)  \:.
\end{split}\end{equation}
Then integrating Eq.\ \eqref{eq:gamma-one-3} over the momentum shell with Eq.\ \eqref{eq:def-g-radiance-alt} and replacing $\bar{\vect{r}}$ by $\vect{r}$, we obtain
\begin{equation}\label{eq:asym-g-one}\begin{split}
& \matr{g}_1(\vect{\Omega},\vect{r}) = \frac{S_d}{\pi\nu} \int\D\vect{R} \abs{G_0^+(\vect{r}, \vect{R})}^2 \\
& \times \Big( \matr{T}_{12}(k\vect{e}_{\vect{r},\vect{R}},\vect{R}) \delta(\vect{\Omega} - \vect{e}_{\vect{r},\vect{R}})   \\
& + \matr{T}_{21}(-k\vect{e}_{\vect{r},\vect{R}},\vect{R}) \delta(\vect{\Omega} + \vect{e}_{\vect{r},\vect{R}}) \Big)  \:.
\end{split}\end{equation}
\par
\begin{figure}[ht]%
\includegraphics{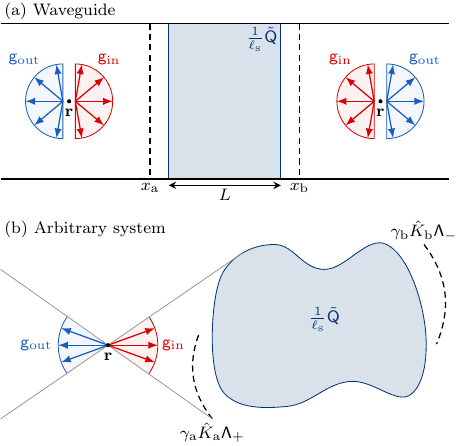}%
\caption{(a) Boundary conditions on the matrix radiance $\matr{g}(\vect{\Omega},\vect{r})$ in a waveguide according to Eq.\ \eqref{eq:boundary-conditions}.
The filled region represents the scattering term $\tfrac{1}{\ell}\tilde{\matr{Q}}(\vect{r})$ in Eq.\ \eqref{eq:full-eilenberger-2}, and the dashed lines are the interaction surfaces.
(b) Boundary conditions for a disordered region and interaction surfaces of arbitrary shape in free space.}%
\label{fig:boundary-conditions}%
\end{figure}

 In summary, the radiance $\matr{g}(\vect{\Omega},\vect{r})$ outside the disordered region is given by Eq.\ \eqref{eq:radiance-from-g0-g1} with Eqs.\ \eqref{eq:asym-g-zero-2} and \eqref{eq:asym-g-one}.
The arguments of the Dirac deltas in Eq.\ \eqref{eq:asym-g-one} show that the rays in the outgoing direction behave as $\lmat_3+g_{12}\lmat_+$ and are thus upper triangular matrices, while the incoming rays behave as $\lmat_3+g_{21}\lmat_-$ and are lower triangular matrices.
Formally, these boundary conditions read
\begin{equation}\label{eq:boundary-conditions}
\matr{g}(\vect{\Omega},\vect{r}) = \begin{cases}
\matr{g}_{\rm out}  & \text{if}~\exists\,\vect{R}\in\mathcal{V} \mid \vect{\Omega} =  \vect{e}_{\vect{r},\vect{R}}  \:,\\
\matr{g}_{\rm in}   & \text{if}~\exists\,\vect{R}\in\mathcal{V} \mid \vect{\Omega} = -\vect{e}_{\vect{r},\vect{R}}  \:,\\
\lmat_3             & \text{otherwise}  \:,
\end{cases}\end{equation}
where $\mathcal{V}$ designates the scattering region containing the disordered region and the contact interactions $\gamma_\porta\op{K}_\porta\lmat_+$ and $\gamma_\portb\op{K}_\portb\lmat_-$.
The matrices $\matr{g}_{\rm out}$ and $\matr{g}_{\rm in}$ are defined by
\begin{equation}\label{eq:g-upper-lower-triangular}
\matr{g}_{\rm out} = \begin{pmatrix}1 & g_{12}\\ 0 & -1\end{pmatrix}  \:,\quad
\matr{g}_{\rm in}  = \begin{pmatrix}1 & 0\\ g_{21} & -1\end{pmatrix}  \:.
\end{equation}
The boundary conditions \eqref{eq:boundary-conditions}--\eqref{eq:g-upper-lower-triangular} are schematically represented in Fig.\ \ref{fig:boundary-conditions}(a) for the waveguide geometry considered in this paper.
It is worth noting that, due to the generality of our derivation, these boundary conditions apply to a disordered region of arbitrary shape, as depicted in Fig.\ \ref{fig:boundary-conditions}(b).
Since the contact interactions $\gamma_\porta\op{K}_\porta\lmat_+$ and $\gamma_\portb\op{K}_\portb\lmat_-$ must be included in the $\matr{T}$ matrix, we have to take them into account in the volume $\mathcal{V}$ of Eq.\ \eqref{eq:boundary-conditions}.
This is emphasized in Fig.\ \ref{fig:boundary-conditions}(b) by the coverage of the interaction surface $\gamma_\porta\op{K}_\porta\lmat_+$ by the boundary conditions.
\par In the waveguide, the boundary conditions \eqref{eq:boundary-conditions}--\eqref{eq:g-upper-lower-triangular} mean that three out of four elements of the $\matr{g}(\vect{\Omega},\vect{r})$ matrix are fixed at infinity, with the last element ($g_{12}$ or $g_{21}$) remaining free.
This degree of freedom in both the incoming and outgoing directions makes the boundary conditions \eqref{eq:boundary-conditions}--\eqref{eq:g-upper-lower-triangular} quite unusual for a transport theory.
Indeed, in general, the incoming radiance is imposed and the outgoing radiance is unknown.
\par The boundary conditions \eqref{eq:boundary-conditions}--\eqref{eq:g-upper-lower-triangular} also have important consequences regarding the properties of $\matr{g}(\vect{\Omega},\vect{r})$ everywhere in space including in the disordered region.
The first consequence is the traceless property
\begin{equation}\label{eq:g-traceless}
\tr\matr{g}(\vect{\Omega},\vect{r}) = 0  \:,\quad\forall \vect{\Omega},\vect{r}  \:,
\end{equation}
which is known for the Eilenberger equation in superconductivity \cite{Eilenberger1968, Usadel1970, Kopnin2001}.
The property \eqref{eq:g-traceless} comes from the tracelessness of the boundary conditions \eqref{eq:boundary-conditions}--\eqref{eq:g-upper-lower-triangular} and from the trace conservation property of the matrix transport equation \eqref{eq:full-eilenberger-2}: $\grad_{\vect{r}}\tr\matr{g}=\vect{0}$.
The second consequence is the remarkable normalization property
\begin{equation}\label{eq:g-normalization}
\matr{g}(\vect{\Omega},\vect{r})^2 = \matr{1}_2  \:,\quad\forall \vect{\Omega},\vect{r}  \:,
\end{equation}
also known in superconductivity \cite{Eilenberger1968, Larkin1968, Usadel1970, Kopnin2001}.
The property \eqref{eq:g-normalization} comes from the corresponding normalization of $\matr{g}(\vect{\Omega},\vect{r})$ in Eq.\ \eqref{eq:boundary-conditions}--\eqref{eq:g-upper-lower-triangular} and from the conservation of $\matr{g}^2$ by the matrix transport equation \eqref{eq:full-eilenberger-2}: $\grad_{\vect{r}}(\matr{g}^2)=\vect{\matr{0}}$.
This less obvious conservation law can be derived from the anticommutator of Eq.\ \eqref{eq:full-eilenberger-2} with $\matr{g}$,
\begin{equation}\label{eq:eilenberger-anticom-1}
\vect{\Omega}\cdot\{\matr{g}, \grad_{\vect{r}}\matr{g}\} + \{\matr{g}, [\matr{P}, \matr{g}]\} = \matr{0}  \:,
\end{equation}
where $\matr{P}$ contains all the matrices other than $\matr{g}$ in Eq.\ \eqref{eq:full-eilenberger-2}.
The second anticommutator in Eq.\ \eqref{eq:eilenberger-anticom-1} is zero,
\begin{equation}\label{eq:eilenberger-anticom-2}
\{\matr{g}, [\matr{P}, \matr{g}]\} = [\matr{P}, \matr{g}^2] = \matr{0}  \:,\quad\forall\matr{P}  \:,
\end{equation}
since $\matr{g}^2$ is proportional to the identity matrix due to Eq.\ \eqref{eq:g-traceless}.
Therefore, Eqs.\ \eqref{eq:eilenberger-anticom-1} and \eqref{eq:eilenberger-anticom-2} show that the quantity $\matr{g}^2$ is conserved,
\begin{equation}\label{eq:eilenberger-anticom-3}
\grad_{\vect{r}}(\matr{g}^2) = \vect{\matr{0}}  \:,
\end{equation}
hence extending the domain of validity of the normalization property $\matr{g}^2=\matr{1}_2$ from the boundaries to the whole space.
It should be noted that the property $\tr\matr{g}=0$ is inherited by the $\tilde{\matr{Q}}$ field and its current,
\begin{equation}\label{eq:qn-jn-traceless}
\tr\tilde{\matr{Q}}(\vect{r}) = 0  \:,\qquad
\tr\tilde{\vect{\matr{J}}}(\vect{r}) = \vect{0}  \:,
\end{equation}
but not the normalization property $\matr{g}^2=\matr{1}_2$ since the latter is not a linear relation.
Despite this, when the radiance $\matr{g}(\vect{\Omega},\vect{r})$ weakly depends on $\vect{\Omega}$ (typically in the multiple-scattering regime $\ell\ll L$), the directional integral \eqref{eq:self-consistency-3} approximately preserves this normalization, hence
\begin{equation}\label{eq:qn-normalization}
\tilde{\matr{Q}}(\vect{r})^2 \simeq \matr{1}_2  \:.
\end{equation}
This explains the origin of the well-known normalization property of the $\tilde{\matr{Q}}(\vect{r})$ field in nonlinear sigma models \cite{Efetov1997, Nazarov2009, Kamenev2023, Lerner2003}.
The approximate nature of the property \eqref{eq:qn-normalization} highlights the important fact that nonlinear sigma models are restricted to the multiple-scattering regime and do not apply to the quasiballistic regime ($L\lesssim\ell$).

\section{Numerical results}\label{sec:results}
In this section, we present some numerical results for the transmission eigenvalue distribution based on the theory of Sec.\ \ref{sec:eilenberger-theory}.

\subsection{Numerical results in a waveguide}\label{sec:numerical-waveguide}
We present here some numerical results based on RFT for the transmission eigenvalue distribution in a disordered waveguide of length $L$ and square cross section $W^{d-1}$.
In this geometry, the matrix transport equation \eqref{eq:full-eilenberger-2} must be modified to take into account the transverse quantization.
Indeed, Eq.\ \eqref{eq:full-eilenberger-2} assumes a continuum of propagation directions, which although valid in the semiclassical limit of a large number of modes, does not apply when the waveguide width is comparable to the wavelength, which we consider in this section.
The modifications of Eq.\ \eqref{eq:full-eilenberger-2} are derived in Appendix \ref{app:modal-eilenberger} resulting in Eqs.\ \eqref{eq:full-eilenberger-modal} and \eqref{eq:qn-modal}.
These equations can be solved numerically using the program {\sc Ebsolve} \cite{GaspardD2025-ebsolve} which is based on the integration algorithm of Appendix \ref{app:integration}.
\par Furthermore, we assume periodic boundary conditions in the transverse directions.
With periodic boundaries, the transverse modes are plane waves, leading to constant Wigner transforms in the transverse direction.
Consequently, the field $\tilde{\matr{Q}}$ becomes constant in this direction, depending only on the longitudinal coordinate $x$.
This translational invariance extends to $\matr{g}$ and $\tilde{\matr{J}}$ as well.
By contrast, Dirichlet boundary conditions would introduce transverse dependencies in all these quantities, complicating the numerical implementation.
Under the assumption of periodic boundaries, the direction cosines of the modes entering Eqs.\ \eqref{eq:full-eilenberger-modal} and \eqref{eq:qn-modal} are given by
\begin{equation}\label{eq:mu-periodic}
\mu_{\vect{n}} = \frac{k_{\para,\vect{n}}}{k} = \sqrt{1 - \left(\frac{2\pi}{kW}\vect{n}\right)^2}  \qquad\forall \vect{n}\in\mathbb{Z}^{d-1}  \:.
\end{equation}
It should be noted that only the propagating modes (such that $\Im\mu_{\vect{n}}=0$) must be retained in Eq.\ \eqref{eq:mu-periodic}.
The contribution of the evanescent modes cannot be taken into account since these modes fall outside the scope of the semiclassical theory.
In practice, the numerical procedure can be improved by solving the matrix transport equation \eqref{eq:full-eilenberger-modal} only for distinct values of $\mu_{\vect{n}}$.
Of course, this should be implemented with care because the directional means \eqref{eq:qn-modal} and \eqref{eq:jn-modal} must be weighted by the appropriate multiplicities.
\par The numerical procedure to compute the transmission eigenvalue density $\rho(T)$ is as follows \cite{GaspardD2025-ebsolve}:
For each transmission value $T$ in the interval $[0,1]$:
\begin{enumerate}%
\item Set $\gamma = T^{-1} + \I 0^+$ and initialize $\tilde{\matr{Q}}(x)=\matr{0}$.
\item Iterate until $\tilde{\matr{Q}}(x)$ converges:
\begin{enumerate}%
\item Solve the matrix transport equation \eqref{eq:full-eilenberger-modal} for $\matr{g}^\pm_n(x)$ along $x$, using the fixed field $\tilde{\matr{Q}}(x)$.
See Appendix \ref{app:integration} for more details on this step.
\item Update $\tilde{\matr{Q}}(x)$ according to Eq.\ \eqref{eq:qn-modal}.
\end{enumerate}%
\item Compute $\rho(T)$ using Eqs.\ \eqref{eq:rho-from-gen-fun}, \eqref{eq:avg-gen-fun-7}, and \eqref{eq:jn-modal}.
\end{enumerate}%
\begin{figure}[ht]%
\includegraphics{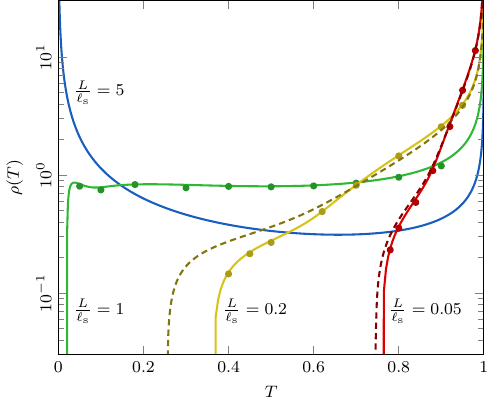}%
\caption{Transmission eigenvalue distribution in a two-dimensional (2D) waveguide of width $W/\lambda=25.5$ and aspect ratio $L/W=1$, with transverse periodic boundary conditions, shown at different optical thicknesses $L/\ell$.
Predictions from the modal matrix transport equation \eqref{eq:full-eilenberger-modal} (solid lines) \cite{GaspardD2025-ebsolve} are compared to results from the saddle-point equations \eqref{eq:saddle-point-2}--\eqref{eq:self-consistency-1} (dots) \cite{GaspardD2025-qfield} and the quasiballistic approximation \eqref{eq:ufa-system-2}--\eqref{eq:ufa-gen-fun} (dashed lines).}%
\label{fig:rho-waveguide}%
\end{figure}%
\par The convergence of this iterative process is discussed in Appendix \ref{app:convergence-and-smoothness}.
This algorithm produces the solid curves in Fig.\ \ref{fig:rho-waveguide}.
For comparison, the dots represent numerical solutions of the saddle-point equation \eqref{eq:saddle-point-2}--\eqref{eq:self-consistency-1}, with $\rho(T)$ computed using Eqs.\ \eqref{eq:rho-from-gen-fun}, \eqref{eq:def-matrix-current}, \eqref{eq:def-normalized-current}, and \eqref{eq:avg-gen-fun-7}.
The agreement with the solid curves confirms the validity of the semiclassical approximation carried out in Secs.\ \ref{sec:wigner} and Appendix \ref{app:modal-eilenberger}.
An analytical approximation of the distribution for the quasiballistic regime is derived in Appendix \ref{app:quasiballistic-solution}.
The result \eqref{eq:ufa-system-2}--\eqref{eq:ufa-gen-fun} is depicted by dashed lines in Fig.\ \ref{fig:rho-waveguide}.
This approximation is valid when the optical thickness is much smaller than one and is particularly accurate in the vicinity of the peak at $T=1$.
Notably, it provides higher accuracy than the DMPK solution [Eqs.\ (203)--(204) of Ref.\ \cite{Beenakker1997}] in this regime.
In the diffusive regime ($L/\ell=5$), we retrieve the expected bimodal law.
\par In the accompanying Letter \cite{GaspardD2024-short}, the transmission eigenvalue distributions in Fig.\ \ref{fig:rho-waveguide} are also compared to numerical simulations based on the wave equation \eqref{eq:wave-equation}.

\subsection{Numerical results in an infinite slab}\label{sec:numerical-slab}
In this section, we show that RFT is also able to predict the transmission eigenvalue distribution in an infinite slab, that is when the waveguide boundaries are sent to infinity ($W\rightarrow\infty$).
In this scenario, momentum is no longer quantized in the transverse direction, so the matrix transport equation \eqref{eq:full-eilenberger-2} must, in principle, be solved over a continuum of directions $\vect{\Omega}$.
This equation is accompanied by the expressions \eqref{eq:qn-jn-from-g-unified}, where the directional mean is defined in Eq.\ \eqref{eq:directional-mean-continuous}.
To evaluate this mean numerically in the program \cite{GaspardD2025-ebsolve}, we resort to the discrete ordinates method \cite{Chandrasekhar1960} and the Gauss-Jacobi quadrature \cite[Sec.~4.6]{Press2007}.
Since $\matr{g}(\vect{\Omega},x)$ is rotationally invariant in $\vect{\Omega}$ around the normal to the slab, the directional mean \eqref{eq:directional-mean-continuous} simplifies to
\begin{equation}\label{eq:slab-mean-1}
\avg{\matr{g}}^\pm_\kappa = \frac{\int_0^1 \mu^\kappa (1-\mu^2)^{\frac{d-3}{2}} \matr{g}(\pm\mu,x) \D{\mu}}{\int_0^1 \mu^\kappa (1-\mu^2)^{\frac{d-3}{2}} \D{\mu}}  \:,
\end{equation}
where we used $\D\vect{\Omega}=S_{d-1}(1-\mu^2)^{\frac{d-3}{2}}\D{\mu}$ and $\mu=\Omega_{\para}=\cos\theta$.
It would be tempting to apply a Gaussian quadrature directly to the integral in the numerator of Eq.\ \eqref{eq:slab-mean-1}.
However, this approach does not work because the radiance $\matr{g}(\mu,x)$ exhibits an accumulation point of poles at $\mu=0$, as evidenced by the functions $\tanh(1/\mu)$ in the approximate solution \eqref{eq:ufa-system-2}.
These poles are intrinsic to the matrix transport equation and do not arise from the transverse momentum quantization in the waveguide geometry.
They pose a challenge for the evaluation of the integral \eqref{eq:slab-mean-1}, as they are located close to the real axis of $\mu$.
In order to overcome this problem, it is necessary to move the integration path in the complex plane of $\mu$.
This can be done with the change of variable
\begin{equation}\label{eq:slab-contour-shift}
\mu(t) = t + \I a t(1-t^2)  \quad\forall t\in[0,1]  \:,
\end{equation}
where $a$ is an arbitrary contour parameter.
This parameter must be positive to avoid crossing the poles which are located in the region $\Im(\mu^2)<0$ according to Eqs.\ \eqref{eq:ufa-system-2}--\eqref{eq:ufa-sigma-2}.
After the change of variable \eqref{eq:slab-contour-shift}, Gaussian quadrature can be applied to the integral \eqref{eq:slab-mean-1}.
As a reminder, Gaussian quadrature is the discretization rule
\begin{equation}\label{eq:def-gauss-quad}
\int_{0}^{1} f(t) \D{t} \simeq \sum_{i=1}^{N_\mu} w_i \frac{f(t_i)}{W(t_i)}  \:,
\end{equation}
where $t_i$ is the $i$-th zero of the orthogonal polynomial of order $N_\mu$ defined with respect to the weight function $W(t)$ on the interval $t\in[0, 1]$, and $w_i$ is the corresponding Gaussian weight.
The number of points, $N_\mu$, is chosen large enough so that the sum converges.
The nodes $t_i$ and weights $w_i$ can be obtained numerically using for instance the Golub-Welsch algorithm \cite[Sec.~4.6.2]{Press2007}.
The quadrature is more accurate when the weight function $W(t)$ compensates for the possible singularities of $f(t)$ at the end points.
In the case of Eq.\ \eqref{eq:slab-mean-1}, an appropriate choice is
\begin{equation}\label{eq:weight-function}
W(t) = (1 - t)^{\frac{d-3}{2}}  \:.
\end{equation}
This choice is well-suited for the evaluation of $\tilde{\matr{Q}}(x)$ for which $\kappa=0$.
Using Eq.\ \eqref{eq:def-gauss-quad}, the directional mean \eqref{eq:slab-mean-1} can be written as
\begin{equation}\label{eq:slab-mean-3}
\avg{\matr{g}}_\kappa^\pm = \frac{\sum_{i=1}^{N_\mu} c_i\mu_i^{\kappa-1} \matr{g}(\pm\mu_i,x)}{\sum_{i=1}^{N_\mu} c_i\mu_i^{\kappa-1}}  \:,
\end{equation}
with the weights
\begin{equation}\label{eq:slab-mean-weights}
c_i = \frac{w_i}{W(t_i)} \mu'_i \mu_i (1-\mu_i^2)^{\frac{d-3}{2}}   \:,
\end{equation}
where $\mu_i=\mu(t_i)$ and $\textstyle\mu'_i=\der{\mu}{t}(t_i)$.
Equation \eqref{eq:slab-mean-3} reproduces the expression of the directional mean in the waveguide case [see Eq.\ \eqref{eq:directional-mean-discrete}].
In this way, only the values of $\mu_i$ and $c_i$ have to be adapted to the geometry.
The weights $c_i$ must be such that, in the continuum limit $N_\mu\rightarrow\infty$, they satisfy the normalization
\begin{equation}\label{eq:slab-mean-c-norm}
\sum_{i=1}^{N_\mu} c_i\mu_i^{\kappa-1} \xrightarrow{N_\mu\rightarrow\infty} \int_0^1 \mu^\kappa (1-\mu^2)^{\frac{d-3}{2}} \D{\mu} = \frac{V_{d+\kappa-2}}{S_{d-1}V_{\kappa-1}}  \:.
\end{equation}
This relation has been used to validate the numerical implementation.
\par
\begin{figure}[t]%
\includegraphics{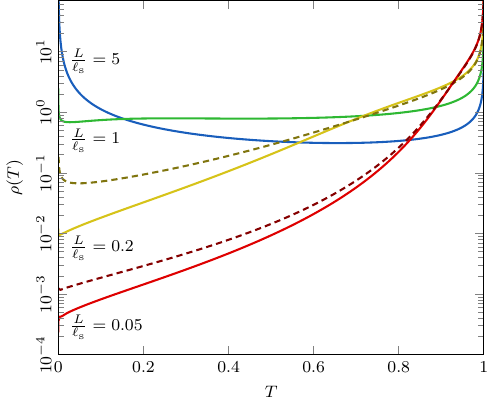}%
\caption{Transmission eigenvalue distribution through a 2D infinite slab, at different optical thicknesses $L/\ell$.
The solid curves are the predictions of the matrix transport equation \eqref{eq:full-eilenberger-2} \cite{GaspardD2025-ebsolve} using the discrete ordinates method \eqref{eq:slab-mean-3}--\eqref{eq:slab-mean-weights} to compute $\tilde{\matr{Q}}(x)$ and $\tilde{\matr{J}}_{\para}(x)$ in Eq.\ \eqref{eq:qn-jn-from-g-unified}.
The dashed curves are the quasiballistic approximation \eqref{eq:ufa-system-2}--\eqref{eq:ufa-gen-fun}.}%
\label{fig:rho-infinite-slab}%
\end{figure}

\begin{figure}[ht]%
\includegraphics{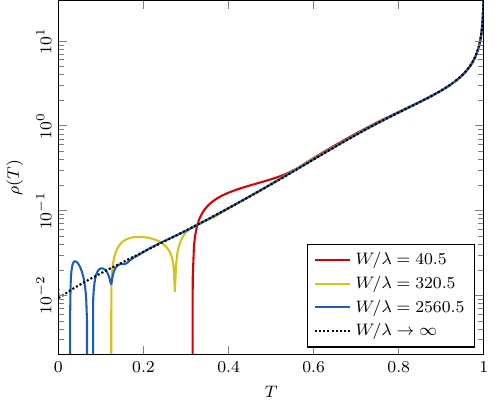}%
\caption{Transmission eigenvalue distribution through a 2D disordered medium of optical thickness $L/\ell=0.2$ predicted by RFT \cite{GaspardD2025-ebsolve}.
The solid curves are the distributions for a waveguide of increasingly large width based on Eqs.\ \eqref{eq:full-eilenberger-modal}, \eqref{eq:qn-modal}, \eqref{eq:jn-modal}, and \eqref{eq:mu-periodic}, and the dotted curve is the infinite-slab limit distribution based on Eqs.\ \eqref{eq:full-eilenberger-2} and \eqref{eq:slab-mean-3}--\eqref{eq:slab-mean-weights}.}%
\label{fig:rho-infinite-limit}%
\end{figure}

 The transmission eigenvalue density for the infinite slab is computed numerically \cite{GaspardD2025-ebsolve} using the same algorithm as for the waveguide in Sec.\ \ref{sec:numerical-waveguide}, except that the directional means in Eq.\ \eqref{eq:qn-jn-from-g-unified} must rely on Eqs.\ \eqref{eq:slab-mean-3}--\eqref{eq:slab-mean-weights}.
This algorithm yields the solid curves in Fig.\ \ref{fig:rho-infinite-slab}.
The main difference with the distributions for the waveguide in Fig.\ \ref{fig:rho-waveguide} is the absence of a minimum cutoff.
Transmission eigenvalues are thus allowed to reach the point $T=0$.
It should be noted, however, that the density in the region $T=0$ is likely to be influenced by strong localization effects because it comes from grazing modes (which propagate in a direction almost parallel to the slab).
Given that the present theory does not take into account strong localization (see Sec.\ \ref{sec:disorder-averaging}), possible changes of the density in the region $T=0$ could be expected.
The theoretical predictions of Fig.\ \ref{fig:rho-infinite-slab} cannot be compared to numerical simulations based on the wave equation \eqref{eq:wave-equation} because the continuum of modes cannot be discretized on complex angles in the same way as in Eqs.\ \eqref{eq:slab-mean-3}--\eqref{eq:slab-mean-weights}.
Nevertheless, as shown in Fig.\ \ref{fig:rho-infinite-limit}, it is possible to check that the distribution for the finite-width waveguide studied in Sec.\ \ref{sec:numerical-waveguide} converges, in the infinite-width limit ($W\rightarrow\infty$), to the infinite-slab distribution based on Eqs.\ \eqref{eq:slab-mean-3}--\eqref{eq:slab-mean-weights}.
We find that, when $W$ is large enough, the transmission eigenvalue density splits into several lobes, which are attributed to the grazing modes, i.e., modes with large transverse momentum ($\norm{\vect{p}_{\perp,n}}\simeq k$).
These modes travel a longer path in the medium and are therefore more strongly affected by scattering, leading to smaller $T$ values.
As $W$ increases, these lobes accumulate near $T=0$, becoming progressively smaller and thinner.
This behavior seems to be consistent with the limiting distribution for $W\rightarrow\infty$ (dotted line in Fig.\ \ref{fig:rho-infinite-limit}).

\section{Conclusions}\label{sec:conclusions}

\par In this paper, we formulated a field-theoretic framework, which we called radiant field theory (RFT), to describe the transmission eigenvalue distribution in disordered media. 
Unlike previous approaches, RFT applies both in the quasiballistic and diffusive regimes and accommodates arbitrary open geometries, including the infinite slab.
In Sec.\ \ref{sec:duplication}, we introduced a field-theoretic partition function $Z$, whose logarithmic derivative generates the sought distribution.
Using the replica method (Secs.\ \ref{sec:disorder-averaging}--\ref{sec:single-replica}), we evaluated the disorder-averaged $\avg{\ln Z}$ and showed that in the nonlocalized regime, the calculation reduces to a single replica.
The saddle-point of the resulting functional integral is given by a self-consistent equation for a $2\times2$ complex matrix field $\tilde{\matr{Q}}(\vect{r})$, as discussed in Sec.\ \ref{sec:saddle-point}.
Since $\tilde{\matr{Q}}(\vect{r})$ varies slowly on the wavelength scale (Appendix \ref{app:q-smoothness}), we employed a semiclassical approximation to derive a transport equation for the matrix radiance $\matr{g}(\vect{\Omega}, \vect{r})$, structurally analogous to the Eilenberger equation from type-II nonequilibrium superconductivity (Sec.\ \ref{sec:wigner}).
The boundary conditions for $\matr{g}$, derived in Sec.\ \ref{sec:boundaries}, are unusual in that they constrain only three of its four components in every direction.
The generality of the derivation in Sec.\ \ref{sec:field-theory} makes the RFT framework adaptable to a wide range of geometries, as illustrated in Fig.\ \ref{fig:boundary-conditions}.

\par In Appendix \ref{app:quasiballistic-solution}, we solved the matrix transport equation analytically in the quasiballistic regime, under the assumption that $\tilde{\matr{Q}}(\vect{r})$ remains spatially uniform in the bulk.
This assumption can be viewed as the spatial counterpart of the isotropy assumption used in the diffusive regime, where the radiance $\matr{g}(\vect{\Omega},\vect{r})$ is assumed to be weakly dependent on direction.
Our solution reveals that the transmission eigenvalue distribution is influenced by the waveguide shape via the weights $\mu_n$ in the directional mean \eqref{eq:directional-mean-discrete}, indicating a lack of universality in the quasiballistic regime.

\par In Sec.\ \ref{sec:results}, we solved the RFT equations numerically for two geometries: the waveguide and the infinite slab.
For waveguides (Sec.\ \ref{sec:numerical-waveguide}), we adapted the transport equation to account for transverse momentum quantization (Appendix \ref{app:modal-eilenberger}).
For the infinite slab (Sec.\ \ref{sec:numerical-slab}), we evaluated the directional integral by deforming the integration contour in the complex plane.
This method circumvents the limitations of standard wave-equation solvers, which require boundary conditions.
The ability of our theory to access the continuum limit is a direct consequence of the semiclassical formulation.
Figure \ref{fig:rho-infinite-limit} shows how the transmission distribution for the infinite slab emerges from wide waveguides, and highlights features such as the lobe structure near $T=0$ caused by grazing modes.
However, these modes may be sensitive to localization, so the infinite-slab distribution should be interpreted with care.

\par The companion Letter \cite{GaspardD2024-short} extends RFT to include experimentally relevant effects such as absorption \cite{Misirpashaev1997, Brouwer1998a} and incomplete channel control \cite{Goetschy2013, Hsu2017}.
Overall, our work demonstrates that it is possible to dispense with the restrictive macroscopic assumptions of random matrix theories like DMPK when modeling coherent wave transport in disordered media.
The microscopic nature of RFT offers a practical tool to investigate unresolved issues in wavefront shaping, including access to observables beyond transmission, such as energy deposition statistics \cite{Bender2022a, McIntosh2024, McIntosh2025} and the spatial structure of transmission eigenchannels \cite{Choi2011, Davy2015b, Yilmaz2019b}.
Its microscopic foundation also makes it well-suited to incorporate more complex physical phenomena, such as anisotropic scattering in biological tissues or correlated disorder \cite{Vynck2023}.
These directions will be explored in future work.

\begin{acknowledgments}%
The authors thank Romain Pierrat for early-stage discussions.
This research has been supported by the ANR project MARS\_light under reference ANR-19-CE30-0026, a grant from the Simons Foundation (No.\ 1027116), and by the program ``Investissements d'Avenir'' launched by the French Government.
\end{acknowledgments}%

\appendix

\section{Reduction to a nonlinear sigma model}\label{app:nonlinear-sigma}
In this appendix, we derive the Lagrangian of the matrix nonlinear sigma model following an approach inspired by Lerner \cite{Lerner2003}.
We start from the full Lagrangian \eqref{eq:full-lagrangian}
\begin{equation}\label{eq:nlsigma-full}
\mathcal{L}[\matr{Q}] = \Tr(\tfrac{\alpha}{2}\op{\matr{Q}}^2) - \Tr\ln(\op{\matr{W}}_0 + \I\alpha\op{\matr{Q}})  \:,
\end{equation}
where $\op{\matr{W}}_0$ is the Hamiltonian defined in Eq.\ \eqref{eq:def-matrix-w-zero}, and $\matr{Q}(\vect{r})$ is the $2R\times 2R$ matrix field, $R$ being the number of replicas.
Note that the results of this appendix are valid for arbitrary $R$, including the special case $R=1$.
\par The numerical simulations of the saddle-point equation \eqref{eq:saddle-point-1} in Fig.\ \ref{fig:qfield1d-dirac}(a) show that $\matr{Q}(\vect{r})$ behaves slowly in space, especially in the multiple-scattering regime ($L\gg\ell$).
In addition, $\matr{Q}(\vect{r})^2$ appears to be approximately proportional to the identity matrix.
We aim to derive these properties of $\matr{Q}(\vect{r})$ in the multiple-scattering regime.
To do so, we momentarily disregard the contact interaction terms $\gamma_\porta\op{K}_\porta\lmat_+$ and $\gamma_\portb\op{K}_\portb\lmat_-$, and assume that $\matr{Q}(\vect{r})$ is constant in the bulk.
When expressed in the momentum basis, the saddle-point condition \eqref{eq:saddle-point-1} reads
\begin{equation}\label{eq:nlsigma-q-norm-1}
\matr{Q} = \int \frac{\D\vect{p}}{(2\pi)^d} \frac{\I}{k^2 - \vect{p}^2 + \I\varepsilon\lmat_3 + \I\alpha\matr{Q}}  \:.
\end{equation}
To achieve the integral in Eq.\ \eqref{eq:nlsigma-q-norm-1} on the momentum shell ($\norm{\vect{p}}\in[k-\delta k,k+\delta k]$), we exploit the result
\begin{equation}\label{eq:on-shell-integration}\begin{split}
\int \frac{\D\vect{p}}{(2\pi)^d} f(\vect{p}^2 - k^2) & = \int_{k-\delta k}^{k+\delta k} \D{p} \frac{S_d p^{d-1}}{(2\pi)^d} f(\vect{p}^2 - k^2)  \:,\\
 & = \int_{-2k\delta k}^{2k\delta k} \D{u} \frac{S_d (k^2 + u)^{\frac{d-2}{2}}}{2(2\pi)^d} f(u)  \:,\\
 & \simeq \nu(k) \dashint_0 \D{u} f(u)  \:,
\end{split}\end{equation}
where $u=\vect{p}^2-k^2$, the dashed integral denotes the Cauchy principal value defined in Eq.\ \eqref{eq:def-cauchy-pv}, and $\nu(k)$ is the density of states \eqref{eq:cavity-dos}.
The result \eqref{eq:on-shell-integration} assumes that the interval is small ($\delta{k}\ll k$), but nevertheless large enough to encompass most of the contribution of $f(u)$.
Using Eq.\ \eqref{eq:on-shell-integration} and the notation $\matr{P}=\varepsilon\lmat_3 + \alpha\matr{Q}$, Eq.\ \eqref{eq:nlsigma-q-norm-1} becomes
\begin{equation}\label{eq:nlsigma-q-norm-2}
\matr{Q} = -\I\nu \dashint_0 \frac{\D{u}}{u - \I\matr{P}}  \:.
\end{equation}
The integral in Eq.\ \eqref{eq:nlsigma-q-norm-2} is given by
\begin{equation}\label{eq:integral-matrix-sign}
\dashint_0 \frac{\D{u}}{u - \I\matr{P}} = \I\pi\sign(\matr{P})   \:,
\end{equation}
where $\sign(\matr{P})$ is the matrix sign function given by eigendecomposition, $\sign(\matr{P})=\matr{U}\sign(\matr{\Pi})\matr{U}^{-1}$, $\matr{\Pi}$ being the diagonalized matrix and $\matr{U}$ a unitary matrix.
Therefore, the condition \eqref{eq:nlsigma-q-norm-2} reads
\begin{equation}\label{eq:nlsigma-q-norm-3}
\matr{Q} = \pi\nu \sign(\matr{P})  \:.
\end{equation}
Since $\sign(\matr{P})^2$ yields the identity matrix whatever the matrix $\matr{P}$, Eq.\ \eqref{eq:nlsigma-q-norm-3} implies that
\begin{equation}\label{eq:nlsigma-q-norm-4}
\matr{Q}^2 = (\pi\nu)^2 \matr{1}_{2R}  \:,
\end{equation}
where $\matr{1}_{2R}$ is the $2R\times 2R$ identity matrix.
We stress that Eq.\ \eqref{eq:nlsigma-q-norm-4} is only an approximation valid in the multiple-scattering regime ($L\gg\ell$).
It is not valid in the more general transport theory considered in Sec.\ \ref{sec:eilenberger-theory}, for instance, because $L$ is then arbitrary.
\par Now, we restore the presence of the contact terms and relax the assumption of constant $\matr{Q}$.
We let this field slowly vary in space, but still under the constraint \eqref{eq:nlsigma-q-norm-4}.
The field has thus the form
\begin{equation}\label{eq:nlsigma-q-manifold}
\matr{Q}(\vect{r}) = \pi\nu\, \matr{U}(\vect{r}) \lmat \matr{U}(\vect{r})^{-1}  \:,
\end{equation}
where $\lmat$ is an arbitrary constant matrix such that $\lmat^2=\matr{1}_{2R}$ but excluding the identity itself ($\lmat\neq\matr{1}_{2R}$), for instance, $\lmat=\lmat_3$.
It is also convenient to use the normalized matrix field $\tilde{\matr{Q}}(\vect{r})$ defined by Eq.\ \eqref{eq:def-normalized-q}.
To deal with the Lagrangian \eqref{eq:nlsigma-full}, we first define a Green's operator based on the most significant terms in the Lagrangian, namely $k^2$, $\op{\vect{p}}^2$, and $\I\alpha\matr{Q}$:
\begin{equation}\label{eq:def-green-l}
\op{\matr{L}} \defeq \frac{1}{k^2 - \op{\vect{p}}^2 + \I\alpha\matr{Q}(\op{\vect{r}})}  \:.
\end{equation}
With this operator, the Lagrangian \eqref{eq:nlsigma-full} can be expanded into
\begin{equation}\label{eq:nlsigma-log-expansion}\begin{split}
\Tr\ln(\op{\matr{W}}_0 + \I\alpha\op{\matr{Q}}) & = \Tr\ln\op{\matr{L}}^{-1} + \Tr\ln(1 + \op{\matr{L}} \I\varepsilon\lmat_3 + \ldots)  \:,\\
 & \simeq \Tr\ln\op{\matr{L}}^{-1} + \I\varepsilon \Tr(\op{\matr{L}}\lmat_3) + \cdots  \:.
\end{split}\end{equation}
The trailing dots represent other terms in the Hamiltonian $\op{\matr{W}}_0$, such as the contact interaction.
In the following calculations, we consider each of the terms of the last line of Eq.\ \eqref{eq:nlsigma-log-expansion}.

\paragraph{Gradient expansion}
We first deal with the first term in the right-hand side of Eq.\ \eqref{eq:nlsigma-log-expansion}.
Applying the similarity transformation $\op{\matr{U}}^{-1}\cdots\op{\matr{U}}$ defined in Eq.\ \eqref{eq:nlsigma-q-manifold} to the logarithm, we get
\begin{equation}\label{eq:trlogl-1}
\Tr\ln(\op{\matr{L}}^{-1}) = \Tr\ln(\op{\matr{L}}_0^{-1} - \op{\matr{U}}^{-1}[\op{\vect{p}}^2, \op{\matr{U}}])  \:,
\end{equation}
where the Green's operator $\op{\matr{L}}_0$ is defined by
\begin{equation}\label{eq:def-green-l-zero}
\op{\matr{L}}_0 \defeq \frac{1}{k^2 - \op{\vect{p}}^2 + \I\epsilonscat\lmat}  \:,
\end{equation}
and the imaginary shift is given by
\begin{equation}\label{eq:def-epsilon-scat}
\epsilonscat \defeq \pi\nu\alpha = \frac{k}{\ell}  \:.
\end{equation}
Expression \eqref{eq:trlogl-1} can be rewritten as
\begin{equation}\label{eq:trlogl-2}
\Tr\ln(\op{\matr{L}}^{-1}) = \Tr\ln(\op{\matr{L}}_0^{-1}) + \Tr\ln(1 - \op{\matr{L}}_0\op{\matr{U}}^{-1}[\op{\vect{p}}^2, \op{\matr{U}}])  \:.
\end{equation}
Since the commutator in Eq.\ \eqref{eq:trlogl-2} can be understood as the gradient of $\matr{U}(\vect{r})$, it is expected to be relatively small compared to the other terms, the logarithm can be expanded in Taylor series
\begin{equation}\label{eq:trlogl-expansion}
\Tr\ln(\op{\matr{L}}^{-1}) = \Tr\ln(\op{\matr{L}}_0^{-1}) - I_1 - \frac{1}{2} I_2 + \cdots  \:,
\end{equation}
where $I_{\beta}$ is defined for all $\beta\in\{1,2,\ldots\}$ by
\begin{equation}\label{eq:def-ibeta}
I_{\beta} \defeq \Tr\left( (\op{\matr{L}}_0\op{\matr{U}}^{-1}[\op{\vect{p}}^2, \op{\matr{U}}])^\beta \right)  \:.
\end{equation}
Since some operators in Eq.\ \eqref{eq:def-ibeta} are diagonal in the momentum basis and others in the position basis, the trace does not evaluate exactly.
To overcome this issue, we switch to the Wigner representation \eqref{eq:def-wigner-transform} in which the trace reads
\begin{equation}\label{eq:trace-wigner}
\Tr\op{A} = \iint \frac{\D\vect{r} \D\vect{p}}{(2\pi)^d} \wigner{\op{A}}  \:.
\end{equation}
Then, an excellent approximation of Eq.\ \eqref{eq:def-ibeta} can be obtained from
\begin{equation}\label{eq:wigner-factorization}
\wigner{\op{\matr{L}}_0\op{\matr{U}}^{-1}[\op{\vect{p}}^2, \op{\matr{U}}]} \simeq \wigner{\op{\matr{L}}_0} \wigner{\op{\matr{U}}^{-1}} \wigner{[\op{\vect{p}}^2, \op{\matr{U}}]}  \:,
\end{equation}
although this factorization holds only for commuting operators.
The first two Wigner terms in the right-hand side of Eq.\ \eqref{eq:wigner-factorization} are trivial, while the last one reads
\begin{equation}\label{eq:commut-u-gradient}
\wigner{[\op{\vect{p}}^2, \op{\matr{U}}]} = -2\I\vect{p}\cdot\grad_{\vect{r}}\matr{U}(\vect{r})  \:.
\end{equation}
We notice that the first integral in Eqs.\ \eqref{eq:trlogl-expansion}--\eqref{eq:def-ibeta} vanishes,
\begin{equation}\label{eq:i1-is-zero}
I_1 = 0  \:,
\end{equation}
since the average momentum is zero: $\textstyle\int\vect{p}\D\vect{p}=\vect{0}$.
However, this is not the case of the second integral, $I_2$, that is
\begin{equation}\label{eq:i2-step-1}
I_2 = -4 \sum_{i,j=1}^{d} \iint \frac{\D\vect{r} \D\vect{p}}{(2\pi)^d} p_i p_j \tr\left( \matr{L}_0 \matr{A}_i \matr{L}_0 \matr{A}_j \right)  \:,
\end{equation}
where
\begin{equation}
\matr{A}_i(\vect{r}) \defeq \matr{U}(\vect{r})^{-1} \partial_{x_i}\matr{U}(\vect{r})  \:.
\end{equation}
The integral over the momentum in Eq.\ \eqref{eq:i2-step-1} can be achieved using a generalization of the property \eqref{eq:on-shell-integration} to dyadic tensors, namely
\begin{equation}\label{eq:on-shell-dyadic}
\int \frac{\D\vect{p}}{(2\pi)^d} p_i p_j f(\vect{p}^2-k^2) = k^2\nu \frac{\delta_{ij}}{d} \dashint_0 \D{u} f(u)  \:.
\end{equation}
Using Eq.\ \eqref{eq:on-shell-dyadic}, Eq.\ \eqref{eq:i2-step-1} can be written as
\begin{equation}\label{eq:i2-step-2}
I_2 = -4 \int \D\vect{r} \frac{k^2\nu}{d} \sum_{i=1}^{d} \dashint_0 \D{u} \tr\left[ \matr{L}_0(u) \matr{A}_i \matr{L}_0(u) \matr{A}_i \right]  \:,
\end{equation}
The integral over the momentum shell $u$ can be evaluated as follows
\begin{equation}\label{eq:i2-integral-1}\begin{split}
& \dashint_0 \D{u} \tr\left( \frac{1}{u - \I\epsilonscat\lmat} \matr{A}_i \frac{1}{u - \I\epsilonscat\lmat} \matr{A}_i \right)  \\
& = \dashint_0 \D{u} \tr\left( \frac{u + \I\epsilonscat\lmat}{u^2 + \epsilonscat^2} \matr{A}_i \frac{u + \I\epsilonscat\lmat}{u^2 + \epsilonscat^2} \matr{A}_i \right)  \\
& = \dashint_0 \D{u} \frac{u^2}{(u^2 + \epsilonscat^2)^2} \tr(\matr{A}_i^2) - \frac{\epsilonscat^2}{(u^2 + \epsilonscat^2)^2} \tr(\lmat\matr{A}_i\lmat\matr{A}_i)  \:.
\end{split}\end{equation}
The two remaining integrals over $u$ in the last line of Eq.\ \eqref{eq:i2-integral-1} turns out to be equal:
\begin{equation}\label{eq:i2-integral-2}
\dashint_0 \frac{u^2 \D{u}}{(u^2 + \epsilonscat^2)^2} = \dashint_0 \frac{\epsilonscat^2 \D{u}}{(u^2 + \epsilonscat^2)^2} = \frac{\pi}{2\epsilonscat}  \:.
\end{equation}
Therefore, Eq.\ \eqref{eq:i2-step-2} becomes
\begin{equation}\label{eq:i2-step-3}
I_2 = -4 \int \D\vect{r} \frac{k^2\nu}{d} \frac{\pi}{2\epsilonscat} \sum_{i=1}^{d} [ \tr(\matr{A}_i^2) - \tr(\lmat\matr{A}_i\lmat\matr{A}_i) ]  \:.
\end{equation}
We recognize in the difference $\matr{A}_i^2-\lmat\matr{A}_i\lmat\matr{A}_i$ the square of a commutator due to the fact that $\lmat^2=\matr{1}_{2R}$:
\begin{equation}\label{eq:i2-step-4}
I_2 = \int \D\vect{r} \frac{k^2\pi\nu}{d\epsilonscat} \sum_{i=1}^{d} \tr([\lmat, \matr{A}_i]^2)  \:.
\end{equation}
In addition, according to Eq.\ \eqref{eq:nlsigma-q-manifold}, the commutator is directly related to the gradient of the $\tilde{\matr{Q}}(\vect{r})$ field:
\begin{equation}
[\lmat, \matr{A}_i(\vect{r})] = -\matr{U}(\vect{r})^{-1} (\partial_{x_i} \tilde{\matr{Q}}(\vect{r})) \matr{U}(\vect{r})  \:.
\end{equation}
Therefore, Eq.\ \eqref{eq:i2-step-4} reduces to
\begin{equation}\label{eq:i2-step-5}
I_2 = \int \D\vect{r}\, \pi\nu D \tr(\grad_{\vect{r}}\tilde{\matr{Q}}\grad_{\vect{r}}\tilde{\matr{Q}})  \:,
\end{equation}
where $D$ is the diffusion constant defined by Eq.\ \eqref{eq:def-diffusivity}.
The sought approximation of Eq.\ \eqref{eq:trlogl-expansion} in the multiple-scattering regime thus reads
\begin{equation}\label{eq:nlsigma-gradient-result}
\Tr\ln(\op{\matr{L}}^{-1}) = \Tr\ln(\op{\matr{L}}_0^{-1}) - \int \D\vect{r}\, \frac{\pi\nu D}{2} \tr(\grad_{\vect{r}}\tilde{\matr{Q}}\grad_{\vect{r}}\tilde{\matr{Q}})  \:.
\end{equation}

\paragraph{Other terms and result}
One still has to deal with the other terms in Eq.\ \eqref{eq:nlsigma-log-expansion}, especially the term $\Tr(\op{\matr{L}}\lmat_3)$.
Using the Wigner representation \eqref{eq:trace-wigner}, we can write
\begin{equation}\label{eq:nlsigma-abso-1}
\Tr(\op{\matr{L}}\lmat_3) = \int \frac{\D\vect{r} \D\vect{p}}{(2\pi)^d} \tr\left( \frac{1}{k^2 - \vect{p}^2 + \I\alpha\matr{Q}(\vect{r})} \lmat_3 \right)  \:,
\end{equation}
which quickly evaluates from Eqs.\ \eqref{eq:nlsigma-q-norm-1}--\eqref{eq:nlsigma-q-norm-3}, the result being
\begin{equation}\label{eq:nlsigma-abso-2}
\Tr(\op{\matr{L}}\lmat_3) = \int \D\vect{r}\, (-\I\pi\nu) \tr( \tilde{\matr{Q}} \lmat_3 )  \:.
\end{equation}
Other terms in the Hamiltonian can be obtained in a similar way.
\par In the end, Eqs.\ \eqref{eq:nlsigma-gradient-result} and \eqref{eq:nlsigma-abso-2} can be substituted into Eq.\ \eqref{eq:nlsigma-log-expansion}, and then back into the Lagrangian \eqref{eq:nlsigma-full}. The result is
\begin{equation}\label{eq:full-lagrangian-from-nlsigma}
\mathcal{L}[\matr{Q}] = \Tr(\tfrac{\alpha}{2}\op{\matr{Q}}^2 + \ln\op{\matr{L}}_0) + \mathcal{M}[\tilde{\matr{Q}}]  \:,
\end{equation}
where $\mathcal{M}[\tilde{\matr{Q}}]$ is the famous nonlinear sigma model Lagrangian \cite{Efetov1997}
\begin{equation}\label{eq:nlsigma-lagrangian}
\mathcal{M}[\tilde{\matr{Q}}] = \int \D{\vect{r}}\, M ,\quad
M = \pi\nu \tr\left( \tfrac{D}{2} (\grad_{\vect{r}}\tilde{\matr{Q}})^2 - \varepsilon\lmat_3\tilde{\matr{Q}} \right) .
\end{equation}
The first terms in the right-hand side of Eq.\ \eqref{eq:full-lagrangian-from-nlsigma} can be ignored because they are responsible for the normalization \eqref{eq:nlsigma-q-norm-4} of $\matr{Q}(\vect{r})$ and are thus redundant under this constraint.
\par Due to the constraint $\tilde{\matr{Q}}(\vect{r})^2=\matr{1}_{2R}$ prescribed by Eq.\ \eqref{eq:nlsigma-q-norm-4}, the saddle-point equation associated to the Lagrangian \eqref{eq:nlsigma-lagrangian} is not given by the usual Euler-Lagrange equation for $\tilde{\matr{Q}}(\vect{r})$ but instead by the commutator equation
\begin{equation}\label{nlsigma-euler-lagrange}
\left[ \tilde{\matr{Q}}, \grad_{\vect{r}}\cdot\left(\pder{M}{\grad_{\vect{r}}\tilde{\matr{Q}}}\right) - \pder{M}{\tilde{\matr{Q}}} \right] = \matr{0} ,
\end{equation}
which for the particular Lagrangian \eqref{eq:nlsigma-lagrangian} reduces to
\begin{equation}\label{eq:nlsigma-usadel-1}
\left[ \tilde{\matr{Q}}, \grad_{\vect{r}}\cdot(D\grad_{\vect{r}}\tilde{\matr{Q}}) + \varepsilon\matr{\Lambda}_3 \right] = \matr{0} ,
\end{equation}
or equivalently according to the anticommutation relation $\tilde{\matr{Q}}\grad_{\vect{r}}\tilde{\matr{Q}} + (\grad_{\vect{r}}\tilde{\matr{Q}})\tilde{\matr{Q}} = \matr{0}$ resulting from the constraint \eqref{eq:nlsigma-q-norm-4},
\begin{equation}\label{eq:nlsigma-usadel-2}
\grad_{\vect{r}}\cdot(D\tilde{\matr{Q}}\grad_{\vect{r}}\tilde{\matr{Q}}) = \tfrac{\varepsilon}{2} [\matr{\Lambda}_3, \tilde{\matr{Q}}] .
\end{equation}
This matrix diffusion equation is reminiscent of the Usadel equation in the superconductivity literature \cite{Usadel1970, Kamenev2023, Nazarov1994a, Nazarov2009} and can be obtained as the diffusive limit of the Eilenberger-type transport equation \eqref{eq:full-eilenberger-2}.

\section{Smoothness of the Q field}\label{app:q-smoothness}
\par In order to approach Eqs.\ \eqref{eq:saddle-point-2}--\eqref{eq:self-consistency-1} semiclassically, we have to assume that the saddle-point field $\matr{Q}(\vect{r})$ slowly varies in space at the wavelength scale.
However, this assumption is anything but trivial.
Indeed, as pointed out in Refs.\ \cite{Zaitsev1984, Nazarov1999b}, the complete phase-envelope decomposition of the Green's function $\matr{\Gamma}(\vect{r}, \vect{r}')$ in a waveguide reads
\begin{equation}\label{eq:zaitsev-decomposition}\begin{split}
& \matr{\Gamma}(\vect{r}, \vect{r}') = \\
& \sum_{n=1}^{\infty} \sum_{\substack{\sigma=\pm\\ \sigma'=\pm}} \matr{C}^{(n)}_{\sigma\sigma'}(x, x') \E^{\I k_{\para,n}(\sigma x + \sigma'x')} \chi_{n}(\vect{y}) \cc{\chi}_{n}(\vect{y}')  \:,
\end{split}\end{equation}
where $\matr{C}_{++},\matr{C}_{+-},\matr{C}_{-+},\matr{C}_{--}$ are the slowly varying envelopes and $n$ is the index of modes.
We notice that the envelopes $\matr{C}_{++}$ and $\matr{C}_{--}$ are responsible for fast oscillations of the field $\matr{Q}(\vect{r})$ at the wavelength scale because of the behavior $\E^{\pm 2\I k_{\para,n}x}$ at the return point ($\vect{r}=\vect{r}'$).
In contrast, the envelopes $\matr{C}_{+-}$ and $\matr{C}_{-+}$ lead to slow variations of $\matr{Q}(\vect{r})$.
The point is that the oscillations corresponding to $\matr{C}_{++}$ and $\matr{C}_{--}$ are typically caused by the presence of obstacles such as interfaces or boundaries \cite{Zaitsev1984}, because these structures generally produce interferences between the incident and the reflected waves.
It can be shown that these interferences exponentially decay in space at the scale of the mean free path $\ell$.
\par Since the edges of the disordered region and the contact interactions $\gamma_\porta\op{K}_\porta$ and $\gamma_\portb\op{K}_\portb$ are likely to cause oscillations of $\matr{Q}(\vect{r})$, we think it necessary to carry out a numerical verification of the smoothness of the field $\matr{Q}(\vect{r})$.
To this end, we restrict the problem to one dimension ($d=1$).
The full saddle-point equation we consider thus reads
\begin{equation}\label{eq:saddle-point-with-obstacle}\begin{split}
\bigg( \partial_{x}^2 & + k^2 + \I\varepsilon\lmat_3 + \I\alpha\matr{Q}(x) + \gamma_\porta\op{K}_\porta\lmat_+ + \gamma_\portb\op{K}_\portb\lmat_-  \\
 & + B(x) \bigg) \matr{\Gamma}(x, x') = \I\matr{1}_2\delta(x-x')  \:,
\end{split}\end{equation}
where $B(x)$ is an additional obstacle which is not considered for the moment ($B=0$), and $\alpha(x)$ is given by
\begin{equation}\label{eq:alpha-qfield1d}
\alpha(x) = \frac{k}{\pi\nu\ell} \frac{\tanh(\frac{x}{\varsigma}) - \tanh(\frac{x - L}{\varsigma})}{2}  \:,
\end{equation}
$\varsigma$ being a smoothing length.
From the numerical point of view, the solution of Eqs.\ \eqref{eq:saddle-point-with-obstacle}--\eqref{eq:alpha-qfield1d} can be obtained iteratively starting from the initial ansatz $\matr{Q}(x)=\matr{0}$.
The question of the convergence of this iterative process is discussed in Appendix \ref{app:convergence-and-smoothness}.
\par
\begin{figure*}[ht]%
\includegraphics{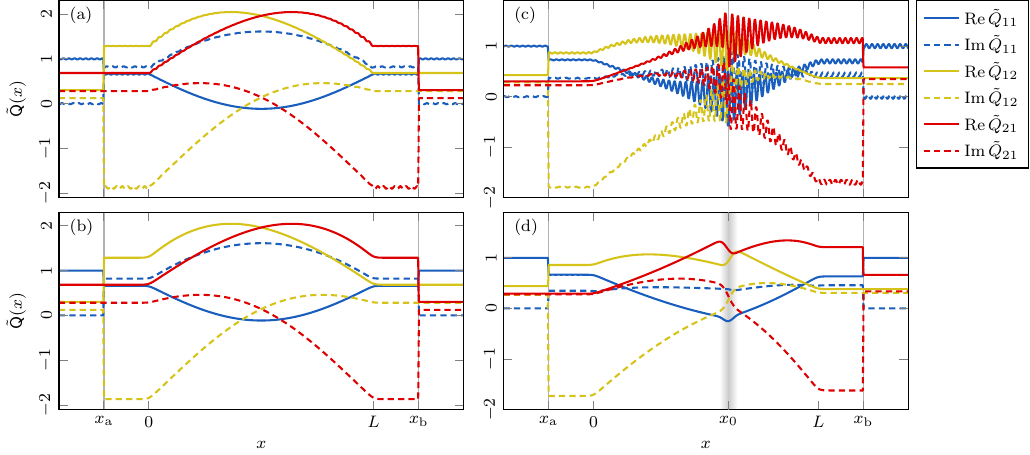}%
\caption{Normalized field $\tilde{\matr{Q}}(x)=\tfrac{1}{\pi\nu}\matr{Q}(x)$, satisfying the one-dimensional saddle-point equation \eqref{eq:saddle-point-with-obstacle}--\eqref{eq:alpha-qfield1d} \cite{GaspardD2025-qfield}, for thickness $L=20\lambda$, optical thickness $L/\ell=5$, and parameters $\gamma_\porta=\gamma_\portb=1.2+10^{-5}\I$.
(a) Field for sharp edges ($\varsigma=0$) without obstacle ($B=0$).
(b) Field for smooth edges ($\varsigma=L/50$) without obstacle ($B=0$).
(c) Field for smooth edges ($\varsigma=L/50$) and sharp obstacle ($\gamma_0=1$, $\sigma_\para=0$).
(d) Field for smooth edges ($\varsigma=L/50$) and smooth obstacle ($\gamma_0=1$, $\sigma_\para=L/50$).}%
\label{fig:qfield1d-dirac}%
\end{figure*}

 Numerical solutions of Eq.\ \eqref{eq:saddle-point-with-obstacle} for the normalized field $\tilde{\matr{Q}}(x)=\tfrac{1}{\pi\nu}\matr{Q}(x)$ are shown in Fig.\ \ref{fig:qfield1d-dirac}.
The omitted matrix element $\tilde{Q}_{22}$ is given by $\tilde{Q}_{22}=-\tilde{Q}_{11}$.
This exact relation comes from the traceless property of the field $\tilde{\matr{Q}}(x)$ which is proved in Sec.\ \ref{sec:boundaries}.
We notice in Fig.\ \ref{fig:qfield1d-dirac}(a) that the sharp edges of the disordered region are responsible for small oscillations of the field.
These oscillations disappear in Fig.\ \ref{fig:qfield1d-dirac}(b) when the smoothing length $\varsigma$ becomes comparable or larger than the wavelength.
Here, the reader may legitimately wonder why no interference is visible in Fig.\ \ref{fig:qfield1d-dirac}(b) although the contact interactions $\gamma_\porta\op{K}_\porta$ and $\gamma_\portb\op{K}_\portb$ behave as Dirac deltas (see Eq.\ \eqref{eq:contact-op-current}).
In fact, it can be shown by solving Eq.\ \eqref{eq:saddle-point-2} in the representation \eqref{eq:zaitsev-decomposition} that these particular interactions do not cause reflection of the incident wave because they are based on the current operator \eqref{eq:contact-op-current}.
However, this calculation is relatively tedious and is not presented here.
The reader may also object that the oscillations in Fig.\ \ref{fig:qfield1d-dirac}(a) are very small compared to the magnitude of the field.
However, this is not trivial because if we insert the Gaussian obstacle
\begin{equation}\label{eq:gaussian-obstacle}
B(x) = \frac{\gamma_0 k}{\sqrt{2\pi}\sigma_\para} \E^{-\frac{1}{2} \left(\frac{x - x_0}{\sigma_\para}\right)^2}
\end{equation}
in Eq.\ \eqref{eq:saddle-point-with-obstacle}, then the oscillations become significant in the Dirac-delta limit ($\sigma_\para\rightarrow 0$) as shown in Fig.\ \ref{fig:qfield1d-dirac}(c).
These oscillations can only be mitigated by smoothing out the obstacle at the wavelength scale as done in Fig.\ \ref{fig:qfield1d-dirac}(d).
This smoothing removes the oscillations without significantly altering the mean field.
\par In the following sections, we will assume that there is no sharp obstacle in the waveguide so that the $\matr{Q}(\vect{r})$ field behaves as a smooth function at the wavelength scale.

\section{Convergence and stability of the Q field}\label{app:convergence-and-smoothness}
\par Here, we demonstrate that the iterative solution of the system made of the saddle-point equation \eqref{eq:saddle-point-2} and the self-consistent condition \eqref{eq:self-consistency-1} actually converges if the initial ansatz is close enough to the solution.
To this end, we first solve the saddle-point equation, and then show that, after a perturbation, the field $\matr{Q}(\vect{r})$ converges back to the saddle point.
We will refer to the saddle-point field as $\matr{Q}^{(0)}$ to distinguish it from the fields at subsequent iterations.
This approach proves the convergence and the stability of the solution under the iterations of the saddle-point equation.
\par The saddle-point equations \eqref{eq:saddle-point-2}--\eqref{eq:self-consistency-1} reads in the basis of the transverse eigenmodes,
\begin{equation}\label{eq:stab-saddle-1}\begin{cases}
\left( \partial_{x}^2 + k_{\para}^2 + \I\alpha\matr{Q}^{(0)} + \op{\matr{Y}} \right) \matr{\Gamma}^{(0)}(\vect{p}_\perp, x, x') = \I\delta(x-x')  \:,\\
\matr{Q}^{(0)}(x) = \int_{\norm{\vect{p}_\perp}<k} \frac{\D\vect{p}_\perp}{(2\pi)^{d-1}} \matr{\Gamma}^{(0)}(\vect{p}_\perp, x, x)  \:,
\end{cases}\end{equation}
where $k_{\para}=(k^2-\vect{p}_\perp^2)^{\frac{1}{2}}$ is the longitudinal wavenumber, $\vect{p}_\perp$ the transverse momentum, and $\op{\matr{Y}}=\gamma_\porta\op{K}_\porta\lmat_+ + \gamma_\portb\op{K}_\portb\lmat_-$ contains the contact interactions.
In the absence of contact interaction ($\op{\matr{Y}}=\matr{0}$), the solution of Eq.\ \eqref{eq:stab-saddle-1} is a constant field $\matr{Q}^{(0)}$ satisfying the normalization condition \eqref{eq:nlsigma-q-norm-3}, i.e.,
\begin{equation}\label{eq:stab-q0-gen-sol}
\matr{Q}^{(0)} = \pi\nu \sign(\matr{Q}^{(0)})  \:,
\end{equation}
$\sign(\cdot)$ being the matrix sign introduced in Eq.\ \eqref{eq:integral-matrix-sign}.
An important consequence of this normalization condition is that the two eigenvalues of $\matr{Q}^{(0)}$ are $+\pi\nu$ and $-\pi\nu$, which are real numbers.
The set of solutions for $\matr{Q}^{(0)}$ is degenerate because any global gauge transformation $\matr{Q}^{(0)}\rightarrow\matr{U}\matr{Q}^{(0)}\matr{U}^{-1}$ preserves the validity of the solution.
However, this gauge symmetry is universally broken by the contact interactions $\gamma_\porta\op{K}_\porta\lmat_+$ and $\gamma_\portb\op{K}_\portb\lmat_-$.
These terms are responsible for the spatial variation of the saddle-point field $\matr{Q}^{(0)}(x)$.
\par Now, we let the matrix field $\matr{Q}^{(0)}(x)$ in the first line of Eq.\ \eqref{eq:stab-saddle-1} be modified by a spatially localized perturbation called $\matr{\Delta}^{(0)}(x)$.
This perturbation is assumed to be much smaller than $\matr{Q}^{(0)}(x)$,
\begin{equation}\label{eq:stab-delta0-smallness}
\norm{\matr{\Delta}^{(0)}(x)} \ll \norm{\matr{Q}^{(0)}} = \pi\nu  \:,
\end{equation}
where $\norm{\cdot}$ stands for the matrix spectral norm defined as the largest eigenvalue in absolute value.
In addition, we assume that the field $\matr{\Delta}^{(0)}(x)$ is summable in norm, meaning that the integral $\textstyle\int\D x\:\norm{\matr{\Delta}^{(0)}(x)}$ should exist.
The perturbation $\matr{\Delta}^{(0)}(x)$ affects the Green's function, denoted $\matr{\Gamma}^{(1)}$, according to the first line of Eq.\ \eqref{eq:stab-saddle-1}, which, in turn, alters the matrix field $\matr{Q}^{(1)}(x)$.
This self-consistent scheme can be iterated \emph{ad infinitum}, leading to the following set of equations for all $t\in\mathbb{Z}_{\geq 0}$:
\begin{equation}\label{eq:stab-saddle-2}\begin{cases}
\begin{aligned}
\Big[ \partial_{x}^2 + k_{\para}^2 & + \I\alpha \Big( \matr{Q}^{(0)} + \matr{\Delta}^{(t)}(x) \Big) \\
 & + \op{\matr{Y}} \Big] \matr{\Gamma}^{(t+1)}(\vect{p}_\perp, x, x') = \I\delta(x-x')
\end{aligned} \\
\matr{Q}^{(t+1)}(x) = \int \frac{\D\vect{p}_\perp}{(2\pi)^{d-1}} \matr{\Gamma}^{(t+1)}(\vect{p}_\perp, x, x)  \:.
\end{cases}\end{equation}
The perturbation $\matr{\Delta}^{(t)}(x)$ at iteration $t$ in the first line of Eq.\ \eqref{eq:stab-saddle-2} is defined by the following relation to the total matrix $\matr{Q}^{(t)}(x)$:
\begin{equation}\label{eq:stab-qtot-from-delta}
\matr{Q}^{(t)}(x) = \matr{Q}^{(0)} + \matr{\Delta}^{(t)}(x)  \qquad\forall t\in\mathbb{Z}_{>0}  \:.
\end{equation}
The first line of Eq.\ \eqref{eq:stab-saddle-2} can be formally solved from perturbation theory treating $\matr{\Delta}^{(t)}(x)$ as the small parameter:
\begin{equation}\label{eq:stab-gamma-perturbation}
\op{\matr{\Gamma}}^{(t+1)} \simeq \op{\matr{\Gamma}}^{(0)} - \alpha\op{\matr{\Gamma}}^{(0)} \matr{\Delta}^{(t)}(\op{x}) \op{\matr{\Gamma}}^{(0)}  \:.
\end{equation}
Projecting in position space with $\bra{x}\cdots\ket{x'}$, integrating over the transverse modes with the second line of Eq.\ \eqref{eq:stab-saddle-2}, and then using Eq.\ \eqref{eq:stab-qtot-from-delta}, we get the iterative equation for the perturbation:
\begin{equation}\label{eq:stab-delta-map}\begin{split}
\matr{\Delta}^{(t+1)}(x) & = -\alpha \int\frac{\D\vect{p}_\perp}{(2\pi)^{d-1}}  \\
\times & \int\D x'\: \matr{\Gamma}^{(0)}(\vect{p}_\perp, x, x') \matr{\Delta}^{(t)}(x') \matr{\Gamma}^{(0)}(\vect{p}_\perp, x', x)  \:.
\end{split}\end{equation}
The matrix Green's function $\matr{\Gamma}^{(0)}$ in Eq.\ \eqref{eq:stab-delta-map} is the solution of Eq.\ \eqref{eq:stab-saddle-1}, i.e., 
\begin{equation}\label{eq:stab-gamma0-sol-2}
\matr{\Gamma}^{(0)}(\vect{p}_\perp, x, x') = \frac{\sign(\matr{Q}^{(0)}) \E^{ \I\sign(\matr{Q}^{(0)}) \sqrt{k_{\para}^2 + \I\alpha\matr{Q}^{(0)}} \abs{x-x'} }}{2\sqrt{k_{\para}^2 + \I\alpha\matr{Q}^{(0)}}}  \:.
\end{equation}
This Green's function exponentially vanishes at the scale of the mean free path for $\abs{x-x'}\rightarrow\infty$. 
Although the solution \eqref{eq:stab-gamma0-sol-2} holds exactly only for a constant field $\matr{Q}^{(0)}$ (in the case $\op{\matr{Y}}=\matr{0}$), we assume that it also holds for a spatially varying field in the diffusive regime ($\ell\ll L$).
Indeed, in this regime, the exponential decay of $\matr{\Gamma}^{(0)}$ makes it insensitive to the possible obstacles that could be found beyond several mean free paths.
We thus expect that the solution \eqref{eq:stab-gamma0-sol-2} holds in the bulk.
\par The linear map \eqref{eq:stab-delta-map} describes the evolution of the perturbation $\matr{\Delta}^{(t)}(x)$ under the iterations of the self-consistent equations \eqref{eq:stab-saddle-2}.
This equation is important because it shows that, however localized the initial perturbation, the function $\matr{\Delta}^{(t)}(x)$ will inevitably undergo spatial diffusion due to the exponential decay of $\matr{\Gamma}^{(0)}$ given by Eq.\ \eqref{eq:stab-gamma0-sol-2}.
In fact, since the Green's function $\matr{\Gamma}^{(0)}$ decreases at the scale of the mean free path, Eq.\ \eqref{eq:stab-delta-map} can be physically interpreted as a kind of evolution equation for $\matr{\Delta}^{(t)}(x)$ where the iteration index $t$ plays the role of time in the units of the mean intercollisional time.
\par In order to continue the demonstration of convergence, we will show that the map \eqref{eq:stab-delta-map} is actually a contraction, that is a map which sends any initial function $\matr{\Delta}^{(0)}(x)$ to zero after enough iterations.
Taking the matrix norm $\norm{\cdot}$ on both sides of Eq.\ \eqref{eq:stab-delta-map}, using the triangle inequality to deal with the integral and exploiting the submultiplicativity of the matrix norm ($\norm{\matr{A}\matr{B}}\leq\norm{\matr{A}}\norm{\matr{B}}$), we get the inequality:
\begin{equation}\label{eq:stab-upper-bound-1}\begin{split}
& \norm{\matr{\Delta}^{(t+1)}(x)} \leq \alpha \int\frac{\D\vect{p}_\perp}{(2\pi)^{d-1}}  \\
& \times \int\D x'\: \norm{\matr{\Gamma}^{(0)}(\vect{p}_\perp, x, x')}^2 \norm{\matr{\Delta}^{(t)}(x')}  \:.
\end{split}\end{equation}
Integrating both sides of Eq.\ \eqref{eq:stab-upper-bound-1} over $x$ and exploiting the translational invariance of the Green's function \eqref{eq:stab-gamma0-sol-2} to change the variable to $\xi=x-x'$, we find
\begin{equation}\label{eq:stab-upper-bound-2}\begin{split}
& \int\D x\:\norm{\matr{\Delta}^{(t+1)}(x)} \leq \alpha \int\frac{\D\vect{p}_\perp}{(2\pi)^{d-1}}  \\
& \times \iint\D\xi\D x'\: \norm{\matr{\Gamma}^{(0)}(\vect{p}_\perp, \xi)}^2 \norm{\matr{\Delta}^{(t)}(x')}  \:.
\end{split}\end{equation}
Note that, although $\matr{\Gamma}^{(0)}(\vect{p}_\perp,\xi)$ also depends on $x'$ due to the variation of $\matr{Q}^{(0)}(x)$ in Eq.\ \eqref{eq:stab-gamma0-sol-2}, its norm does not because it is fixed by the constant eigenvalues $\pm\pi\nu$ of $\matr{Q}^{(0)}(x)$.
In this way, the ratio between the functions $\matr{\Delta}^{(t+1)}(x)$ and $\matr{\Delta}^{(t)}(x)$ can be isolated,
\begin{equation}\label{eq:stab-upper-bound-3}
\frac{\int\D x\:\norm{\matr{\Delta}^{(t+1)}(x)}}{\int\D x\:\norm{\matr{\Delta}^{(t)}(x)}} \leq R  \:,
\end{equation}
providing the upper bound $R$ defined by
\begin{equation}\label{eq:stab-def-r-integral}
R \defeq \alpha \int\frac{\D\vect{p}_\perp}{(2\pi)^{d-1}} I(\vect{p}_\perp)  \:,
\end{equation}
where $I(\vect{p}_\perp)$ is the following integral over $x$:
\begin{equation}\label{eq:stab-def-i-integral}
I(\vect{p}_\perp) \defeq \int\D x \norm{\matr{\Gamma}^{(0)}(\vect{p}_\perp, x)}^2  \:.
\end{equation}
The matrix norm in Eq.\ \eqref{eq:stab-def-i-integral} can be evaluated from Eq.\ \eqref{eq:stab-gamma0-sol-2} and the fact that the eigenvalues of $\matr{Q}^{(0)}$ are just $\pm\pi\nu$ according to Eq.\ \eqref{eq:stab-q0-gen-sol}.
Therefore, we find
\begin{equation}\label{eq:stab-gamma0-norm}
\norm{\matr{\Gamma}^{(0)}(\vect{p}_\perp,x)}^2 = \frac{\E^{ -2\Im\sqrt{k_{\para}^2 + \I\pi\nu\alpha}\abs{x} }}{4\abs{k_{\para}^2 + \I\pi\nu\alpha}}   \:.
\end{equation}
Integrating Eq.~\eqref{eq:stab-def-i-integral} with Eq.~\eqref{eq:stab-gamma0-norm} yields
\begin{equation}\label{eq:stab-i-1}
I(\vect{p}_\perp) = \frac{1}{4\abs{k_{\para}^2 + \I\pi\nu\alpha} \Im\sqrt{k_{\para}^2 + \I\pi\nu\alpha}}  \:.
\end{equation}
The integral \eqref{eq:stab-def-r-integral} is not known in closed form for the integrand \eqref{eq:stab-i-1}.
To overcome this problem, we consider instead the following upper bound
\begin{equation}\label{eq:stab-i-2}
I(\vect{p}_\perp) < \frac{1}{2k_{\para}\pi\nu\alpha}  \:,
\end{equation}
which turns out to be a good approximation of the function \eqref{eq:stab-i-1} in the weak disorder regime ($k\ell\gg 1$).
Substituting Eq.\ \eqref{eq:stab-i-2} into Eq.\ \eqref{eq:stab-def-r-integral}, we get
\begin{equation}\label{eq:stab-r-1}
R < \frac{1}{\pi\nu} \int\frac{\D\vect{p}_\perp}{(2\pi)^{d-1}} \frac{1}{2k_{\para}}  \:.
\end{equation}
Finally, the integral over the transverse modes in Eq.\ \eqref{eq:stab-r-1} is elementary and leads to
\begin{equation}\label{eq:stab-r-2}
R < 1  \:.
\end{equation}
The result \eqref{eq:stab-r-2} shows that the ratio \eqref{eq:stab-upper-bound-3} is strictly less than one.
Therefore, we deduce that the map \eqref{eq:stab-delta-map} is indeed a contraction.
This implies that the perturbation $\matr{\Delta}^{(t)}(x)$ progressively vanishes for $t\rightarrow\infty$ due to the iteration of the self-consistent equations \eqref{eq:stab-saddle-2},
\begin{equation}\label{eq:stab-delta-to-zero}
\matr{\Delta}^{(t)}(x) \xrightarrow{t\rightarrow\infty} \matr{0}  \:,
\end{equation}
and that the field $\matr{Q}^{(t)}(x)$ converges to the saddle-point field $\matr{Q}^{(0)}(x)$.

\section{Solution in the quasiballistic regime}\label{app:quasiballistic-solution}
It is possible to obtain an approximate analytical result for the transmission eigenvalue distribution by solving the matrix transport equation \eqref{eq:full-eilenberger-2} in the quasiballistic regime ($L\lesssim\ell$).
Since this equation is translationally invariant in the transverse direction, the radiance $\matr{g}$ only depends on $x$ and the direction cosine, which we rename $\mu$ (instead of $\Omega_{\para}$) for the occasion.
The key simplification in the quasiballistic regime is the fact that the matrix field $\tilde{\matr{Q}}(x)$ is almost uniform in the bulk of the disordered region ($x\in[0,L]$).
Indeed, in the quasiballistic regime, the radiance varies at the scale of the mean free path, in contrast to the diffusive regime where it is governed by the system size.
Therefore, we can write
\begin{equation}\label{eq:uniform-field-approx}
\tilde{\matr{Q}}(x) \simeq \tilde{\matr{Q}}_0  \quad\forall x\in[0,L]  \:.
\end{equation}
As a consequence of this assumption, the matrix transport equation \eqref{eq:full-eilenberger-2} can be solved without resorting to path-ordered exponential.
One way to obtain a closed equation for the generating function \eqref{eq:avg-gen-fun-7} is to relate the radiances $\matr{g}(\mu,x)$ at the two contact points $x_\porta$ and $x_\portb$.
According to the matrix transport equation \eqref{eq:full-eilenberger-2}, these two points are related by
\begin{equation}\label{eq:ufa-g-transfer-1}
\matr{g}(\mu,x_\portb^+) = \matr{M} \,\matr{g}(\mu,x_\porta^-) \,\matr{M}^{-1}  \:,
\end{equation}
where $\matr{M}$ is a transfer matrix given in the approximation \eqref{eq:uniform-field-approx} by
\begin{equation}\label{eq:ufa-m-from-q0}
\matr{M} = \E^{\I\gamma_\portb\matr{\Lambda}_-} \E^{-\beta\tilde{\matr{Q}}_0} \E^{\I\gamma_\porta\matr{\Lambda}_+}  \:,
\end{equation}
and
\begin{equation}\label{eq:def-beta-depth}
\beta \defeq \frac{L}{2\ell\mu}  \:.
\end{equation}
Nothing prevents $\beta$ from being negative for backward-propagating modes (those for which $\mu<0$).
Furthermore, the boundary conditions \eqref{eq:boundary-conditions}--\eqref{eq:g-upper-lower-triangular} constrain the expression of $\matr{g}(\mu,x_\portb^+)$ and $\matr{g}(\mu,x_\porta^-)$ in Eq.\ \eqref{eq:ufa-g-transfer-1}:
\begin{equation}\label{eq:ufa-g-transfer-2}\begin{aligned}
& \underbrace{\begin{pmatrix}1 & g^+_{12}\\ 0 & -1\end{pmatrix}}_{\matr{g}(\mu>0,x_\portb^+)} = \matr{M} \underbrace{\begin{pmatrix}1 & 0\\ g^+_{21} & -1\end{pmatrix}}_{\matr{g}(\mu>0,x_\porta^-)} \matr{M}^{-1} \:,\\
& \underbrace{\begin{pmatrix}1 & 0\\ g^-_{21} & -1\end{pmatrix}}_{\matr{g}(\mu<0,x_\portb^+)} = \matr{M} \underbrace{\begin{pmatrix}1 & g^-_{12}\\ 0 & -1\end{pmatrix}}_{\matr{g}(\mu<0,x_\porta^-)} \matr{M}^{-1} \:.
\end{aligned}\end{equation}
In the following equations, the arguments of $g^\pm_{12}$ and $g^\pm_{21}$ are implied.
The two lines of Eq.\ \eqref{eq:ufa-g-transfer-2} can be solved independently for $g^\pm_{12}$ and $g^\pm_{21}$, leading to
\begin{equation}\label{eq:g-from-m-ratios}\begin{aligned}
g^+_{12} & = -\frac{2M_{12}}{M_{22}}  \:,\\
g^+_{21} & = -\frac{2M_{21}}{M_{22}}  \:,
\end{aligned}\qquad\begin{aligned}
g^-_{12} & =  \frac{2M_{12}}{M_{11}}  \:,\\
g^-_{21} & =  \frac{2M_{21}}{M_{11}}  \:.
\end{aligned}\end{equation}
In order to close the equations for the matrix elements of $\matr{g}(\mu,x_\porta^-)$, we need to express $\matr{M}$ and thus $\tilde{\matr{Q}}_0$ in terms of these matrix elements only.
To this end, we use Eq.\ \eqref{eq:qn-jn-from-g-unified} and the boundary conditions \eqref{eq:boundary-conditions}--\eqref{eq:g-upper-lower-triangular} at $x_\porta^-$:
\begin{equation}\label{eq:ufa-qinf-from-g}
\tilde{\matr{Q}}_\porta = \tilde{\matr{Q}}(x_\porta^-) = \begin{pmatrix}1 & \tfrac{1}{2}\avg{g^-_{12}}^-_0\\ \tfrac{1}{2}\avg{g^+_{21}}^+_0 & -1\end{pmatrix}  \:.
\end{equation}
Note that this field does not satisfy the normalization condition of the nonlinear sigma model ($\tilde{\matr{Q}}_\porta^2\neq\matr{1}$) due to the quasiballistic regime.
Since $\tilde{\matr{Q}}_\porta$ and $\tilde{\matr{Q}}_0$ are related by $\tilde{\matr{Q}}_0=\E^{\I\gamma_\porta\matr{\Lambda}_+}\tilde{\matr{Q}}_\porta\E^{-\I\gamma_\porta\matr{\Lambda}_+}$ [see Eq.\ \eqref{eq:contact-cond-qn}], the transfer matrix \eqref{eq:ufa-m-from-q0} can be written directly in terms of $\tilde{\matr{Q}}_\porta$:
\begin{equation}\label{eq:ufa-m-from-qinf}
\matr{M} = \E^{\I\gamma_\portb\matr{\Lambda}_-} \E^{\I\gamma_\porta\matr{\Lambda}_+} \E^{-\beta\tilde{\matr{Q}}_\porta}  \:.
\end{equation}
The transfer matrix \eqref{eq:ufa-m-from-qinf} can now be evaluated explicitly from the matrix exponential of Eq.\ \eqref{eq:ufa-qinf-from-g}. The result is
\begin{equation}\label{eq:ufa-transfer-matrix}\begin{aligned}
M_{11} & = \cosh(\beta\sigma) - \left(1 + \tfrac{\I\gamma_\porta}{2}\avg{g^+_{21}}^+_0\right) \frac{\sinh(\beta\sigma)}{\sigma}  \:,\\
M_{12} & = \left(\I\gamma_\porta - \tfrac{1}{2}\avg{g^-_{12}}^-_0\right) \frac{\sinh(\beta\sigma)}{\sigma} + \I\gamma_\porta \cosh(\beta\sigma)  \:,\\
M_{21} & = \left(\tfrac{1}{2}\avg{g^+_{21}}^+_0 (\gamma_\porta\gamma_\portb - 1) - \I\gamma_\portb\right) \frac{\sinh(\beta\sigma)}{\sigma}  \\
 & + \I\gamma_\portb \cosh(\beta\sigma)  \:,\\
M_{22} & = (1-\gamma_\porta\gamma_\portb) \cosh(\beta\sigma)  \\
 & - \left(\gamma_\porta\gamma_\portb + \tfrac{\I\gamma_\portb}{2}\avg{g^-_{12}}^-_0 - 1\right) \frac{\sinh(\beta\sigma)}{\sigma}  \:,
\end{aligned}\end{equation}
where $\sigma$ is defined as
\begin{equation}\label{eq:ufa-sigma-1}
\sigma \defeq \sqrt{-\det\tilde{\matr{Q}}_\porta} = \sqrt{1 + \frac{1}{4}\avg{g^+_{21}}^+_0\avg{g^-_{12}}^-_0}  \:.
\end{equation}
Inserting Eq.\ \eqref{eq:ufa-transfer-matrix} into Eq.\ \eqref{eq:g-from-m-ratios}, we get a self-consistent system of equations closed for $g^+_{21}$ and $g^-_{12}$.
In order to shorten the expressions, it is appropriate to redefine the unknowns:
\begin{equation}\label{eq:def-ufa-f-g}
g^+_{21}(\mu) = \frac{-2\I\gamma_\portb}{1 - \gamma_\porta\gamma_\portb} f(\mu)   \:,\quad
g^-_{12}(\mu) = 2\I\gamma_\porta g(\mu)  \:.
\end{equation}
Using in addition the relation \eqref{eq:gamma-product}, the system becomes
\begin{equation}\label{eq:ufa-system-1}\begin{cases}
g(\mu) = \frac{1 + \left(1 - \avg{g}^-_0\right) \frac{\tanh(\beta\sigma)}{\sigma}}{1 - \left(1 + \frac{\gamma}{1-\gamma}\avg{f}^+_0\right) \frac{\tanh(\beta\sigma)}{\sigma}}  \:,\\
f(\mu) = \frac{1 - \left(1 - \avg{f}^+_0\right) \frac{\tanh(\beta\sigma)}{\sigma}}{1 + \left(1 + \frac{\gamma}{1-\gamma}\avg{g}^-_0\right) \frac{\tanh(\beta\sigma)}{\sigma}}  \:.
\end{cases}\end{equation}
This system can be simplified further.
Indeed, we notice that the two lines of Eq.\ \eqref{eq:ufa-system-1} are equivalent if we flip the sign of $\mu$ (and then of $\beta$) in the first one.
This implies that $g(-\mu)=f(\mu)$ and thus
\begin{equation}\label{eq:ufa-symmetry}
\avg{g}_0^- = \avg{f}_0^+ 
\end{equation}
satisfy the system of equations.
Therefore, the system \eqref{eq:ufa-system-1} reduces to a single self-consistent equation closed for $f$:
\begin{equation}\label{eq:ufa-system-2}
f(\mu) = \frac{1 - \left(1 - \avg{f}^+_0\right) \frac{\tanh(\beta\sigma)}{\sigma}}{1 + \left(1 + \frac{\gamma}{1-\gamma}\avg{f}^+_0\right) \frac{\tanh(\beta\sigma)}{\sigma}}  \:,
\end{equation}
where $\sigma$ now reads
\begin{equation}\label{eq:ufa-sigma-2}
\sigma = \sqrt{1 + \frac{\gamma}{1 - \gamma} \left(\avg{f}_0^+\right)^2 }  \:.
\end{equation}
Finally, according to Eqs.\ \eqref{eq:avg-gen-fun-7}, \eqref{eq:qn-jn-from-g-unified}, and \eqref{eq:def-ufa-f-g}, the sought generating function is given by
\begin{equation}\label{eq:ufa-gen-fun}
F(\gamma) = \frac{\I}{\gamma_\portb} \tilde{J}_{\para}^{21}(x_\porta)
 = \frac{\I}{2\gamma_\portb} \avg{g^+_{21}}^+_1 
 = \frac{\avg{f}^+_1}{1 - \gamma}   \:.
\end{equation}
Note that in Eq.\ \eqref{eq:ufa-gen-fun}, we used the convention $(\gamma_\porta,\gamma_\portb)=(\gamma,1)$ mentioned earlier below Eq.\ \eqref{eq:gamma-primes-simu}.
As shown in Secs.\ \ref{sec:numerical-waveguide} and \ref{sec:numerical-slab}, Eqs.\ \eqref{eq:ufa-system-2}--\eqref{eq:ufa-gen-fun} provide a relevant approximation of the transmission eigenvalue distribution in the quasiballistic regime.

\section{Matrix transport equation for a waveguide}\label{app:modal-eilenberger}
A key implication of the semiclassical approximation used in Sec.\ \ref{sec:wigner} is that it assumes an infinite number of transverse modes ($N_{\rm p}\rightarrow\infty$), effectively extending the transverse boundaries of the waveguide to infinity.
This assumption is beneficial in that it validates the matrix transport equation in Eq.\ \eqref{eq:full-eilenberger-2} for broader geometries beyond waveguides.
However, it does not accurately describe waveguides of finite width, so the matrix transport equation \eqref{eq:full-eilenberger-2} must be modified accordingly to account for a finite number of modes.
To this end, we project the saddle-point equation \eqref{eq:saddle-point-2} on the basis of transverse modes before performing the Wigner transform in the variable $x$.
We introduce for this purpose the modal Wigner transform
\begin{equation}\label{eq:gamma-wigner-modal}
\matr{\Gamma}_{n}(p_{\para},x) = \int_{\mathbb{R}} \D{s} \bra{\chi_{n},x+\tfrac{s}{2}}\op{\matr{\Gamma}}\ket{\chi_{n},x-\tfrac{s}{2}} \E^{-\I p_{\para}s}  \:,
\end{equation}
where $\chi_{n}(\vect{y})$ are the transverse modes initially defined in Eq.\ \eqref{eq:def-transverse-modes}.
Note that $\op{\matr{\Gamma}}$ is diagonal in the modal space due to the translational symmetry of the saddle-point Hamiltonian \eqref{eq:saddle-point-2} in the transverse direction.
By analogy with Eq.\ \eqref{eq:def-g-radiance-alt}, we define the modal radiance
\begin{equation}\label{eq:def-g-radiance-modal}
\matr{g}^\pm_{n}(x) \defeq \frac{2k_{\para,n}}{\pi} \dashint_{\pm k_{\para,n}} \D{p_{\para}}\, \matr{\Gamma}_{n}\!\left(p_{\para}, x\right)  \:.
\end{equation}
The $\matr{g}^+_{n}(x)$ radiance propagates in the direction of increasing $x$ and the $\matr{g}^-_{n}(x)$ radiance in the direction of decreasing $x$.
The definition \eqref{eq:def-g-radiance-modal} ensures the normalization property $\matr{g}^\pm_{n}(x)^2=\matr{1}_2$ equivalent to Eq.\ \eqref{eq:g-normalization} but for the modes.
It is thus different from the previous definition \eqref{eq:def-g-radiance-alt}.
In the representation \eqref{eq:gamma-wigner-modal}--\eqref{eq:def-g-radiance-modal}, the saddle-point equation \eqref{eq:saddle-point-2} reduces to a matrix transport equation very similar to Eq.\ \eqref{eq:full-eilenberger-2}, that is
\begin{equation}\label{eq:full-eilenberger-modal}\begin{split}
\mu_n\partial_{x}\matr{g}^\pm_n = & \mp\tfrac{1}{2\ell} [\tilde{\matr{Q}}(x), \matr{g}^\pm_n] \mp \tfrac{\varepsilon}{2k}[\lmat_3, \matr{g}^\pm_n]  \\
 & + \I\gamma_\porta\mu_n\delta(x-x_\porta) [\lmat_+, \matr{g}^\pm_n]  \\
 & + \I\gamma_\portb\mu_n\delta(x-x_\portb) [\lmat_-, \matr{g}^\pm_n]  \:,
\end{split}\end{equation}
where $\mu_n=k_{\para,n}/k$ is the direction cosine of mode $n$ (defined as positive).
Of course, Eq.\ \eqref{eq:full-eilenberger-modal} must be supplemented by the boundary conditions \eqref{eq:boundary-conditions}--\eqref{eq:g-upper-lower-triangular} which remain the same in the modal basis.
However, it also needs the self-consistent condition \eqref{eq:self-consistency-3} which possesses a noticeably different expression in this basis.
In order to derive this expression, we consider the modal expansion
\begin{equation}\label{eq:gamma-modal-wigner}
\matr{\Gamma}(\vect{p},\vect{r}) = \sum_{n=1}^{N_{\rm p}} \matr{\Gamma}_{n}(p_{\para},x) W_{\chi_n}(\vect{p}_\perp,\vect{y})  \:,
\end{equation}
where $W_{\chi_n}$ is the transverse Wigner transform of the modes given by $W_{\chi_n}(\vect{p}_\perp,\vect{y})=\wigner{\ket{\chi_{n}}\bra{\chi_{n}}}_{\vect{p}_\perp,\vect{y}}$.
We will assume that this function is approximately independent of the transverse coordinate $\vect{y}$ so that $\textstyle\int \D\vect{p}_\perp \,W_{\chi_n}(\vect{p}_\perp,\vect{y}) \simeq (2\pi)^{d-1}/S_\perp$.
This independence property is exact only when the transverse boundary conditions are periodic.
Furthermore, it is worth noting that the restriction of the sum in Eq.\ \eqref{eq:gamma-modal-wigner} to the propagating modes is an approximation.
In principle, this sum should cover all the transverse modes including the evanescent modes for which $\norm{\vect{p}_{\perp,n}}>k$.
However, these evanescent modes cannot be described semiclassically, because of the absence of a carrier wave, and must therefore be neglected in this theory.
\par Substituting Eq.\ \eqref{eq:gamma-modal-wigner} into Eq.\ \eqref{eq:self-consistency-2} and using Eq.\ \eqref{eq:def-g-radiance-modal}, we get
\begin{equation}\label{eq:q-modal-2}
\matr{Q}(x) = \frac{1}{2S_\perp} \sum_{n=1}^{N_{\rm p}} \frac{\matr{g}^+_{n}(x) + \matr{g}^-_{n}(x)}{2k_{\para,n}}  \:.
\end{equation}
However, in the matrix transport equation \eqref{eq:full-eilenberger-modal} we need the normalized field $\tilde{\matr{Q}}$ defined by Eq.\ \eqref{eq:def-normalized-q} and not $\matr{Q}$.
Therefore, a division by the density of states is required. In a finite-width waveguide, it is given by
\begin{equation}\label{eq:dos-waveguide}
\nu = \frac{1}{2\pi S_\perp} \sum_{n=1}^{N_{\rm p}} \frac{1}{k_{\para,n}}  \:.
\end{equation}
In the semiclassical limit ($N_{\rm p}\rightarrow\infty$), one can verify that the density of states \eqref{eq:dos-waveguide} tends indeed to the free-space result \eqref{eq:free-dos}.
The normalized field thus reads
\begin{equation}\label{eq:qn-modal}
\tilde{\matr{Q}}(x) = \frac{1}{\sum_{n=1}^{N_{\rm p}} \frac{1}{\mu_n}} \sum_{n=1}^{N_{\rm p}} \frac{\matr{g}^+_n(x) + \matr{g}^-_n(x)}{2\mu_n}  \:.
\end{equation}
Following the same procedure for the normalized matrix current given by Eqs.\ \eqref{eq:j-matrix-from-g} and \eqref{eq:def-normalized-current} yields
\begin{equation}\label{eq:jn-modal}
\tilde{\matr{J}}_{\para}(x) = \frac{1}{N_{\rm p}} \sum_{n=1}^{N_{\rm p}} \frac{\matr{g}^+_{n}(x) - \matr{g}^-_{n}(x)}{2}  \:.
\end{equation}
Equations \eqref{eq:qn-modal} and \eqref{eq:jn-modal} are compatible with the general expressions \eqref{eq:qn-jn-from-g-unified} up to a redefinition of the directional mean $\avg{\cdot}^\pm_\kappa$ by
\begin{equation}\label{eq:directional-mean-discrete}
\avg{A}^\pm_\kappa = \frac{\sum_{n=1}^{N_{\rm p}} \mu_n^{\kappa-1} A^\pm_n}{\sum_{n=1}^{N_{\rm p}} \mu_n^{\kappa-1}}  \:.
\end{equation}
Note that the exponent in Eq.\ \eqref{eq:directional-mean-discrete} is not $\kappa$ as in the continuous definition \eqref{eq:directional-mean-continuous} but $\kappa-1$.
This difference arises from the distinct definitions used for the modal and continuous radiances in Eqs.\ \eqref{eq:def-g-radiance-modal} and \eqref{eq:def-g-radiance-alt}, respectively.

\section{Numerical integration of the matrix transport equation}\label{app:integration}
\par In this appendix, we detail the numerical integration method of the matrix transport equation \eqref{eq:full-eilenberger-2} implemented in {\sc Ebsolve} \cite{GaspardD2025-ebsolve}.
We assume that, at one single step of the iterative procedure, the matrix field $\tilde{\matr{Q}}(\vect{r})$ is fixed, so that the problem reduces to an integration over space.
This spatial integration is far from trivial for at least two reasons: First, because of the peculiar boundary conditions \eqref{eq:boundary-conditions}--\eqref{eq:g-upper-lower-triangular}, and second because of the well-known numerical instability of the Eilenberger equation in the superconductivity literature \cite{Schopohl1995, Schopohl1998-arxiv, Belzig1999, Nagai2012}.
\par
\begin{figure}[ht]%
\includegraphics{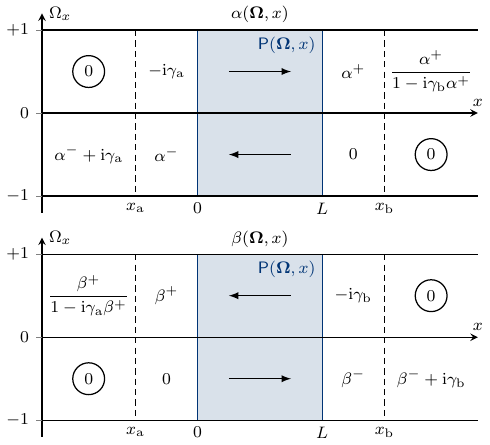}%
\caption{Behavior of the Schopohl parameters $\alpha(\vect{\Omega},x)$ and $\beta(\vect{\Omega},x)$ in the waveguide geometry, with a disordered region depicted in blue.
The circles highlight the boundary conditions \eqref{eq:schopohl-boundary-cond}, and the arrows the direction of numerically stable integration.
The values $\alpha^\pm$ and $\beta^\pm$ self-consistently depend on the $\tilde{\matr{Q}}(x)$ field in the disordered region so that they are not known in general.}%
\label{fig:schopohl-param}%
\end{figure}

 The first source of numerical instability in Eq.\ \eqref{eq:full-eilenberger-2} comes from the existence of constraints imposed by Eqs.\ \eqref{eq:g-traceless} and \eqref{eq:g-normalization} which reduce the number of degrees of freedom of the matrix radiance $\matr{g}(\vect{\Omega},\vect{r})$ from four complex numbers to only two.
If precautions are not taken, the naive integration of Eq.\ \eqref{eq:full-eilenberger-2} may result in a violation of these constraints due to the accumulation of roundoff errors.
This problem can be avoided by an appropriate parametrization of the matrix radiance.
Of course, there are many different ways to parametrize $\matr{g}(\vect{\Omega},\vect{r})$ with two parameters such that the constraints \eqref{eq:g-traceless} and \eqref{eq:g-normalization} are automatically satisfied.
We first consider a well-known parametrization in the superconductivity literature, namely the Schopohl parametrization \cite{Schopohl1995, Schopohl1998-arxiv},
\begin{equation}\label{eq:def-schopohl-param}
\matr{g}(\vect{\Omega},\vect{r}) = \frac{1}{1 + \alpha\beta} \begin{pmatrix}1 - \alpha\beta & 2\alpha\\ 2\beta & -1 + \alpha\beta\end{pmatrix}  \:,
\end{equation}
where $\alpha(\vect{\Omega},\vect{r})$ and $\beta(\vect{\Omega},\vect{r})$ are two complex functions of the direction and the position.
In the following calculations, we will establish the equations of propagation for $\alpha$ and $\beta$ in the waveguide geometry so that these functions only depend on the longitudinal coordinate $x$.
Although not perfect as we will see, the special parametrization \eqref{eq:def-schopohl-param} offers many computational advantages, in particular the expressions of the boundary conditions at infinity, Eqs.\ \eqref{eq:boundary-conditions}--\eqref{eq:g-upper-lower-triangular}, reduce to simple cancellations of the Schopohl parameters in a symmetrical fashion:
\begin{equation}\label{eq:schopohl-boundary-cond}
\begin{aligned}
\alpha^+_{x\rightarrow-\infty} = 0  \:,\\
 \beta^-_{x\rightarrow-\infty} = 0  \:,
\end{aligned}\qquad%
\begin{aligned}
\alpha^-_{x\rightarrow+\infty} = 0  \:,\\
 \beta^+_{x\rightarrow+\infty} = 0  \:,
\end{aligned}
\end{equation}
where the superscripts refer to the sign of $\Omega_{\para}$, i.e., the direction.
These boundary conditions are schematically depicted in Fig.\ \ref{fig:schopohl-param}.
The boundary conditions \eqref{eq:schopohl-boundary-cond} strongly suggest to integrate the functions $\alpha(\vect{\Omega},x)$ and $\beta(\vect{\Omega},x)$ starting from the points where they are zero.
This approach is not only the most convenient, it is also the most numerically stable because the functions are expected to increase in absolute value.
Integrating in the decreasing direction is often numerically unstable due to the accumulation of roundoff errors which prevent reaching the true zero.
The consequence of this approach is that $\alpha(\vect{\Omega},x)$ must be integrated in the same direction as $\vect{\Omega}$, while $\beta(\vect{\Omega},x)$ must be integrated in the opposite direction \cite{Schopohl1998-arxiv}.
\par We have to determine the equations of propagation for $\alpha$ and $\beta$.
It can be shown by substituting Eq.\ \eqref{eq:def-schopohl-param} into the matrix transport equation \eqref{eq:full-eilenberger-2} that, along the longitudinal axis of the waveguide, the Schopohl parameters $a$ and $b$ obey two apparently uncoupled Riccati equations \cite{Schopohl1995, Schopohl1998-arxiv, Belzig1999, Nagai2012},
\begin{equation}\label{eq:eilenberger-riccati}
\begin{cases}
\partial_{x}\alpha = +P_{21} \alpha^2 + 2P_{11} \alpha - P_{12}  \:,\\
\partial_{x}\beta  = -P_{12} \beta^2  - 2P_{11} \beta  + P_{21}  \:,
\end{cases}
\end{equation}
where the matrix $\matr{P}$ is defined by
\begin{equation}\label{eq:def-p-matrix}\begin{split}
\matr{P}(\vect{\Omega},x) \defeq & -\tfrac{1}{2\ell\Omega_{\para}} \tilde{\matr{Q}}(x) - \tfrac{\varepsilon}{2k\Omega_{\para}} \lmat_3  \\
& + \I\gamma_\porta\delta(x-x_\porta) \lmat_+ + \I\gamma_\portb\delta(x-x_\portb) \lmat_-  \:.
\end{split}\end{equation}
In fact, the two equations \eqref{eq:eilenberger-riccati} are coupled under the hood by the $\tilde{\matr{Q}}(x)$ field according to the self-consistent condition \eqref{eq:self-consistency-3} and the parametrization \eqref{eq:def-schopohl-param}.
The issue with the Riccati equations \eqref{eq:eilenberger-riccati} is that they cannot be integrated using traditional integrators (such as the Runge-Kutta methods) because they are intrinsically unstable due to the presence of singularities in the solution for $\det(\matr{P})\in\mathbb{R}^+$.
Indeed, the solution of the simple Riccati equation $\partial_{x}f(x) = f(x)^2$ is $f(x)=\tfrac{1}{x_0-x}$ which has a distinct pole at $x_0$ (the integration constant).
Fortunately, the instability of Eq.\ \eqref{eq:eilenberger-riccati} can be overcome through a specific integration method that we now present.
\par As proposed by Schopohl \cite{Schopohl1998-arxiv, Nagai2012}, a more appropriate way to integrate the matrix transport equation in the parametrization \eqref{eq:def-schopohl-param} is to introduce vector parameters
\begin{equation}\label{eq:def-schopohl-vectors}
\vect{a}(\vect{\Omega},\vect{r}) \defeq \begin{pmatrix}a_1\\ a_2\end{pmatrix}  \:,\qquad
\vect{b}(\vect{\Omega},\vect{r}) \defeq \begin{pmatrix}b_1\\ b_2\end{pmatrix}  \:.
\end{equation}
such that
\begin{equation}\label{eq:schopohl-from-a-b}
\alpha = -\frac{a_1}{a_2}  \:,\qquad
\beta  =  \frac{b_2}{b_1}  \:.
\end{equation}
The interest of $\vect{a}$ and $\vect{b}$ is that they obey much simpler equations than $\alpha$ and $\beta$.
Indeed, if we substitute Eq.\ \eqref{eq:schopohl-from-a-b} into the Riccati equations \eqref{eq:eilenberger-riccati}, we get
\begin{equation}\label{eq:eilenberger-vector}\begin{cases}
\partial_{x}\vect{a}(\vect{\Omega},\vect{r}) = \matr{P}(\vect{\Omega},\vect{r}) \vect{a}(\vect{\Omega},\vect{r})  \:,\\
\partial_{x}\vect{b}(\vect{\Omega},\vect{r}) = \matr{P}(\vect{\Omega},\vect{r}) \vect{b}(\vect{\Omega},\vect{r})  \:,
\end{cases}\end{equation}
The reappearance of the matrix $\matr{P}$ in Eq.\ \eqref{eq:eilenberger-vector} is not due to chance.
In fact, the Schopohl parametrization \eqref{eq:def-schopohl-param} can be constructed by considering an arbitrary similarity transformation of the vacuum state $\lmat_3$:
\begin{equation}\label{eq:schopohl-transfo}
\matr{g} = \matr{M} \lmat_3 \matr{M}^{-1}  \:,\qquad
\matr{M} = \begin{pmatrix}b_1 & a_1\\ b_2 & a_2\end{pmatrix}  \:.
\end{equation}
The expression of $\matr{g}$ in Eq.\ \eqref{eq:schopohl-transfo} can be substituted into the matrix transport equation \eqref{eq:full-eilenberger-2} to obtain the system \eqref{eq:eilenberger-vector}.
The important point with the system \eqref{eq:eilenberger-vector} is that, in contrast to the Riccati equations \eqref{eq:eilenberger-riccati}, it is linear (if we forget about the self-consistency of the $\matr{Q}$ field) in such way that this differential problem is numerically stable as far as we respect the direction of integration imposed by the boundary conditions \eqref{eq:schopohl-boundary-cond}.
The latter now read
\begin{equation}\label{eq:vector-boundary-cond}
\begin{aligned}
\vect{a}^+_{x\rightarrow-\infty} = \begin{pmatrix}0\\ 1\end{pmatrix}  \:,\\
\vect{b}^-_{x\rightarrow-\infty} = \begin{pmatrix}1\\ 0\end{pmatrix}  \:,
\end{aligned}\qquad%
\begin{aligned}
\vect{a}^-_{x\rightarrow+\infty} = \begin{pmatrix}0\\ 1\end{pmatrix}  \:,\\
\vect{b}^+_{x\rightarrow+\infty} = \begin{pmatrix}1\\ 0\end{pmatrix}  \:.
\end{aligned}
\end{equation}
According to these boundary conditions, $\vect{a}^+$ and $\vect{b}^-$ must be integrated in the forward direction and $\vect{a}^+$ and $\vect{b}^-$ in the backward direction to ensure numerical stability.
\par Regarding the numerical integration of Eq.\ \eqref{eq:eilenberger-vector}, many methods may prove satisfactory, including the Runge-Kutta methods.
However, the structure of the equation encourages the use of an exponential integrator, which is a method based on the exact solution for constant $\matr{P}$ between two consecutive points on the regular mesh $\{x_i=x_1+(i-1)\Delta{x}\}_{i=1,\ldots,N_x}$,
\begin{equation}\label{eq:exp-integrator-1}
\vect{v}(\vect{\Omega},x_{i+1}) = \E^{\matr{P}_{i}\Delta{x}} \vect{v}(\vect{\Omega},x_{i})  \:,
\end{equation}
whether $\vect{v}$ is equal to $\vect{a}$ or $\vect{b}$.
Furthermore, we known that the exponential of the traceless $2\times 2$ matrix $\matr{P}\Delta{x}$ is given by
\begin{equation}\label{eq:matrix-exp-traceless}
\E^{\matr{P}\Delta{x}} = \cosh(\sigma) \matr{1}_2 + \frac{\sinh(\sigma)}{\sigma} \matr{P}\Delta{x}  \:,
\end{equation}
where $\sigma$ reads
\begin{equation}\label{eq:matrix-exp-sigma}
\sigma = \sqrt{-\det(\matr{P})} \Delta{x} = \sqrt{P_{11}^2 + P_{12}P_{21}} \Delta{x}  \:.
\end{equation}
In addition, it is interesting to note that, in numerical calculations, the vectors $\vect{a}$ and $\vect{b}$ can be integrated up to a global factor since the latter disappears anyway in the calculation of $\alpha$ and $\beta$ by Eq.\ \eqref{eq:schopohl-from-a-b}.
Therefore, it is possible to reduce the number of hyperbolic functions in Eqs.\ \eqref{eq:exp-integrator-1}--\eqref{eq:matrix-exp-traceless}:
\begin{equation}\label{eq:exp-integrator-2}
\vect{v}(\vect{\Omega},x_{i+1}) = \left( \matr{1}_2 + \frac{\tanh(\sigma)}{\sigma} \matr{P}_{i}\Delta{x} \right) \vect{v}(\vect{\Omega},x_{i})  \:.
\end{equation}
In practice, it is convenient to separate the contribution of the contact interactions from the rest of the bulk integration due to the singular nature of the Dirac deltas in Eq.\ \eqref{eq:def-p-matrix}.
This contribution is given by Eq.\ \eqref{eq:contact-cond-g}:
\begin{equation}\label{eq:contact-cond-transfer-m}\begin{cases}
\vect{v}(\vect{\Omega},x_\porta^+) = \E^{\I\gamma_\porta\lmat_+} \vect{v}(\vect{\Omega},x_\porta^-) \:,\\
\vect{v}(\vect{\Omega},x_\portb^+) = \E^{\I\gamma_\portb\lmat_-} \vect{v}(\vect{\Omega},x_\portb^-) \:.
\end{cases}\end{equation}
Equations \eqref{eq:exp-integrator-2} and \eqref{eq:contact-cond-transfer-m} lie at the core of {\sc Ebsolve}'s integrator \cite{GaspardD2025-ebsolve}.
\par We end this appendix with two comments.
First, the integration method of the Riccati equation \eqref{eq:eilenberger-riccati} for $\alpha$ and $\beta$ based on Eqs.\ \eqref{eq:def-schopohl-vectors}--\eqref{eq:eilenberger-vector} is equivalent to the integrator based on Möbius transformations proposed by Ref.\ \cite{Schiff1999} to solve numerically Riccati-type equations.
Indeed, it can be shown that the exact solution of Eq.\ \eqref{eq:eilenberger-riccati} for piecewise-constant $\matr{P}$ can be expressed as a succession of Möbius transformations for $\alpha$ and $\beta$.
The matrix representation of these transformations is precisely given by Eq.\ \eqref{eq:exp-integrator-1}.
\par Second, we stress that all the equations of this appendix are more general than what we have suggested and are not restricted to the waveguide geometry.
In fact, they can be interpreted more generally as the equations along individual rays (of fixed direction $\vect{\Omega}$) if we replace $x/\Omega_{\para}$ by the traveled path $s$ \cite{Schopohl1998-arxiv, Nagai2012}.


\begin{thebibliography}{76}
\bibitem [{Gas()}]{GaspardD2025-qfield}%
  \BibitemOpen
  \bibinfo {note} {The codes to solve the saddle-point equation in one- and
  two-dimensional waveguides are available upon request from the
  authors.}\BibitemShut {Stop}%
\end{thebibliography}
\end{document}